\documentclass[aps,onecolumn]{revtex4}

\usepackage{psfig}
\begin{document}

Published in {\em Modern Nonlinear Optics}, eds. M. Evans and
S. Kielich,\\ {\em Advances in Chemical Physics}, vol. {\bf
85}(III) (Wiley, New York, 1994) p.531--626.

\title{Quantum-statistical theory of Raman scattering
processes}

\author{A. Miranowicz and S. Kielich}

\affiliation{Nonlinear Optics Division, Institute of Physics,
Adam Mickiewicz University, 60-780 Pozna\'n, Poland}

\begin{abstract}
Raman scattering from a great number of phonon modes is
described from a quantum-statistical point of view within the
standing-wave model. The master equation for the completely
quantum case, including laser pump depletion and stochastic
coupling of Stokes and anti-Stokes modes, is derived and
converted to classical equations: either into a generalized
Fokker-Planck equation and an equation of motion for the
characteristic function or into the master equation in Fock
representation. These two approaches are developed both in
linear and nonlinear r\'egime. A detailed analysis of
scattering into Stokes and anti-Stokes modes in linear
r\'egime, i.e., under parametric approximation, is presented.
The existence of $s$-parametrized quasiprobability
distributions, in particular the Glauber-Sudarshan
$P$-function, is investigated. An analysis of Raman
scattering into separate Stokes and anti-Stokes modes in
nonlinear r\'egime, thus including pump depletion, is given.
The master equation in Fock representation is solved exactly
for the complete density matrix using the Laplace transform
method. Short-time solutions, steady-state solutions and
approximate compact form solutions are obtained. Relations
between the quasidistribution approach based on the
Fokker-Planck equation and the density matrix approach based
on the master equation in Fock representation are presented.
The photocount distribution and its factorial moments as well
as variances and extremal variances of quadratures are
calculated in both approaches giving the basis for the
analysis of the quantum properties of radiation such as
sub-Poissonian photon-counting statistics and squeezing. A
comparison of various statistical moments obtained from
numerical calculations utilizing our exact solution of the
master equation and from the approximate relations for short
times, as well as obtained under parametric approximation, is
presented graphically.
\end{abstract}

\maketitle

\tableofcontents

\section{Introduction: Historical developments}

Almost simultaneously in 1928 Raman and
Krishnan~\cite{r193a,r193b} and Landsberg and
Mandel'stamm~\cite{r194} observed a new kind of scattering,
now referred to as (spontaneous) Raman scattering. For the
last 65 years Raman scattering has unceasingly been in the
forefront of both scientific and experimental investigations,
particularly after the first observation of stimulated Raman
scattering by Woodbury and Ng~\cite{r195} (see also
Ref.~\cite{r196}). Without exaggeration one can say that
Raman scattering and spectroscopy constitute a completely
autonomous discipline.

The literature on Raman scattering is quite prodigious. The
theoretical principles and milestone experiments describing
the Raman effect are summarized in a number of excellent
monographs and reviews, for instance, by
Bloembergen~\cite{r12}, Kaiser and Maier~\cite{r184},
Koningstein~\cite{r189}, Grasyuk~\cite{r185,r186}, Wang
\cite{r136,r137}, Cardona~\cite{r191}, Long~\cite{r190},
Hayes and Loudon~\cite{r192}, Penzkofer et al. \cite{r188},
Kielich~\cite{r46,r294,r44}, Shen~\cite{r154}, D'yakov and
Nikitin~\cite{r187}, and the most recent reviews by Raymer
and Walmsley~\cite{r91}, Pe\v{r}ina~\cite{r79}, and Mostowski
and Raymer~\cite{r91a}.  We also refer the Reader to the
special issue of the {\em Journal of the Optical Society of
America  B}~\cite{r258} which is devoted entirely to Raman
scattering. Although an extensive literature has accumulated
dealing with Raman scattering, it should be emphasized that
the understanding of the fundamental principles that govern
the process is still incomplete.

There are several major groups of theories treating the Raman
effect in the semiclassical and quantum approaches, and
theories for standing waves and spatially propagating waves.
Here, we discuss in detail the quantum theory of Raman
scattering for several radiation modes only; this implies
that the theory is the best suited for scattering in a tuned
cavity.  Nevertheless, some predictions from the standing
wave model also can be applied for traveling wave
models~\cite{r97,r110,r3,r4}.

Various methods have been applied to the Raman effect in each
of the above theories.  Taking into account the equation of
motion as the basis for classification, we can distinguish
the following approaches, based on the photon rate equation,
the Schr\"odinger equation, the Heisenberg equation
(Heisenberg-Langevin equation), the master equation
(generalized Fokker-Planck equation), and the
Maxwell-Heisenberg equation (Maxwell-Bloch equation); we
refer to Refs.~\cite{r91,r79,r57,r341}. The above
classification is obviously oversimplified. Firstly, there
are many relations bridging these approaches. For instance,
we shall apply the master equation approach from which we
shall derive the Fokker-Planck equation and the photon rate
equation. Secondly, there exist other alternative methods,
which do not fit into our classification. Let us mention, for
example, those developed by
Mavroyannis~\cite{r197a,r197b,r197c} and
Freedhoff~\cite{r198}. Thirdly, one can classify the Raman
effect theories in many other ways (see, e.g.,
Ref.~\cite{r91}).

We shall be considering the incident laser photons to be
scattered by chaotic phonons or quantized chaotic vibrations
in a crystal. The process leads to Stokes and anti-Stokes
photons. To the description of Raman scattering, we use two
trilinear Hamiltonians coupled via an infinite number of
phonon modes; one Hamiltonian describes Stokes radiation, and
the other describes anti-Stokes radiation. The problem of
coupled Stokes and anti-Stokes modes were been studied
previously by Bloembergen and Shen~\cite{r173,r174,r99} who
applied the coupled wave theory of nonlinear optics
formulated by Armstrong et al.~\cite{r99a}. Later, Mishkin
and Walls~\cite{r70} quantized the Stokes and anti-Stokes
modes, but dealt with the laser mode as a constant amplitude
(so-called the parametric approximation). In fact, they
considered two bilinear Hamiltonians, coupled by way of a
phonon mode. Stokes scattering was treated as a parametric
amplifier, whereas anti-Stokes scattering was treated as a
parametric frequency converter. A detailed study of quantum
statistics of the bilinear Hamiltonians, proposed by Louisell
et al.~\cite{r344}, has been extensively carried out e.g.,
Ref.~\cite{r71,r71a,r178,r71b,r345,r85a,r85b,r18,r19} and
applied to Raman scattering in particular by Mishkin and
Walls~\cite{r70}, Walls~\cite{r129,r132a},
Pe\v{r}ina~\cite{r77,r78,r163}, K\'arsk\'a and
Pe\v{r}ina~\cite{r41} and others. Walls~\cite{r130} (see also
Ref.~\cite{r178}) has extended the bilinear Hamiltonian to a
trilinear form to describe Raman scattering. The dynamics of
Raman processes with trilinear Hamiltonian has been studied
by Szlachetka et al.~\cite{r114,r114a,r113}, Szlachetka and
Kielich~\cite{r114b}, Szlachetka~\cite{r110}, Trung and
Sch\"utte~\cite{r118a}, T\"anzler and Sch\"utte~\cite{r118},
Reis and Sharma~\cite{r92}, Pe\v{r}ina et
al.~\cite{r82,r264}, Pe\v{r}ina and
K\v{r}epelka~\cite{r81,r81a}, Levenson et al.~\cite{r53} and
others (for general analyses see also
Refs.~\cite{r6,r96,r79}).

We shall describe Raman scattering from phonons as collective
phenomena involving the interaction of many molecules. Much
attention has also been drawn to a microscopic picture of the
Raman effect by considering the interaction with individual
molecules. Shen~\cite{r97,r98}, in his quantum-statistical
theory of nonlinear phenomena, proposed the general $m+n$
photon Hamiltonian, describing $m$ emissions and $n$
absorptions, and atomic transitions of an ensemble of $N$
$f$-level atoms. This microscopically correct Hamiltonian
contains Bose operators of a field and Fermi operators for
optically active electrons and therefore describes a variety
of nonlinear phenomena, in particular Raman scattering (for
two- or three-level atoms). The same general Hamiltonian has
been used by Walls~\cite{r132a} and McNeil and
Walls~\cite{r66}. Raman scattering from a two-level molecular
(atomic)
system~\cite{r131,r100,r342,r244,r318,r319,r255,r256,r34,r33,r8}
and in a three-level molecular
system~\cite{r131,r131a,r343,r5,r33a,r298,r148,r126,r339} has
been extensively studied by various authors.
Walls~\cite{r131} has shown that a description of Raman
scattering from two-level molecules with a large cooperation
number (coherent molecular coupling) is markedly similar to
the results for Raman scattering from phonons. This is
because for coherent molecular coupling sums of the Fermi
operators for the individual molecules can be replaced by the
collective operators approximately satisfying boson
commutation relations.

Nonclassical properties of radiation, such as squeezing,
sub-Poissonian photon statistics and photon antibunching,
remain central topics in quantum optics.  The literature in
this area is truly prodigious.  The Reader is referred to the
articles published in Vol. 85 of this series and references
therein, for instance Refs.~\cite{r299,r300,r301}, as well as
the reviews by Kielich et al.~\cite{r295},
Leuchs~\cite{r292}, Loudon and Knight~\cite{r58}, Teich and
Saleh~\cite{r121,r180}, Zaheer and Zubairy~\cite{r143} and
the topical issues of the {\em Journal of the Optical Society
of America B}~\cite{r1} and the {\em Journal of  Modern
Optics}~\cite{r2}.

Squeezing properties of Raman scattering have been studied by
Pe\v{r}inov\'a and et al. Pe\v{r}ina, K\'arsk\'a and
Pe\v{r}ina~\cite{r41}, Levenson et al. and Pe\v{r}ina and
K\v{r}epelka~\cite{r81,r81a}. Sub-Poissonian photon-counting
statistics and/or photon anticorrelations (in particular
antibunching) have been investigated within various
approaches to the standing-wave Raman effect by
Loudon~\cite{r57}, Simaan~\cite{r100}, Agarwal and
Jha~\cite{r5}, Trung and Sch\"utte~\cite{r118b}, Szlachetka
and Kielich~\cite{r114b}, Szlachetka et
al.~\cite{r114,r114a,r113}, Gupta and
Mohanty~\cite{r318,r319}, Pe\v{r}ina~\cite{r77,r78},
T\"anzler and Sch\"utte~\cite{r118}, Germey et
al.~\cite{r30}, Mohanty et al.~\cite{r255},
Kr\'al~\cite{r289}, Gupta and Dash~\cite{r148,r144}, Ritsch
et al.~\cite{r339} and in papers already
mentioned~\cite{r82,r264,r81a,r163,r41,r81} (see also
Ref.~\cite{r79}). We note that photons scattered in the
hyper-Raman effect can also exhibit nonclassical
photon-counting
correlations~\cite{r144,r101,r310,r340,r87,r118a,r320,r30a,r149,r161,r162},~\cite{r113,r110,r82}
or squeezing~\cite{r82,r30a,r162}.

We shall analyze, in particular, cross-fluctuations
(cross-correlations) in quadratures and in photocount
statistics between different radiation modes. The theory of
coherent light scattering within the consistent multipole
tensor formalism, developed by Kielich~\cite{r293} (see also
Refs.~\cite{r110,r113,r114}) was successfully applied to
disclose a novel cross-fluctuation mechanism. Here, an
analysis of cross-correlations is presented along the lines
of Szlachetka et al.~\cite{r114,r114a} (see also
Ref.~\cite{r79}), as well as Loudon~\cite{r57}.

We shall be studying sub- or super-Poissonian photon-counting
statistics.  We shall not analyze photon antibunching or
bunching. The inclusion, in our Raman scattering model, of
standard (i.e., temporal) photon antibunching would pose no
problem.  Let us mention the difference between
sub-Poissonian statistics and anti-bunching pointed out in
Refs.~\cite{r265}and~\cite{r266}, which enables us to claim
that these are distinct phenomena, and definitions should not
be confused.  The Raman scattering model is not suitable for
investigations of spatial antibunching as defined by Le
Berre-Rousseau et al.~\cite{r51} and Bia{\l}ynicka-Birula et
al.~\cite{r10} in terms of negative angular correlations of
photons.

As stated above, we shall be considering the quantum
statistics of Raman scattering from phonons. We shall
concentrate on a statistical analysis within the master
equation approach to the Raman effect proposed by
Shen~\cite{r97}, Walls~\cite{r132b} and McNeil and
Walls~\cite{r66}. This approach has been studied by various
authors, e.g.,  Simaan~\cite{r100}, Schenzle and
Brand~\cite{r341}, Pe\v{r}ina~\cite{r77,r78,r79}, Germey et
al.~\cite{r30}, Gupta and Dash~\cite{r144}, Bogolubov et
al.~\cite{r33a}, Grygiel~\cite{r33}, Miranowicz~\cite{r8},
and K\'arsk\'a and Pe\v{r}ina~\cite{r41}. Usually a master
equation is converted to a classical differential equation.
Here, we shall apply a transformation to a Fokker-Planck
equation (FPE) for $s$-parametrized quasidistributions using
the coherent-state technique and an alternative method of a
master equation in terms of Fock states (or a rate equation
for the conditional photon-number probabilities).

Walls~\cite{r132b} was the first to apply the FPE technique
to Raman scattering. This approach was extensively developed
by Pe\v{r}ina and coworkers (Ref.~\cite{r79} and references
therein). Unfortunately, a FPE for the Raman effect has been
solved exactly under parametric approximation only; i.e., a
pump depletion was not included. It means that Raman
scattering is described as a competing process of parametric
amplification (Stokes scattering) and parametric frequency
conversion of light (anti-Stokes scattering) in a nonlinear
crystal. This approximation seems to be a real shortcoming of
the FPE approach. A problem of the existence of a solution of
the FPE also arises. A diffusion matrix of the FPE for the
$s$-parametrized quasidistributions (with $s\approx 1$) in
many cases is not positive or positive semidefinite.
Therefore, such a FPE cannot be interpreted as an equation of
motion describing the Brownian motion under the influence of
a suitable force~\cite{r106}. For this reason the term
pseudo- or generalized-FPE is used in the literature. It is
sometimes argued that equations of this type are unphysical.
However by doubling the phase space, it is possible to
introduce a generalized $P$-representation (the positive
$P$-representation)~\cite{r331,r333a}. The equation of motion
for this generalized $P$-representation is a FPE with a
positive or positive semidefinite diffusion matrix. The
nonpositive definite diffusion matrix plays an essential role
in the production of nonclassical fields~\cite{r26}.

The second method of an equation of motion in Fock
representation has been applied to various multi-photon Raman
processes~\cite{r57,r66,r100,r101,r318,r319,r255,r320,r144,r148,r149,r161,r162}.
The master equation (in terms of  Fock states) for
first-order Stokes scattering can be solved by applying the
Laplace transform method. Solutions obtained by McNeil and
Walls~\cite{r66}, Simaan~\cite{r100}, and others apply only
to the diagonal elements of the density matrix $\hat{\rho}$,
which is a serious drawback of these formulations~\cite{r79}.
The photocount statistics (sub-Poissonian photon statistics,
antibunching, or anticorrelation) can be fully analyzed using
the diagonal, in Fock representation, matrix elements of
$\hat{\rho}$ only. However, the phase properties of the
fields~\cite{r311,r304,r305}, or squeezing
properties~\cite{r58,r303} (which are sensitive to the phase
of the field) require the availability of the non-diagonal
terms of the density matrix.  We shall derive, for Raman
scattering including depletion of the pump field, an exact
solution of the master equation for the complete density
matrix in Fock representation $\langle n,m|\rho|n',m'\rangle$
with arbitrary $n,m,n',m'$.

The classical description of Raman scattering into both the
Stokes and anti-Stokes fields seems to be well
understood~\cite{r12,r154} contrary to quantum description,
which is hampered by the complexity of the underlying
Hamiltonians and hence the complex structure of the equations
of motion. One of the simplest nontrivial models describing
the coupling of the Stokes and anti-Stokes scattering was
proposed by Knight~\cite{r156}. This Raman-coupled model, in
which a single atom is coupled to a single-cavity mode by
Raman type transitions, has attracted some attention and has
been generalized to much more realistic experimental
conditions in the subsequent papers by Phoenix and
Knight~\cite{r157}, Schoendorff and Risken~\cite{r145},
Agarwal and Puri~\cite{r7,r214} as well as Gerry and
Eberly~\cite{r215}, Gerry~\cite{r158}, Gerry and
Huang~\cite{r215a},and Gangopadhyay and Ray~\cite{r336}. It
is quite remarkable that there exists a strict {\em operator}
solution~\cite{r7,r214} of the master equation describing the
evolution of the generalized Knight model, which describes
the system of an atom undergoing Raman transitions between
two degenerate levels on interaction with a quantized field
in a lossy cavity driven by an external field including the
effects of atomic dephasing collisions. An extension of the
propagation theory of Raman effect~\cite{r72,r90} to include
anti-Stokes scattering has been developed by
Kilin~\cite{r336a} and independently by Li et
al.~\cite{r147}. As mentioned, we shall analyze another model
of the Stokes-anti-Stokes coupling within the framework of
the temporal theory of Raman effect proposed by
Walls~\cite{r132b} and extensively studied by Pe\v{r}ina and
coworkers (see Refs.~\cite{r79} and~\cite{r81a}).

We shall discuss only temporal variations of fields instead
of full temporal and spatial analysis. The assumption of
monochromatic pump, Stokes, and anti-Stokes fields restricts
the validity of our theory to a cavity problem.  However, a
temporal evolution in a cavity problem can usually be
converted to a corresponding steady-state propagation in a
dispersionless medium by simply replacing the time variable
$t$ by $-z/c$, a ``normalized'' space variable $z$. This
procedure permits us to address nonlinear optical phenomena,
in particular Raman scattering, in a manner analogous to
their classical treatment~\cite{r97,r154,r3,r4}. Formal
space-time analogies have also been pointed out in the
differential equations for the propagation of short light
pulses~\cite{r334}. Obviously, a full quantum space-time
description is considerably more complex and resides in
solving equations of motion for an infinite number of
creation (annihilation) operators of the single-mode
radiation fields.  The total spatially dependent field is a
sum of the single-mode solutions.

Here we mention only some spatial propagation theories of
Raman scattering. For a detailed analysis we refer the Reader
to the review by Raymer and Walmsley~\cite{r91} and
references therein. The temporal and spatial evolution of the
radiation fields (laser, Stokes and anti-Stokes fields) in
Raman scattering was successfully described within the
framework of the classical coupled wave theory developed by
Bloembergen and Shen~\cite{r173,r174,r99} (see also
Ref.~\cite{r154}). The first quantum theories of Raman
scattering including spatial propagation were proposed
independently by von Foerster and Glauber~\cite{r128} and
Akhmanov et al.~\cite{r218} using the analogy of Raman effect
and optical parametric amplification processes. Another
method was proposed by Emel'yanov and Seminogov~\cite{r220},
and Mostowski and Raymer~\cite{r72,r90} using the analogy
between Raman scattering and superfluorescent
processes~\cite{r221}. Spectacular predictions of the latter
theory have been, in particular, macroscopic pulse-energy
fluctuations of the emitted radiation in a manner reflecting
the underlying spontaneous initiation~\cite{r222,r223,r224}.
The negative-exponential probability distribution (NEPD),
derived by Raymer et al.~\cite{r222}, describes the
macroscopic fluctuations of the scattered radiation.
K\'arsk\'a and Pe\v{r}ina~\cite{r41} pointed out that the
NEPD corresponds to the generating function of the integrated
intensity extensively used in this paper (see Sect. 5.2). The
standing-wave theory of Raman scattering properly describes
the macroscopic fluctuations in the low-gain and high-gain
regime (see Refs.~\cite{r79,r147,r91} and references
therein). In the latter limit the quantum fluctuations of the
generated fields can be thought of as arising from a
classical noise process, contrary to the low-gain limit,
where certain nonclassical effects occur.

Finally we should mention certain crucial experiments
revealing some manifestations of Raman scattering.  For more
details, see the review article by Raymer and
Walmsley~\cite{r91}.  Experiments on the detection of
fluctuations of Stokes pulse energies were carried out by
Walmsley and Raymer~\cite{r199,r200}, and Fabricius et
al.~\cite{r201}. The temporal and spatial fluctuations of the
Stokes beam profile, the spectrum, and delay have been
investigated in a number of experiments both for depleted and
undepleted pump pulse (for references see Ref.~\cite{r91}).
We mention these experiments because the theory of Raman
scattering for cavity modes, to be presented here, correctly
predicts the existence of macroscopic quantum fluctuations of
the Stokes pulses.

Generation of Raman solitons in the heavily depleted pump
pulse has recently been observed by MacPherson et
al.~\cite{r202} and Swanson et al.~\cite{r203} as predicted
by Englund and Bowden~\cite{r204a,r204b}. Cooperative effects
in Raman scattering, referred to as cooperative Raman
scattering, which is analogous to two-level
superfluorescence~\cite{r206}, occurs for a laser pump not
significantly depleted.  The effect was first observed by
Kirin et al.~\cite{r207} and then re-examined under fully
convincing experimental conditions by Pivtsov et
al.~\cite{r208}. In our analysis we clearly distinguish the
two cases of the depleted and undepleted pump field, and
therefore have listed some effects and experiments in which
this condition for the intensity pump is crucial.

Hyper-Raman scattering, i.e., the three-photon analog of
Raman scattering, was discovered in 1965 by Terhune et
al.~\cite{r347} (see also Ref.~\cite{r348}). This effect was
predicted theoretically by Neugebauer~\cite{r42a},
Kielich~\cite{r42b,r42c} and Li~\cite{r42d} prior to its
experimental detection.  Since the discovery of hyper-Raman
scattering, numerous papers have appeared reporting
theoretical investigations and observations of the process in
a variety of solids, liquids and vapours. Here, we shall not
discuss higher-order Raman scattering processes.  We refer to
the reviews of Refs.~\cite{r349,r110,r46,r79} for details and
literature.

This paper is organized as follows. In Section 2, the
standing-wave model of Raman scattering is constructed and
the basic equation of motion (master equation) is derived. In
Section 3, we give a short account of multimode
$s$-parametrized quasidistributions and $s$-parametrized
characteristic functions. In Section 4, we introduce
definitions of nonclassical properties of radiation such as
quadrature (``usual'' and principal) squeezing and
sub-Poissonian photon statistics. In Section 5, we present
the $s$-parametrized quasidistribution formalism of Raman
scattering either including (in Sect.  5.1) or neglecting (in
Sect.  5.2) depletion of the pump laser beam in the process
of scattering.  In Section 6 we develop the density matrix
formalism of Raman scattering. We derive exact solutions of
the master equation in Fock representation in Sect. 6.1.2. We
also give short-time (in Sect. 6.1.1) and long-time (in Sect.
6.1.3) solutions of the master equation. In Sect. 6.2 we
present approximate solutions valid under parametric
approximation, i.e., when pump depletion is neglected.

\section{Model and master equation}

Let us analyze Raman scattering starting from a completely
quantum Hamiltonian but describing phenomenologically only
the net effect, i.e., ignoring the details of the scattering
mechanism. We describe the interaction of three single-mode
radiation fields: an incident laser beam at the frequency
$w_L$, a Stokes field at the frequency $w_S$ and an
anti-Stokes field at the frequency $w_A$ through an infinite
phonon system at frequencies $w_{Vj}$, after
Walls~\cite{r132b} (see also Refs.~\cite{r77,r78,r163}), by
the effective Hamiltonians:
\begin{eqnarray}
\hat{H}_0 = \hbar \omega_L \hat{a}^+_L \hat{a}_L + \hbar \omega_S
\hat{a}^+_S \hat{a}_S + \hbar \omega_A \hat{a}^+_A \hat{a}_A +
\hbar \sum_{j} \omega_{Vj}\, \hat{a}^+_{Vj}\, \hat{a}_{Vj},
\label{001}
\end{eqnarray}
\begin{eqnarray}
\hat{H}_S = \hbar \sum_{j} \lambda_{Sj} \hat{a}_L \hat{a}_S^+
\hat{a}_{Vj}^+ \:+\: h.c.,
\nonumber\\
\hat{H}_A = \hbar \sum_{j} \lambda_{Aj}^* \hat{a}_L \hat{a}^+_A
\hat{a}_{Vj} \:+\: h.c., \label{002}
\end{eqnarray}
\begin{eqnarray}
\hat{H}_T = \hat{H}_0  - \hat{H}_S  - \hat{H}_A, \label{003}
\end{eqnarray}
where $\hat{H}_S$ ($\hat{H}_A$) is the trilinear interaction
Hamiltonian for Stokes (anti-Stokes) scattering and $H_0$ is
the unperturbed Hamiltonian. For simplicity, we have dropped
the zero-point contributions. The annihilation operators for
the laser, Stokes, anti-Stokes, and phonon fields are denoted
by $\hat{a}_L$, $\hat{a}_S$, $\hat{a}_A$ and $\hat{a}_{Vi}$,
respectively (we label all Hilbert space operators with
caret). The coupling coefficient $\lambda_{Sj}$
($\lambda_{Aj}$) denotes the strength of the coupling between
the Stokes (anti-Stokes) mode and the optical phonon at the
frequency $\omega_{Vj}$. These coefficients depend on the
actual interaction mechanism. In the Hamiltonians~(\ref{002})
we neglect terms describing higher-order Stokes
scattering~\cite{r130,r131,r92}, as well as terms describing
hyper-Raman scattering~\cite{r110,r82,r144}. In Section 6.1
in the analysis of the Raman effect without parametric
approximation we also neglect anti-Stokes production.

In our model we take into account only the radiation modes
appropriate for a cavity. It should be kept in mind that the
several radiation mode description is applied to the waves
involved in the whole course of the interaction, not only at
the beginning of the interaction process. This approximation
is a shortcoming from the experimental point of view, since
it is not very suitable for describing the most common
experimental arrangements used when measuring stimulated
Raman scattering~\cite{r154,r187,r91}.

We apply the rotating wave approximation since in the
interaction Hamiltonians~(\ref{002}) we have omitted terms of
the form $\hat{a}_{Vj}\hat{a}_S^+\hat{a}_L \:+\: h.c.$ and
$\hat{a}_{Vj}^{+}\hat{a}_A^+\hat{a}_L \:+\: h.c.$. For weak
coupling these terms are negligible because they vary rapidly
as $\exp[\pm i (\omega_{Vj} +|\omega_L -\omega_{S,A}|)t]$,
which implies that their average is approximately zero for
times of evolution much greater than
$|\omega_L-\omega_{S,A}|^{-1}$, contrary to the interaction
Hamiltonians $H_S$ and $H_A$~(\ref{002}), which vary as
$\exp[\pm i (\omega_{Vj}-|\omega_L-\omega_{S,A}|)t]$ giving
unity for $\omega_{Vj}\approx |\omega_L-\omega_{S,A}|$. We
have also neglected terms of the form
$\hat{a}_{Vj}^±\hat{a}_S\hat{a}_L \:+\: h.c.$ and
$\hat{a}_{Vj}^±\hat{a}_A\hat{a}_L \:+\: h.c.$. These terms,
if included, would describe a process in which both the
Stokes (anti-Stokes) and laser photons are annihilated and
created in the scattering act.

The Hamiltonians~(\ref{002}) describe Raman scattering under
the long wavelength approximation, which has several
important implications~\cite{r128,r209,r73}. Firstly, we can
neglect the intermolecular interactions. Each optical
vibrational mode of the medium is equivalent to a simple
harmonic oscillator. Secondly, the optical phonon dispersion
is negligible. A typical dispersion curve for optical
phonons, $\omega_{V}(k_V)$, is almost flat for wave vectors
$k_V$ from the interval $(-1/\lambda,1/\lambda)$, where
$\lambda$ is an optical wavelength. In other words, optical
wave vectors occupy only a very small volume about the origin
of the reciprocal lattice.  Thirdly, a crystal can  be
treated as a continuum; thus, from the mathematical point of
view, sums over lattice sites can be replaced by integrals
over a volume of the crystal. This long wave approximation is
quite realistic for optical processes, in particular Raman
scattering.

A detailed derivation  of the Hamiltonian from first
principles has been given by von Foerster and
Glauber~\cite{r128} in their quantum propagation theory of
Raman scattering from phonons. Although we deal with modes in
a cavity, many aspects of their theory recur in our approach.

In the case of an unbounded medium the momentum is conserved in the
interaction, i.e., the sum (difference) of the wave vectors
$\vec{k}_S$ ($\vec{k}_A$) of the Stokes (anti-Stokes) photon and
$\vec{k}_V$ of the photon involved in the scattering act is
exactly equal to the laser light wave vector  $\vec{k}_L$,
\begin{eqnarray}
\vec{k}_L = \vec{k}_S + \vec{k}_V, \hspace{1cm} \vec{k}_L =
\vec{k}_A - \vec{k}_V. \label{004}
\end{eqnarray}
This means that each laser mode interacts strongly only with
phonons having a single wave vector (one and only one
vibrational mode). This is the requirement of translational
invariance. Momentum is no longer strictly conserved for
interactions in a finite medium, since the introduction of
boundaries destroys the translational invariance of the
medium. The strongest interaction is still for those modes,
which conserve momentum~(\ref{004}) and energy
($\omega_{Vj}\approx |\omega_L-\omega_{S,A}|$); nevertheless,
in this case the radiation modes are coupled to a certain
range of optical phonons whose wave vectors may not satisfy
the adequate conditions~(\ref{004}) by amounts of the order
of the reciprocal of the dimensions of the
medium~\cite{r128}. The coupling constants $\lambda_{Sj}$,
and $\lambda_{Aj}$ contain these momentum mismatches via
phase integrals~\cite{r132b,r110}:
\begin{eqnarray}
\lambda_{Sj} \sim \int\limits_{V} \exp
\left[
-i \left( \vec{k}_L - \vec{k}_S - \vec{k}_{Vj} \right) . \vec{r}
\right] {\rm d}^3r,
\nonumber\\
\lambda_{Aj} \sim \int\limits_{V} \exp
\left[
-i \left( \vec{k}_L - \vec{k}_A + \vec{k}_{Vj} \right) . \vec{r}
\right] {\rm d}^3r.
\label{005}
\end{eqnarray}
Hence, the interaction Hamiltonians~(\ref{002}) are
represented by sums over all optical vibrational modes that
may scatter into or out of the desired mode.  This means that
the coupling of the radiation fields (in particular Stokes
and anti-Stokes) through a large number of optical phonons is
treated stochastically.

In the Hamiltonians~(\ref{001})--(\ref{003}) we have assumed
all the radiation fields and phonons to be polarized linearly
in the same direction. We have not included explicitly the
polarization states of those photons, which might affect the
photon-counting statistics~\cite{r166,r167},
squeezing~\cite{r168a,r168b} and other properties
(Ref.~\cite{r165} and references therein). Obviously, this
would require the discussion of correlation tensors, in place
of correlation functions, involving the photon polarization
states~\cite{r169,r94,r44}.

The model under discussion is restricted to the approximation
of electric-dipole transitions. In previous
papers~\cite{r43,r297}, Kielich has proposed and extensively
developed the formal quantum theory of first-, second-, and
higher-order processes (in particular Raman scattering)
taking into account multipolar electric and magnetic quantum
transitions.

A lot of attention has been devoted to a simpler completely
boson Hamiltonian  applied to the description of the
statistical properties of Raman scattering by phonons treated
as a single monochromatic mode
(Refs.~\cite{r178,r130,r114,r53} and~\cite{r79} with
references therein). It is clear  that the use of a large
number of phonon modes  (a phonon bath) in the model
Hamiltonians (\ref{002}) provides a fuller picture of the
scattering processes. In particular, the model describes the
stochastic coupling of the Stokes and anti-Stokes modes
through a phonon bath. The assumption of a single phonon mode
implies that the Stokes and anti-Stokes fields are coupled in
a deterministic manner, which seems to be a rather serious
drawback~\cite{r132b}.

As a digression, let us mention that the same
phenomenological Hamiltonians~(\ref{002}) have been used in
the description of Brillouin scattering (see, for example,
Refs.~\cite{r164,r30} and~\cite{r79} and references therein).
The main difference between Brillouin and Raman scattering
lies in different kinds of the scatterers responsible for
these effects: acoustic phonons in the Brillouin effect, and
optical phonons in the Raman effect. This difference is
included in the frequencies, the coupling constants
$\lambda_{Sj}$, $\lambda_{Aj}$ and the reservoir spectrum.
More important, acoustic phonons exhibit much greater
dispersion than optical phonons.  In our approach to Raman
scattering we neglect dispersion.  This assumption applied to
Brillouin scattering has considerably less validity.

We are interested only in the statistical properties of the
radiation fields (the pump and scattered beams) considered as
a system. We therefore remove the unnecessary information
about the infinite system of optical phonons, treated as a
reservoir (heat bath). The procedure leading to the master
equation is widely used in quantum optics. For a general
review of the master equation methods and the extensive
bibliography see Refs.~\cite{r59,r29,r79,r129}. We rewrite
the interaction Hamiltonians $H_S$ and $H_A$ in the
interaction picture as:
\begin{eqnarray}
\hat{H}_S  + \hat{H}_A  = \hbar \sum_{k=1}^{4} \hat{F}_k
\hat{Q}_k, \label{006}
\end{eqnarray}
where
\begin{eqnarray}
\hat{F}_1 = \hat{F}_2^+ &=& \sum_{j} \lambda_{Sj} \hat{a}_{Vj}^{+}
\exp[i \omega_{Vj}(t-t_0)],
\nonumber\\
\hat{F}_3 = \hat{F}_4^+ &=& \sum_{j} \lambda_{Aj}^* \hat{a}_{Vj}
\exp[-i \omega_{Vj}(t-t_0)],
\nonumber\\
\hat{Q}_1 = \hat{Q}_2^+ &=& \hat{a}_L \hat{a}_{S}^+ \exp[-i
\Omega_{S}(t-t_0)],
\nonumber\\
\hat{Q}_3 = \hat{Q}_4^+ &=& \hat{a}_L \hat{a}_{A}^+ \exp[i
\Omega_{A}(t-t_0)]. \label{007}
\end{eqnarray}
The $\hat{Q}_i$ ($\hat{F}_i$) are respectively functions of the system
(reservoir) operators only. The `cavity' frequencies $\Omega_s$,
$\Omega_A$ are equal to
\begin{eqnarray}
\Omega_{S,A} = \left| \omega_L-\omega_{S,A}\right|. \label{008}
\end{eqnarray}
Since the system and the reservoir variables are mutually
independent as it follows from
\begin{eqnarray}
\left[ \hat{a}_i, \hat{a}_{j}^{+}\right] = \delta_{ij}
\hspace{1cm}\hbox{for}\hspace{5mm} i,j=L,S,A,V1,V2,...,
\label{009}
\end{eqnarray}
we may trace, in standard manner, the complete density matrix
over the reservoir leading to the reduced density matrix
$\hat{\rho}(t)$. Obviously, we cannot obtain any reservoir
averages from $\hat{\rho}(t)$. There are some Raman
scattering models (e.g. Refs.~\cite{r128,r30,r79}), where
optical phonons are included in the system, whereas other
crystal excitation modes, such as acoustical phonons,
electric excitations, and other species of molecular
vibrations, serve as a thermal reservoir.

The radiation fields are weakly coupled to the thermal
reservoir. The anti-Stokes mode loses energy to the
reservoir. The fluctuations in the reservoir also couple back
into the system introducing noise from the reservoir.
However, we apply the Markov approximation, a condition
sufficient to ensure that energy that goes into the reservoir
will not return to the radiation fields. This conclusion
follows from the definition of the Markovian system as one
that cannot develop memory -- the future of the system is
determined by the present and not its past~\cite{r59,r29}.
The importance of this assumption is sometimes stressed in
the concept of a Schr\"odinger-Markov (or Heisenberg-Markov)
picture, meaning the standard pictures under Markov
approximation~\cite{r59}.  The importance of non-Markovian
effects in Raman scattering has been recently studied by,
e.g., Sugawara et al.~\cite{r126a} and Villaeys et
al.~\cite{r126}. Obviously, the system operators,
$\hat{Q}_i$, obey the same commutation relations under this
approximation as they did originally.

To obtain the equation of motion for the reduced density
matrix $\hat{\rho}(t)$, one has to compute the reservoir
spectral densities
\begin{eqnarray}
w_{ij}^{+} = \int\limits_{0}^{\infty} e^{i \omega_i \tau} \left<
\hat{F}_i(\tau) \hat{F}_j \right>_R {\rm d}\tau,
\nonumber\\
w_{ji}^{-} = \int\limits_{0}^{\infty} e^{i \omega_i \tau} \left<
\hat{F}_j \hat{F}_i(\tau) \right>_R {\rm d}\tau, \label{010}
\end{eqnarray}
where $\langle...\rangle_R$ is the average over all reservoir
operators; $\omega_i$ takes the values $\pm \Omega_{S,A}$.
The infinite system of optical phonons is assumed to be densely
spaced with the number of modes between $\omega_i$ and
$\omega_i+d\omega_i$ equal to $g(\omega_i)d\omega_i$, so we may
replace the sums over the optical vibrational modes by integrals
\begin{eqnarray}
\sum_{j} (...) &\approx & \int\limits_{0}^{\infty}
{\rm d}\omega_j g(\omega_j) (...).
\label{011}
\end{eqnarray}
Let us introduce two quantities: $\Delta\Omega$ -- the
frequency mismatch and $\Omega$ -- the medium ``cavity''
frequency, defined by
\begin{eqnarray}
\Delta\Omega = \frac{\Omega_S - \Omega_A}{2},
\nonumber\\
\Omega = \frac{\Omega_S + \Omega_A}{2}. \label{012}
\end{eqnarray}
The frequency mismatch $\Delta\Omega$, in general, is not equal to
zero.  It is quite realistic for optical phonons to assume that the
coupling constants $\lambda_{S,A}(\omega_j)$ and the phonon density of
$g(\omega_j)$ are flat in the vicinity of $\Omega$, so that we can
write
\begin{eqnarray}
g(\Omega \pm \Delta\Omega) &\approx& g(\Omega),
\nonumber\\
\lambda_k(\Omega \pm \Delta\Omega) &\approx& \lambda_k(\Omega),
\hspace{2cm} k=S,A.
\label{013}
\end{eqnarray}
The reservoir is supposed to be at thermal equilibrium. The
phonons are unaffected by interaction with the radiation
fields. In the classical sense this means that the phonons
are so quickly damped that they remain in their steady
state~\cite{r174,r99,r154}. The mean number of phonons in the
reservoir mode at thermal equilibrium is defined by the
Bose-Einstein distribution
\begin{eqnarray}
\langle \hat{n}(\omega_{Vj}) \rangle = \left[ \exp \left(
\frac{\hbar \omega_{Vj}}{k_B T}\right)-1 \right]^{-1}, \label{014}
\end{eqnarray}
where $k_B$ is the Boltzmann constant and $T$ is the
temperature of the reservoir. Obviously, as the reservoir
temperature approaches absolute zero, the mean number of
phonons $\langle \hat{n}(\omega_{Vj}) \rangle$ tends to zero
as well. In Sect. 5.2, we analyze Raman scattering in a
parametric approximation for a ``noisy'' reservoir ($\langle
\hat{n}(\omega_{Vj}) \rangle\neq 0$), whereas we study Raman
scattering including the pump depletion for ``quiet''
reservoir ($\langle \hat{n}(\omega_{Vj}) \rangle\approx 0$).
After some algebra  one obtains from~(\ref{010}),
\begin{eqnarray}
w_{21}^{+} &=& \left( \frac{\gamma_S}{2} + i \Delta\omega_S
\right) \left( \langle \hat{n}_{V} \rangle +1 \right),
\nonumber\\
w_{12}^{+} &=& \left( \frac{\gamma_S}{2} - i \Delta\omega_S
\right) \langle \hat{n}_{V} \rangle,
\nonumber\\
w_{43}^{+} &=& \left( \frac{\gamma_A}{2} - i \Delta\omega_A
\right) \langle \hat{n}_{V} \rangle,
\nonumber\\
w_{34}^{+} &=& \left( \frac{\gamma_A}{2} + i \Delta\omega_A
\right) \left( \langle \hat{n}_{V} \rangle +1 \right),
\nonumber\\
w_{31}^{+} &=& \left( \frac{\gamma_{AS}}{2} + i \Delta\omega_{AS}
\right) \left( \langle \hat{n}_{V} \rangle +1 \right),
\nonumber\\
w_{13}^{+} &=& \left( \frac{\gamma_{SA}}{2} - i \Delta\omega_{SA}
\right) \langle \hat{n}_{V} \rangle,
\nonumber\\
w_{42}^{+} &=& \left( \frac{\gamma_{AS}}{2} - i \Delta\omega_{AS}
\right) \langle \hat{n}_{V} \rangle,
\nonumber\\
w_{24}^{+} &=& \left( \frac{\gamma_{SA}}{2} + i \Delta\omega_{SA}
\right) \left( \langle \hat{n}_{V} \rangle +1 \right)
\nonumber\\
w_{ij}^{-} &=& \left( w_{ij}^{+}\right)^*. \label{015}
\end{eqnarray}
All other reservoir spectral densities, in particular the
diagonal densities $\omega_{ii}^±$ (for $i=1,...,4$), vanish.
For simplicity we have denoted the mean number of phonons at
frequency $\Omega$ by $\langle \hat{n}_V \rangle=\langle
\hat{n}(\Omega_V) \rangle$.  The gain constant for the Stokes
mode, $\gamma_S$, the damping constant for the anti-Stokes
model $\gamma_A$, and the mutual damping constants for both
scattered fields, $\gamma_{SA}$, $\gamma_{AS}$, are
\begin{eqnarray}
\gamma_k &=& 2\pi g(\Omega) \left| \lambda_k(\Omega)\right|^2
\hspace{1cm} (k=S,A),
\nonumber\\
\gamma_{SA} &=& \gamma_{AS}^* \:=\: 2\pi g(\Omega)
\lambda_S(\Omega) \lambda_A^*(\Omega), \label{016}
\end{eqnarray}
where $g(\Omega)$, as earlier, denotes the density of the
optical phonon modes (the reservoir spectrum) at frequency
$\Omega$. It is seen that the following simple relation
between the single and mutual damping constants holds:
$\left| \gamma_{SA}\right|^2 =\left| \gamma_{AS}\right|^2 =
\gamma_A \gamma_S$. The frequency shifts, representing the
Lamb shift in the frequency $\Omega\approx \Omega_j$, are
expressed by the Cauchy principle value, $\cal{P}$ of the
integrals:
\begin{eqnarray}
\Delta\omega_k  = - {\cal P} \int\limits_{0}^{\infty}
\frac{g(\omega_j) \left| \lambda_k(\omega_j) \right|^2}{\omega_j
-\Omega_k} {\rm d}\omega_j \hspace{1cm} (k=S,A),
\nonumber\\
\Delta\omega_{SA}  = (\Delta\omega_{AS})^* \:=\: - {\cal P}
\int\limits_{0}^{\infty} \frac{g(\omega_j) \lambda_S(\omega_j)
\lambda_A^*(\omega_j)}{\omega_j -\Omega} {\rm d}\omega_j.
\label{017}
\end{eqnarray}
The only effect of the $\Delta\omega_i$ is to change slightly the
frequency $\Omega$, so we neglect them. Having calculated the
reservoir spectral densities we can write the master equation for the
reduced density matrix $\hat{\rho}=\hat{\rho}(\hat{a}_L,\hat{a}_S,
\hat{a}_A,t)$ as:
\begin{eqnarray}
\frac{\partial}{\partial t}\hat{\rho} &=& \frac{1}{2} \gamma_S
\left( \left[
\hat{a}_L\hat{a}_S^+,\hat{\rho}\hat{a}_L^+\hat{a}_S\right] \:+\:
h.c.\right) \nonumber \\ &&+ \frac{1}{2} \gamma_A \left( \left[
\hat{a}_L^+\hat{a}_A,\hat{\rho}\hat{a}_L\hat{a}_A^+\right] \:+\:
h.c.\right)
\nonumber\\
&&+ \frac{1}{2} \gamma_{SA} e^{-2i\Delta\Omega\Delta t} \left(
\left[ \hat{a}_L\hat{a}_S^+,\hat{\rho}\hat{a}_L\hat{a}_A^+\right]
+\left[ \hat{a}_L\hat{a}_S^+\hat{\rho},\hat{a}_L\hat{a}_A^+\right]
\right)
\nonumber\\
&&+ \frac{1}{2} \gamma_{AS} e^{2i\Delta\Omega\Delta t} \left(
\left[ \hat{a}_L^+\hat{a}_A,\hat{\rho}\hat{a}_L^+\hat{a}_S\right]
+\left[ \hat{a}_L^+\hat{a}_A\hat{\rho},\hat{a}_L^+\hat{a}_S\right]
\right)
\nonumber\\
&&- \langle \hat{n}_V\rangle \Big\{ \frac{1}{2} \gamma_S \left(
\left[ \hat{a}_L^+\hat{a}_S, \left[
\hat{a}_L\hat{a}_S^+,\hat{\rho}\right] \right] \:+\: h.c.\right)
\nonumber\\
&&+ \frac{1}{2}\gamma_A \left( \left[ \hat{a}_L\hat{a}_A^+, \left[
\hat{a}_L^+\hat{a}_A,\hat{\rho}\right] \right] \:+\: h.c. \right)
\nonumber\\
&&+ \gamma_{SA} e^{-2i\Delta\Omega \Delta t} \left[
\hat{a}_L^+\hat{a}_S, \left[
\hat{a}_L^+\hat{a}_A,\hat{\rho}\right] \right] \nonumber \\ &&+
\gamma_{AS} e^{2i\Delta\Omega\Delta t} \left[
\hat{a}_L\hat{a}_A^+, \left[
\hat{a}_L\hat{a}_S^+,\hat{\rho}\right] \right] \Big\}. \label{018}
\end{eqnarray}
The term in $\gamma_S$ represents the amplification of the
Stokes mode; the term in $\gamma_A$ describes the loss of
energy from the anti-Stokes mode into the reservoir; the
$\gamma_{AS}$ and $\gamma_{SA}$-terms represent the
stochastic coupling between the Stokes and anti-Stokes modes
through the reservoir; the remaining terms in $\langle
\hat{n}_V\rangle\gamma_{i}$ represent the diffusion of
fluctuations of the reservoir into the system modes. Eq.
(\ref{018}) describes, moreover, the evolution of the laser
beam, i.e., the depletion of the laser field, the coupling of
the field with the Stokes and anti-Stokes fields, as well as
the diffusion of the reservoir fluctuations into the laser
field. The interpretation of the $\gamma_S$ ($\gamma_A$)
terms as the amplification (attenuation) of the radiation
fields is as yet intuitive, but will gain in precision on
solution  of the generalized Fokker-Planck equation.  The
master equation~(\ref{018}) could have been written in more
compact form; albeit for purposes of interpretation the above
form is more convenient.

The master equation~(\ref{018}), in the particular case of
parametric approximation, reduces to the equation obtained by
Walls~\cite{r132b} and Pe\v{r}ina~\cite{r77}, and reduces to
that of McNeil and Walls for Stokes scattering alone but with
no need for the parametric approximation~\cite{r66}. Our
master equation~(\ref{018}) differs but slightly in the
diffusion terms $\langle \hat{n}_V \rangle \gamma_i$ only
from the special case of the master equation given by
Agarwal~\cite{r210} (see also Ref.~\cite{r66}).

The master equation may be solved by various techniques
presented in standard
textbooks~\cite{r183,r59,r302,r29,r155,r79}. Here, we apply
two methods.  We convert the master equation to an associated
classical equation. On the one hand, expressing the quantum
equation in s-ordered form one obtains the generalized
Fokker-Planck equation for the $s$-parametrized
quasi-probability distribution, which can be exactly solved
for a class of Ornstein-Uhlenbeck processes~\cite{r296,r155}.
On the other hand, one can express the master equation in
Fock representation, which can be solved, for instance, by
the Laplace transform method~\cite{r66,r100,r57}.

In the following Sections we analyze three cases. Firstly, we
briefly describe coupling of the three quantum radiation
fields: the laser, Stokes, and anti-Stokes beams. The problem
simplifies considerably if one assumes narrow
quasi-probability distributions.  Secondly, we apply the
parametric approximation, which means that the pump field is
treated classically. We include the coupling of the Stokes
and anti-Stokes field through the phonon bath.  Thirdly, we
separately describe either the laser and Stokes mode or the
laser and anti-Stokes mode, but include the depletion of the
pump laser light. In this case we assume the heat bath to be
``quiet''.

\section{Multimode $s$-parametrized quasidistributions}

A description of the multimode fields via quasiprobability
distributions (quasidistributions, QPDs) or equivalently via
characteristic functions was first proposed by
Glauber,~\cite{r169,r31,r274}, Cahill~\cite{r276} and Klauder
et al.~\cite{r277}.  General ordering theorems have been
given by Agarwal and Wolf~\cite{r279}. The $s$-parametrized
single-mode quasidistributions and characteristic functions
were introduced by Cahill and Glauber~\cite{r278}, who
extensively studied various ways of defining correspondences
between the operators and functions. For a recent review of
the multimode $s$-parametrized functional formalism we refer
the Reader to Ref.~\cite{r79}. Here, we list the basic
definitions and properties of the $s$-parametrized multimode
quasidistributions and characteristic functions useful for
our further investigations.

To solve the master equation~(\ref{018}), i.e., the operator
equation, we use the c-number representations ${\cal
W}^{(s)}( \{\alpha_k \})$ and ${\cal C}^{(s)}(\{\beta_k\})$
of the density operator introduced by Cahill and
Glauber~\cite{r278}. These representations not only are
useful as a calculation tool, but also provide insight into
the interrelations between classical and quantum mechanics.
By virtue of the multimode $s$-parametrized displacement
operator
\begin{eqnarray}
\hat{D}^{(s)} (\{ \beta_k\}) = \prod_{k} \hat{D}^{(s)} (\beta_k)=
\prod_{k} \exp \left( \beta_k \hat{a}_k^+ - \beta_k^* \hat{a}_k
+\frac{s}{2} \left| \beta_k \right|^2 \right), \label{019}
\end{eqnarray}
where the continuous parameter $s$ belongs to the interval
$\langle -1,1\rangle$, one can define the $s$-parametrized
multimode characteristic function as the mean value of
$\hat{D}^{(s)}(\{\beta_k\})$,
\begin{eqnarray}
{\cal C}^{(s)} (\{ \beta_k\}) = {\rm Tr} \left[ \hat{\rho}
\hat{D}^{(s)} (\{ \beta_k\}) \right]. \label{020}
\end{eqnarray}
In our situation involving the  three radiation modes laser
($k=L$), Stokes ($S$) and anti-Stokes ($A$), the simplified
notation in Eqs.~(\ref{019}) and~(\ref{020}) stands for $(\{
\beta_k\})=(\beta_L,\beta_S,\beta_A)$. The Fourier transform
of the characteristic function ${\cal C}^{(s)}(\{\beta_k\})$
(\ref{020}) readily gives the $s$-parametrized multimode
quasidistribution ${\cal W}^{(s)}(\{\alpha_k\})$,
\begin{eqnarray}
{\cal W}^{(s)} (\{ \alpha_k\}) = \int {\cal C}^{(s)} (\{
\beta_k\}) \exp \left[ \sum_{k} \left( \alpha_k \beta_k^* -
\alpha_k^* \beta_k \right) \right] {\rm d}^2 \left\{ \beta_k/
\pi\right\}. \label{021}
\end{eqnarray}
For completeness we write the inverse Fourier transform,
which enables us to determine ${\cal C}^{(s)}$ from ${\cal
W}^{(s)}$, namely,
\begin{eqnarray}
{\cal C}^{(s)} (\{ \beta_k\}) = \int {\cal W}^{(s)} (\{
\alpha_k\}) \exp \left[ \sum_{k} \left( \alpha_k^* \beta_k -
\alpha_k \beta_k^* \right) \right] {\rm d}^2 \left\{ \alpha_k/
\pi\right\}, \label{022}
\end{eqnarray}
where integration extends over $\alpha_k$ in the
following sense:
\begin{eqnarray*}
d^2 \left\{ \alpha_k/ \pi\right\} = \prod_{k=L,S,A} \pi^{-1} d^2
\alpha_k \:=\: \pi^{-3} \prod_{k=L,S,A} d({\rm Re} \alpha_k)
d({\rm Im} \alpha_k). \label{022a}
\end{eqnarray*}
or over $\beta_k$ similarly. The normalization is chosen to satisfy
\begin{eqnarray}
\int {\cal W}^{(s)} (\{ \alpha_k\})
{\rm d}^2 \left\{ \alpha_k/ \pi\right\} &=
{\cal C}^{(s)}(0) &= 1.
\label{023}
\end{eqnarray}
In the three special cases of $s=-1,0,1$ one recognizes the
well-known QPDs~\cite{r278,r280,r283,r282,r69,r79}, namely
the $Q$ function, the Wigner function, and the
Glauber-Sudarshan $P$-function, respectively:
\begin{eqnarray}
Q(\{ \alpha_k\}) &=& \langle \{ \alpha_k\} \left|
\hat{\rho}\right| \{ \alpha_k\} \rangle \:=\: {\cal W}^{(-1)} (\{
\alpha_k\}),
\nonumber\\
W(\{ \alpha_k\}) &=& {\cal W}^{(0)} (\{ \alpha_k\}),
\nonumber\\
P(\{ \alpha_k\}) &=& \pi^{-M} {\cal W}^{(1)} (\{ \alpha_k\}),
\label{024}
\end{eqnarray}
with $M$ denoting the number of modes (in our analysis $M$
will be equal to 3, 2, or 1). One can say that the
$s$-parametrized quasidistribution ${\cal W}^{(s)}$ (with $s$
from the interval $\langle -1,1\rangle$) is a continuous
interpolation between the $P$- and $Q$-functions. The
$Q$-function directly determines antinormally ordered
expectation values, the $P$-function determines normally
ordered averages, and the Wigner function can be used
directly to calculate the averages of symmetrically ordered
operators. The following relations hold for any parameter
$s$:
\begin{eqnarray} \left<
\prod_{k} \left(\hat{a}_k^+\right)^{m_k}
\left(\hat{a}_k\right)^{n_k} \right>_{(s)} &=& {\rm Tr} \left[
\hat{\rho} \left\{ \prod_{k} \left(\hat{a}_k^+\right)^{m_k}
\left(\hat{a}_k\right)^{n_k} \right\}_{(s)} \right]
\nonumber\\
&=& \int {\cal W}^{(s)} (\{ \alpha_k\}) \prod_{k}
\left(\alpha_k^*\right)^{m_k} \left(\alpha_k\right)^{n_k} {\rm
d}^2 \left\{ \alpha_k/ \pi\right\}
\nonumber\\
&=& \left. \prod_{k} \frac{\partial^{m_k}}{\partial\beta^{m_k}_k}
\frac{\partial^{n_k}}{\partial\left(-\beta^*_k\right)^{n_k}} {\cal
C}^{(s)} (\{ \beta_k\}) \right|_{\{\beta_k\}=0}, \label{025}
\end{eqnarray}
where $\{\beta_k\}=0$, in the three-mode case, means that
$\beta_L=\beta_S=\beta_A=0$. The generally accepted criterion
for the definition of a nonclassical field resides in the
existence of a positive $P$-function, i.e., a classical state
is one whose $P$-function is no more singular than a
$\delta$-function and is nonnegative definite (e.g.
Refs.~\cite{r346,r58,r79,r338}). This means that the quantum
statistical properties of the nonclassical field cannot be
described completely within the framework of a classical
probability theory.  A detailed discussion of the existence
of quasidistributions ${\cal W}^{(s)}(\{\alpha_k\})$ for the
Raman scattering model under consideration is presented in
Section 5.2.  The Wigner function always exists as a
nonsingular function, but may assume negative values, and in
this sense is not a classical probability distribution
(nevertheless, as was shown by Stenholm~\cite{r281},
experiments always give a positive Wigner function).  The
$Q$-function has the properties of a well behaved (bounded,
nonnegative and infinitely differentiable) classical
probability distribution.

Let us write down the relation between two  $s_1$- and
$s_2$-parametrized quasidistributions:
\begin{eqnarray}
{\cal W}^{(s_2)} (\{ \alpha_k\},t) \:=\:
\left( \frac{2}{s_1-s_2} \right)^M
\int \exp\left( - \frac{2}{s_1-s_2}
\sum_{k} \left| \alpha_k - \beta_k \right|^2
\right)
\nonumber\\
\times {\cal W}^{(s_1)} (\{ \beta_k\},t)
{\rm d}^2 \left\{ \beta_k/ \pi\right\},
\label{026}
\end{eqnarray}
where $s_2<s_1$. It is seen that the quasidistribution ${\cal
W}^{(s_2)}$ is given by the convolution of ${\cal W}^{(s_1)}$
with the multidimensional Gaussian distribution. The
analogous relation for characteristic functions~(\ref{020})
is simpler and valid for any $s_1$ and  $s_2$,
\begin{eqnarray}
{\cal C}^{(s_2)} (\{ \beta_k\},t) =\:
{\cal C}^{(s_1)} (\{ \beta_k\},t)
\exp \left( \frac{s_2-s_1}{2}
\sum_{k} \left|\beta_k \right|^2\right).
\label{027}
\end{eqnarray}
Even in the case when $s_1$-parametrized QPDs  do not exist,
the calculation of the expectation values $\langle
\hat{a}^{+m}\hat{a}^n \rangle_{(s_1)}$ in $s_1$ order poses
no problem. They can be obtained from the corresponding
$s_1$-parametrized characteristic function ${\cal C}^{(s_1)}$
in view of Eq.~(\ref{025}) or, equivalently, from a
$s_2$-parameterized quasidistribution ${\cal W}^{(s_2)}$,
which does exist, by means of the relation
\begin{eqnarray}
\langle \prod_{k} \left(\hat{a}_k^+\right)^{m_k}
\left(\hat{a}_k\right)^{n_k} \rangle_{(s_1)} =& \int \prod_{k}
m_k! \left( \frac{s_2-s_1}{2}\right)^{m_k}\alpha_{k}^{n_k-m_k}
L_{m_k}^{n_k-m_k} \left( \frac{2 |\alpha_k|^2}{s_1-s_2}\right)
\nonumber\\
&\times {\cal W}^{(s_2)} (\{ \alpha_k\}) {\rm d}^2 \left\{
\alpha_k/ \pi\right\} \label{028},
\end{eqnarray}
where $L_m^{n}(x)$ is the generalized Laguerre polynomial.
Alternatively, to obtain the $s_1$-ordered moments
$\langle\hat{a}^{+m}\hat{a}^n\rangle_{(s_1)}$ one can use the
generalized $P$-representation (positive
$P$-representation)~\cite{r331,r333a,r26,r332,r333}.

\section{Photon-counting statistics and squeezing: definitions}

To investigate nonclassical phenomena such as sub-Poissonian
photon-counting statistics or photon antibunching, one needs
to know the diagonal matrix elements in Fock representation
of the density matrix $\hat{\rho}(\{\hat{a}_k\})$ only. We
start from the probability distribution $p(n)$  of the photon
number $n$ in the $k$-mode field within a given volume $V$
of space at the time $t$, defined by
\begin{eqnarray}
p(n) \:=\: \sum_{\{ n_k \}}
\left< \{ n_k \} \left|\hat{\rho} \right| \{ n_k \}\right>
\delta_{n,\sum n_k},
\label{029}
\end{eqnarray}
where $n_k=|\alpha_k|^2$. The $s$-parametrized
quasidistribution ${\cal W}^{(s)}(\{\alpha_k\},t)$
(\ref{021}) can be readily transformed to the following
$s$-parametrized integrated quasidistribution (intensity
distribution) ${\cal W}^{(s)}(W,t)$ by means of the
$\delta$-function,
\begin{eqnarray}
{\cal W}^{(s)}(W,t) \:=\: \int {\cal W}^{(s)} (\{ \alpha_k\},t)
\delta \left( \sum_{k} \left| \alpha_k\right|^2 -W\right)
{\rm d}^2 \left\{ \alpha_k/ \pi\right\},
\label{030}
\end{eqnarray}
where the variable $W$ can be interpreted as the integrated
intensity. The  photodetection equation gives a connection
between the continuous integrated quasidistribution ${\cal
W}^{(1)}(W,t)$ and the discrete photon-number distribution
first derived by Mandel~\cite{r269,r271}. This photodetection
equation states that the photocount distribution $p(n)$ is
the Poisson transform of the integrated quasidistribution
${\cal W}^{(1)}(W,t)$. A generalized photodetection equation
for ${\cal W}^{(s)}(\{\alpha_k\},t)$ or for ${\cal
W}^{(s)}(W,t)$ can be written as~\cite{r80a,r79}
\begin{eqnarray}
p(n) &=& \left(\frac{2}{1+s}\right)^M
\left(\frac{s-1}{1+s}\right)^n \int {\cal W}^{(s)} (\{ \alpha_k\})
\nonumber\\
&&\times\exp \left( -\frac{2}{1+s}\sum_{k} \left|
\alpha_k\right|^2 \right) L_{n}^{M-1} \left(
\frac{4}{1-s^2}\sum_{k} \left| \alpha_k\right|^2 \right) {\rm d}^2
\left\{ \alpha_k/ \pi\right\}
\nonumber\\
&=& \left(\frac{2}{1+s}\right)^M \left(\frac{s-1}{1+s}\right)^n
\int {\cal W}^{(s)}(W)\:
\nonumber\\
&&\times\exp \left( -\frac{2 W}{1+s} \right) L_{n}^{M-1}\left(
\frac{4 W}{1-s^2} \right) {\rm d}^2W, \label{031}
\end{eqnarray}
with $L_n^{M-1}(x)$ denoting the generalized Laguerre
polynomial. We formally identify the photon-number
distribution~(\ref{029}) with the photocount distribution
(\ref{030}). There is some slight difference in their
physical interpretation, since the former distribution
describes the probability of having $n$ photons in the mode
volume $V$, whereas the latter distribution describes the
probability of detecting $n$ photons in the detector volume
$V_{det}$, defined by its parameters (sensitive area,
response time, quantum efficiency, etc.).  It can be argued,
however, (e.g. Refs.~\cite{r290,r273,r75}), that there is
perfect physical equivalence between the photon-number
moments obtained from~(\ref{029}) and the photocount-number
moments calculated from~(\ref{031}), under the assumption of
ideal detectors.

The $s$-parametrized  time-dependent generating function
$\langle\exp(-\lambda W)\rangle_{(s)}$, defined by the Fourier
transform of the $s$-parametrized quasidistribution ${\cal
W}^{(s)}(\{ \alpha_k\},t)$ or characteristic function ${\cal
C}^{(s)} (\{ \beta_k\},t)$:
\begin{eqnarray}
\left< \exp(-\lambda W(t))\right>_{(s)} = \int {\cal W}^{(s)} (\{
\alpha_k\},t) \exp \left( -\lambda\sum_{k} \left|
\alpha_k\right|^2 \right) {\rm d}^2 \left\{ \alpha_k/ \pi\right\}
\nonumber \\ = \lambda^{-M} \int {\cal C}^{(s)} (\{ \beta_k\},t)
\exp \left( -\frac{1}{\lambda} \sum_{k} \left| \beta_k\right|^2
\right) {\rm d}^2 \left\{ \beta_k/ \pi\right\} \label{032}
\end{eqnarray}
enables us to  calculate the photon-number distribution
$p(n,t)$ and the s-ordered photon-number moments $\langle
\hat{n}^k\rangle_{(s)}$ in a particularly simple manner:
\begin{eqnarray}
p(n) = \frac{(-1)^n}{n!} \left. \frac{d^n}{d\lambda^n} \left( 1 +
\frac{s-1}{2}\lambda \right)^{-M} \left<
\exp\left(-\frac{\lambda}{1+\frac{s-1}{2}\lambda}W
\right)\right>_{(s)} \right|_{\lambda=1}, \label{033}
\end{eqnarray}
\begin{eqnarray}
\left< W^k \right>_{(s)} &=\:
(-1)^k \frac{d^k}{d\lambda^k}
\left.
\left< \exp(-\lambda W)\right>_{(s)}
\right|_{\lambda=0}.
\label{034}
\end{eqnarray}
Eq.~(\ref{033}) takes the simplest form for $s=1$. Several
parameters are widely used in the literature to describe the
photon-number statistics, e.g.; the Mandel Q parameter, the
Fano factor, or the normalized second-order correlation
function. In our analysis we employ the normalized
second-order factorial moment of the photon-number operators
(or integrated intensity) (~\cite{r75},~\cite{r79} and
references therein)
\begin{eqnarray}
\gamma_{k}^{(2)} = \frac{\left< (\Delta \hat{n}_k)^2\right>_{(1)}
} {\left< \hat{n}_k \right>^2} \:=\: \frac{\left<
\hat{n}_k^2\right>_{(1)} } {\left< \hat{n}_k \right>^2}-1 \:=\:
\frac{\left< \hat{n}_k(\hat{n}_k-1)\right>} {\left< \hat{n}_k
\right>^2}-1, \label{035}
\end{eqnarray}
and  its generalization, the normalized $p$-th order
factorial moment of the $k$-th and $l$-th mode   (the
normalized two-mode cross-correlation function of $p$-th
order)
\begin{eqnarray}
\gamma_{kl}^{(p)}
&=\: \frac{\left< \hat{n}_{kl}^p\right>_{(1)} }
{\left< \hat{n}_{kl} \right>^p }-1
&=\:
\frac{\left< \hat{n}_{kl}(\hat{n}_{kl}-1)
\ldots (\hat{n}_{kl}-p+1)\right>}
{\left<
\hat{n}_{kl} \right>^p}-1,
\label{036}
\end{eqnarray}
where $\hat{n}_{kl}=\hat{n}_{k}+\hat{n}_{l}$. The
higher-order factorial moments~(\ref{036}) by comparison with
the second-order moments~(\ref{035}) provide us with further
information concerning the photon-number distributions.  In
view of the fact that $\hat{n}_{kl}$ is the sum of the
single-mode photon-number operators, the factorial moment
$\gamma_{kl}^{(2)}$ can be written as
\begin{eqnarray}
\gamma_{kl}^{(2)} = \frac{\left< (\Delta
\hat{n}_{kl})^2\right>_{(1)} } {\left< \hat{n}_{kl} \right>^2} =
\frac{ \left< (\Delta \hat{n}_{k})^2\right>_{(1)} + \left< (\Delta
\hat{n}_{l})^2\right>_{(1)} + 2\left<\Delta \hat{n}_{k} \Delta
\hat{n}_{l}\right> } { \left< \hat{n}_{k} \right>^2 +\left<
\hat{n}_{l} \right>^2 +2\left< \hat{n}_{k}\left>\right<
\hat{n}_{l}\right> }. \label{037}
\end{eqnarray}
The Mandel Q parameter for the mode $k$ is equal to
$\gamma_k^{(2)}\langle \hat{n}_k \rangle$, whereas the Fano
factor F  is ($\gamma_k^{(2)}\langle \hat{n}_k \rangle+1$) (the
photoefficiency $\eta$ of the photodetector is assumed to be
$\eta=1$).

Light with photon-number fluctuations smaller than those of
the Poisson distribution is called sub-Poissonian (or
photon-number squeezed) and is described by a negative value
of $\gamma^{(2)}$, both for $\gamma^{(2)}_{k}$ in the
single-mode case and for $\gamma^{(2)}_{kl}$ in the two-mode
case. In Section 6 we analyze the two-mode model of the Raman
effect that comprises the laser ($L$) and the Stokes mode
($S$).  We show that the sum of photon-number operators in
both modes is a constant of motion, which implies that the
factorial moments $\gamma_{LS}^{(p)}$ are constant as well.
Henceforth we shall be applying another definition to
investigate two-mode cross-correlation, referred to as the
interbeam degree of second-order coherence, given by
(Ref.~\cite{r57} and references therein)
\begin{eqnarray}
g^{(2)}_{kl}= \frac{\langle \Delta \hat{n}_k\Delta
\hat{n}_l\rangle} {\langle \hat{n}_k\rangle \langle
\hat{n}_l\rangle} \:=\:  \frac{\langle \hat{n}_k \hat{n}_l\rangle}
{\langle \hat{n}_k\rangle \langle \hat{n}_l\rangle} -1 \label{038}
\end{eqnarray} (our definition deviates from those of Ref.~\cite{r57} by the extra term $-1$).

To investigate squeezing properties of light we introduce the
Hermitian single- and two-mode operators:
\begin{eqnarray}
\hat{X}_k(\theta) = \hat{a}_k e^{-i\theta} + \hat{a}_k^+
e^{i\theta}, \label{039}
\end{eqnarray}
\begin{eqnarray}
\hat{X}_{jk}(\theta) &=\:
\hat{a}_{jk} e^{-i\theta} + \hat{a}_{jk}^+ e^{i\theta}
&=\: \hat{X}_j(\theta) + \hat{X}_k(\theta),
\label{040}
\end{eqnarray}
where $\hat{a}_{kl}=\hat{a}_k+\hat{a}_l$. The operator $\hat{X}_k
(\hat{X}_{kl}$) for $\theta=0$ corresponds to the in-phase quadrature
component of the $k$-th  ($k$-th and $l$-th) mode (modes) of the
field, whereas for $\theta=\pi/2$ it corresponds  to the out-of-phase
component.  For brevity, we use the notation:
$\hat{X}_{k1}=\hat{X}_{k}(0)$, $\hat{X}_{k2}= \hat{X}_{k}(\pi/2)$, as
well as $\hat{X}_{kl1}= \hat{X}_{kl}(0)$ and $\hat{X}_{kl2}=
\hat{X}_{kl}(\pi/2)$. The following commutation rules hold:
\begin{eqnarray}
\left[ \hat{X}_{k1}, \hat{X}_{k2}\right] = 2i, \label{041}
\end{eqnarray}
\begin{eqnarray}
\left[ \hat{X}_{kl1}, \hat{X}_{kl2}\right] = 4i. \label{042}
\end{eqnarray}
Firstly, we shall discuss in brief the single-mode case. The
variances of the $\theta$-dependent quadrature~(\ref{039})
are
\begin{eqnarray}
\langle ( \Delta \hat{X}_k(\theta))^2\rangle = 2 {\rm Re}\left[
e^{-2i\theta} \langle ( \Delta \hat{a}_{k})^2\rangle\right]
+\langle \left\{ \Delta
\hat{a}^+_{k},\Delta\hat{a}_k\right\}\rangle, \label{043}
\end{eqnarray}
which obviously give $\hat{X}_{k1}$ and $\hat{X}_{k2}$ in special
cases. The Heisenberg uncertainty relation for quadratures,
\begin{eqnarray}
\langle (\Delta \hat{X}_{k1} )^2\rangle
\langle ( \Delta \hat{X}_{k2} )^2\rangle \geq 1,
\label{044}
\end{eqnarray}
lays the basis for the definition of ``usual'' (``standard'')
squeezing. The state of the field is said to be squeezed if
the variance $\hat{X}_{k1}$ or $\hat{X}_{k2}$ becomes smaller
than unity (in general, smaller than the square root of the
right side of the uncertainty relation for the quadratures).
Equivalently, light whose quantum fluctuations in the one
quadrature are smaller than those associated with coherent
light (minimizing the uncertainty relation) is called
squeezed (in the usual meaning).  Since, for a given quantum
state, the variance~(\ref{043}) is still dependent on
$\theta$, the angle $\theta$ can be chosen in a way to
minimize (or maximize) the variance.  Differentiation with
respect to $\theta$ leads to the angles $\theta_+$ and
$\theta_-$ for the maximal and minimal variances,
respectively, given by the relation~\cite{r284,r286}
\begin{eqnarray}
\exp\left( 2i\theta_±\right) = \pm \left( \frac{ \langle \left(
\Delta \hat{a}_{k}\right)^2\rangle }{ \langle \left( \Delta
\hat{a}^+_{k}\right)^2\rangle } \right)^{1/2}, \label{045}
\end{eqnarray}
where the difference between the angles $\theta_+$ and
$\theta_-$ is $\pi/2$. On inserting~(\ref{045}) into
(\ref{043}) one obtains the extremal variances
\begin{eqnarray}
\langle ( \Delta \hat{X}_{k\pm})^2\rangle &\equiv&
\langle ( \Delta \hat{X}_{k}(\theta_±))^2\rangle
\nonumber\\
&=& \pm 2 \left| \langle ( \Delta \hat{a}_{k})^2\rangle \right|
+\langle \left\{ \Delta \hat{a}^+_{k},\Delta \hat{a}_{k} \right\}
\rangle. \label{046}
\end{eqnarray}
It is noteworthy that the $\theta$-dependent
variance~(\ref{043}) can be expressed in terms of the
extremal variances
\begin{eqnarray}
\langle ( \Delta \hat{X}_{k}(\theta) )^2\rangle &=& \langle (
\Delta \hat{X}_{k-} )^2\rangle \cos^2(\theta-\theta_-)
\nonumber\\
&&+\langle ( \Delta \hat{X}_{k+})^2\rangle \sin^2(\theta-\theta_-)
\label{047},
\end{eqnarray}
which is the equation for Booth's elliptical lemniscate in
polar coordinates~\cite{r286}. The principal squeezing,
introduced by Luk\v{s} et al.~\cite{r284,r285}, occurs if the
minimum variance is less than unity:
\begin{eqnarray}
\langle ( \Delta
\hat{X}_{k-})^2\rangle &\leq& 1.
\label{048}
\end{eqnarray}
From~(\ref{046}) it follows that the principal squeezing
requires the fulfillment of the condition
\begin{eqnarray}
\langle \Delta
\hat{a}^+_{k}\Delta \hat{a}_{k} \rangle &<& \left| \langle (\Delta
\hat{a}_{k})^2\rangle \right|,
\label{049}
\end{eqnarray}
whereas the condition for standard squeezing, in view of
(\ref{043}), is
\begin{eqnarray}
{\rm min}
\left\{
\langle \Delta \hat{a}^+_{k}\Delta \hat{a}_{k} \rangle \pm
{\rm Re}\left[ \langle(\Delta \hat{a}_{k})^2\rangle \right]
\right\} &<& 0.
\label{050}
\end{eqnarray}
The mathematically elegant formalism of principal squeezing
(in particular other equivalent conditions for principal
squeezing) can be formulated using the generalized Heisenberg
uncertainty relation (the Schr\"odinger uncertainty
relation)~\cite{r287,r284,r316}:
\begin{eqnarray}
\langle ( \Delta \hat{X}_{k1})^2\rangle
\langle ( \Delta \hat{X}_{k2})^2\rangle
&\geq& \frac{1}{4}
\langle \left\{ \Delta \hat{X}_{k1},
\Delta \hat{X}_{k2}\right\}\rangle^2 +1,
\label{051}
\end{eqnarray}
which includes the Wigner covariance (cross-correlation) of the
quadratures $\hat{X}_{k1}$ and $\hat{X}_{k2}$ equal to
\begin{eqnarray}
\langle \left\{ \Delta \hat{X}_{k1}, \Delta \hat{X}_{k2}
\right\}\rangle= 4 {\rm Im} \left[\langle ( \Delta
\hat{a}_{k})^2\rangle \right]. \label{052}
\end{eqnarray}
For extremal variances  $\langle(\Delta\hat{X}_{k\pm})^2\rangle$ the
generalized Heisenberg relation reduces to the standard
uncertainty relation.

The generalization of the above definitions for the two-mode
case is straightforward.  By virtue of the commutator
(\ref{042}), twice as great as for the single-mode case
(\ref{041}), the standard and principal squeezing can be
defined, respectively, as
\begin{eqnarray}
{\rm min}\left\{
\langle ( \Delta \hat{X}_{kl1})^2\rangle,
\langle ( \Delta \hat{X}_{kl2})^2\rangle
\right\} &\leq& 2,
\label{053}
\end{eqnarray}
\begin{eqnarray}
\langle ( \Delta \hat{X}_{kl-})^2\rangle
&\leq& 2.
\label{054}
\end{eqnarray}
We express the two-mode variances and the Wigner covariances
in terms of the single-mode moments:
\begin{eqnarray}
\langle ( \Delta \hat{X}_{kl})^2\rangle = \langle ( \Delta
\hat{X}_{k})^2\rangle  + \langle ( \Delta \hat{X}_{l})^2\rangle  +
2\langle \Delta \hat{X}_{k}\Delta \hat{X}_{l}\rangle, \label{055}
\end{eqnarray}
\begin{eqnarray}
\langle \left\{\Delta \hat{X}_{kl1},
\Delta \hat{X}_{kl2}\right\}\rangle
=
\langle \left\{\Delta \hat{X}_{k1},\Delta \hat{X}_{k2}\right\}\rangle
+\langle \left\{\Delta \hat{X}_{l1},\Delta \hat{X}_{l2}\right\}\rangle
\nonumber\\
+2\langle \Delta \hat{X}_{k1}\Delta \hat{X}_{l2}\rangle +2\langle
\Delta \hat{X}_{k2}\Delta \hat{X}_{l1}\rangle, \label{056}
\end{eqnarray}
where $\hat{X}_{kl}$ stands for $\hat{X}_{kl}(\theta)$ (in
particular the quadratures). Relations such as~(\ref{055})
and~(\ref{056}) for the quadratures hold for the two-mode
creation and annihilation operators $\hat{a}_{kl}^±$.  The
moments $\langle(\hat{X}_k)^2\rangle$ and
$\langle\{\Delta\hat{X}_{k1}, \Delta\hat{X}_{k2}\}\rangle$
are given by~(\ref{043}) and~(\ref{052}). The remaining
cross-correlations have the following form in terms of the
annihilation and creation operators:
\begin{eqnarray}
\langle \Delta \hat{X}_{k1}\Delta \hat{X}_{l1}\rangle &=& 2 {\rm
Re}\left[ \langle \Delta \hat{a}_{k}\Delta \hat{a}_{l}\rangle
+\langle \Delta \hat{a}^+_{k}\Delta \hat{a}_{l}\rangle \right],
\nonumber\\
\langle \Delta \hat{X}_{k2}\Delta \hat{X}_{l2}\rangle &=& 2 {\rm
Re}\left[ \:-\langle \Delta \hat{a}_{k}\Delta \hat{a}_{l}\rangle
+\langle \Delta \hat{a}^+_{k}\Delta \hat{a}_{l}\rangle \right],
\nonumber\\
\langle \Delta \hat{X}_{k1}\Delta \hat{X}_{l2}\rangle &=& 2 {\rm
Im}\left[ \langle \Delta \hat{a}_{k}\Delta \hat{a}_{l}\rangle
+\langle \Delta \hat{a}^+_{k}\Delta \hat{a}_{l}\rangle \right],
\nonumber\\
\langle \Delta \hat{X}_{k2}\Delta \hat{X}_{l1}\rangle &=& 2 {\rm
Im}\left[ \langle \Delta \hat{a}_{k}\Delta \hat{a}_{l}\rangle
-\langle \Delta \hat{a}^+_{k}\Delta \hat{a}_{l}\rangle \right].
\label{057}
\end{eqnarray}
Substituting Eqs.~(\ref{043}),~(\ref{052}) and~(\ref{057})
into~(\ref{055}) and~(\ref{056}) one obtains explicit
dependencies of the two-mode quadrature moments on the
annihilation operators~\cite{r285}.

Alternatively, the single-mode moments~(\ref{043}),
(\ref{046}) and~(\ref{052}) and the conditions~(\ref{049})
and~(\ref{050}) for single-mode squeezing can be generalized
to a two-mode case by simple replacement of $\hat{a}_k$,
$\hat{X}_k(\theta)$ by $\hat{a}_{kl}$,
$\hat{X}_{kl}(\theta)$, showing complete analogy between the
single- and two-mode descriptions. In particular, the
two-mode extremal variances are
\begin{eqnarray}
\langle ( \Delta \hat{X}_{kl\pm})^2\rangle = \pm 2 \left| \langle
( \Delta \hat{a}_{kl})^2\rangle \right| +\langle \left\{ \Delta
\hat{a}^+_{kl},\Delta \hat{a}_{kl} \right\} \rangle, \label{058}
\end{eqnarray}
by analogy to~(\ref{046}).

\section{Fokker-Planck equation}

\subsection{Raman scattering including pump depletion}

The master equation (ME) is the quantum equation of motion
for operators and hence it is possible to solve it  directly
only for a small class of models. As an example we cite the
Raman-coupled model of Knight~\cite{r156} and its
generalizations (Ref.~\cite{r214} and references therein).
Usually the quantum master equation is converted to a
classical differential equation.  Then, standard methods of
mathematical analysis  can be applied. In this Section we
present one of the most popular methods: transformation to a
generalized Fokker-Planck equation (FPE) or equivalently to
an equation of motion for characteristic functions.  This
method is extensively studied in a number of
textbooks~\cite{r59,r29,r155,r79} and consists of performing
s-ordering of the field operators in the ME~(\ref{018}) and
then applying the quantum-classical number correspondence of
coherent-state technique. The rules for the transformation of
the ME into Fokker-Planck equations for the $s$-parametrized
quasidistribution ${\cal W}^{(s)}(\alpha,\alpha^*,\bar{A})$
are the following (e.g. Refs.~\cite{r216} and~\cite{r217}):
\begin{eqnarray} \left\{
\begin{array}{l} \hat{A}\hat{a}\\
\hat{a}\hat{A}
\end{array}
\right\} &\mapsto& \left( \alpha -\frac{s\pm 1}{2}
\frac{\partial}{\partial \alpha^*}\right) {\cal
W}^{(s)}(\alpha,\alpha^*,\bar{A}),
\nonumber\\
\left\{ \begin{array}{l}
\hat{a}^+\hat{A}\\
\hat{A}\hat{a}^+
\end{array}
\right\} &\mapsto& \left( \alpha^* -\frac{s\pm 1}{2}
\frac{\partial}{\partial \alpha}\right) {\cal
W}^{(s)}(\alpha,\alpha^*,\bar{A}), \label{059}
\end{eqnarray}
where $\hat{A}$ is an arbitrary operator; in particular,
$\hat{A}$ can be the density matrix $\hat{\rho}$; $\bar{A}$
is the classical function associated with the operator $A$;
the parameter $s$ takes arbitrary values in the range
$\langle -1,1\rangle$. If necessary, these rules can be
applied repeatedly. Similarly, we list the rules of
transformation of the master equation~(\ref{018}) to the
equation of motion for the $s$-parametrized characteristic
function ${\cal C}^{(s)}(\beta,\beta^*,\bar {A})$  (e.g.
Ref.~\cite{r216}):
\begin{eqnarray}
\left\{ \begin{array}{l}
\hat{A}\hat{a}\\
\hat{a}\hat{A}
\end{array}
\right\} &\mapsto& \left( -\frac{\partial}{\partial\beta^*}
+\frac{s\pm 1}{2}\beta\right) {\cal
C}^{(s)}(\beta,\beta^*,\bar{A}),
\nonumber\\
\left\{ \begin{array}{l}
\hat{a}^+\hat{A}\\
\hat{A}\hat{a}^+
\end{array}
\right\} &\mapsto& \left( \frac{\partial}{\partial\beta}
-\frac{s\pm 1}{2}\beta^*\right) {\cal
C}^{(s)}(\beta,\beta^*,\bar{A}). \label{060}
\end{eqnarray}
Applying repeatedly the rule~(\ref{059}) of transformation to
the master equation~(\ref{018}) and after some lengthy
algebra we finally arrive at the generalized Fokker-Planck
equation for the $s$-parametrized quasidistribution ${\cal
W}^{(s)}\equiv {\cal W}^{(s)}(\alpha_L,\alpha_S,\alpha_A,t)$:
\begin{eqnarray}
\lefteqn{ \frac{\partial }{\partial t} {\cal W}^{(s)} \:=\:
\frac{1}{2} \gamma_S \mbox{\Huge \{ \huge[}
-\frac{\partial}{\partial \alpha_L}\alpha_L
\frac{\partial}{\partial \alpha_S}\alpha_S \null} \nonumber\\
&&\null + \left( |\alpha_S|^2 +\frac{1+s}{2}\right)
\frac{\partial}{\partial \alpha_L}\alpha_L - \left( |\alpha_L|^2
-\frac{1-s}{2}\right) \frac{\partial}{\partial \alpha_S}\alpha_S
\nonumber\\ &&\null +\frac{1-s^2}{4} \left(
\frac{\partial}{\partial \alpha_L}\alpha_L
\frac{\partial}{\partial \alpha_S} \frac{\partial}{\partial
\alpha_S^*} - \frac{\partial}{\partial \alpha_L}
\frac{\partial}{\partial \alpha_L^*} \frac{\partial}{\partial
\alpha_S}\alpha_S \right) + c.c. \mbox{ \huge]} \nonumber\\
&&\null +\left[ (1-s) |\alpha_S|^2 +\frac{1-s^2}{2}\right]
\frac{\partial}{\partial \alpha_L} \frac{\partial}{\partial
\alpha_L^*} \nonumber\\ &&\null + \left[ (1+s) |\alpha_L|^2
-\frac{1-s^2}{2}\right] \frac{\partial}{\partial \alpha_S}
\frac{\partial}{\partial \alpha_S^*} \mbox{ \Huge\}} {\cal
W}^{(s)}
\nonumber\\ &&\null +\frac{1}{2} \gamma_A \mbox{\Huge \{ \huge[}
-\frac{\partial}{\partial \alpha_L}\alpha_L
\frac{\partial}{\partial \alpha_A}\alpha_A \nonumber\\ &&\null -
\left( |\alpha_A|^2 -\frac{1-s}{2}\right) \frac{\partial}{\partial
\alpha_L}\alpha_L +\left( |\alpha_L|^2 +\frac{1+s}{2}\right)
\frac{\partial}{\partial \alpha_A}\alpha_A \nonumber\\ &&\null
-\frac{1-s^2}{4} \left( \frac{\partial}{\partial \alpha_L}\alpha_L
\frac{\partial}{\partial \alpha_A} \frac{\partial}{\partial
\alpha_A^*} - \frac{\partial}{\partial \alpha_L}
\frac{\partial}{\partial \alpha_L^*} \frac{\partial}{\partial
\alpha_A}\alpha_A \right) + c.c. \mbox{ \huge]} \nonumber\\
&&\null +\left[ (1+s) |\alpha_A|^2 -\frac{1-s^2}{2}\right]
\frac{\partial}{\partial \alpha_L} \frac{\partial}{\partial
\alpha_L^*} \nonumber\\ &&\null +\left[ (1-s) |\alpha_L|^2
+\frac{1-s^2}{2}\right] \frac{\partial}{\partial \alpha_A}
\frac{\partial}{\partial \alpha_A^*} \mbox{ \Huge\}} {\cal
W}^{(s)} \nonumber\\ &&\null
+\mbox{ \Huge\{} \frac{1}{2} \gamma_{SA} \exp(-2 i
\Delta\Omega\Delta t) \mbox{ \huge[} -\alpha_L^2 \left(
\alpha_A^*\frac{\partial}{\partial \alpha_S}
+\frac{\partial}{\partial \alpha_S} \frac{\partial}{\partial
\alpha_A} -\alpha_S^*\frac{\partial}{\partial \alpha_A} \right)
\nonumber\\ &&\null +\alpha_L \left(
(1+s)\alpha_A^*\frac{\partial}{\partial \alpha_S}
+(1-s)\alpha_S^*\frac{\partial}{\partial \alpha_A} \right)
\frac{\partial}{\partial \alpha_L^*} \nonumber\\ &&\null +\left(
\frac{1-s^2}{4} \alpha_A^*\frac{\partial}{\partial \alpha_S}
-\alpha_S^*\alpha_A^* -\frac{1-s^2}{4}
\alpha_S^*\frac{\partial}{\partial \alpha_A} \right)
\frac{\partial^2}{\partial \alpha_L^{*2}} \mbox{\huge ]} + c.c.
\mbox{\Huge \}}  {\cal W}^{(s)} \nonumber\\ &&\null
+\gamma_S \langle \hat{n}_V \rangle \mbox{\Huge \{} \left(
\frac{1}{2}\frac{\partial}{\partial \alpha_L}\alpha_L
-\frac{\partial}{\partial \alpha_L}\alpha_L
\frac{\partial}{\partial \alpha_S}\alpha_S
+\frac{1}{2}\frac{\partial}{\partial \alpha_S}\alpha_S + c.c.
\right) \nonumber\\ &&\null +\left| \alpha_S\right|^2
\frac{\partial}{\partial \alpha_L} \frac{\partial}{\partial
\alpha_L^*}\ +\left| \alpha_L\right|^2 \frac{\partial}{\partial
\alpha_S}\ \frac{\partial}{\partial \alpha_S^*} \mbox{\Huge \}}
{\cal W}^{(s)} \nonumber\\ &&\null
+\gamma_A \langle \hat{n}_V \rangle \mbox{\Huge \{} \left(
\frac{1}{2}\frac{\partial}{\partial \alpha_L}\alpha_L
-\frac{\partial}{\partial \alpha_L}\alpha_L
\frac{\partial}{\partial \alpha_A}\alpha_A
+\frac{1}{2}\frac{\partial}{\partial \alpha_A}\alpha_A + c.c.
\right) \nonumber\\ &&\null +\left| \alpha_A\right|^2
\frac{\partial}{\partial \alpha_L} \frac{\partial}{\partial
\alpha_L^*}\ +\left| \alpha_L\right|^2 \frac{\partial}{\partial
\alpha_A}\ \frac{\partial}{\partial \alpha_A^*} \mbox{\Huge \}}
{\cal W}^{(s)} \nonumber\\ &&\null
-\mbox{ \Huge \{} \gamma_{AS} \langle \hat{n}_V \rangle \exp(2 i
\Delta\Omega\Delta t) \mbox{\huge[} \alpha_S^*\alpha_A^*
\frac{\partial^2}{\partial \alpha_L^{*2}}
+\alpha_L^{2}\frac{\partial}{\partial \alpha_S}
\frac{\partial}{\partial \alpha_A} \nonumber\\ &&\null -\alpha_L
\left( \alpha_A^*\frac{\partial^2}{\partial \alpha_S}
+\alpha_S^*\frac{\partial^2}{\partial \alpha_A} \right)
\frac{\partial}{\partial \alpha_L^*} \mbox{\huge ]} + c.c.
\mbox{\Huge \}}  {\cal W}^{(s)}, \label{061}
\end{eqnarray}
which is a generalization of our former relation for
$\gamma_A,\gamma_{SA}\neq 0$ and arbitrary parameter $s$
(Ref.~\cite{r68}). For brevity, we refer to the generalized
Fokker-Planck equation~\cite{r79} simply as the Fokker-Planck
equation (FPE).  The physical interpretation of~(\ref{061})
can be given in the same manner as the interpretation of the
appropriate terms in the ME~(\ref{018}) given in Section~2.
The FPE~(\ref{061}) exhibits a highly complicated structure.
Nonetheless, the equations of motion for the mean values
$\langle\alpha_k\rangle$, $\langle\alpha_k\alpha_l\rangle$,
$\langle\alpha^*_k\alpha_l\rangle$, (with $k,l=L,S,A$) can be
calculated. In particular, we obtain
\begin{eqnarray}
\frac{d}{dt}(\langle\hat{n}_L(t)\rangle +
\langle\hat{n}_S(t)\rangle+\langle\hat{n}_A(t)\rangle) =0,
\label{ex1}
\end{eqnarray}
with $\langle\hat{n}_k(t)\rangle
=\langle\alpha^*_k\alpha_k\rangle$. Eq.~(\ref{ex1}) states
that the total mean number of photons (in all radiation
modes) is a constant of motion.

The FPE~(\ref{061}) contains terms of the form
$\frac{\partial}{\partial \alpha_i} \alpha_j \alpha_k
\alpha_l {\cal W}^{(s)}$, $\frac{\partial}{\partial \alpha_i}
\frac{\partial}{\partial \alpha_j} \alpha_k \alpha_l {\cal
W}^{(s)}$, $\frac{\partial}{\partial \alpha_i}
\frac{\partial}{\partial \alpha_j} \frac{\partial}{\partial
\alpha_k} \alpha_l {\cal W}^{(s)}$, where $\alpha_i,
\alpha_j, \alpha_k, \alpha_l= \alpha_L, \alpha^*_L, \alpha_S,
\alpha^*_S, \alpha_A, \alpha_A^*$. It is seen that most
components of the drift vector are nonlinear to the third
order in $\alpha$, and most components of the diffusion
matrix are nonlinear up to the second order. It is
particularly difficult to solve a differential equation with
such nonlinear diffusion and drift coefficients. Besides, the
FPE~(\ref{061}) for ${\cal W}^{(s)}\equiv {\cal
W}^{(s)}(\alpha_L,\alpha_S, \alpha_A,t)$ with the parameter
$s\neq\pm 1$ contains third-order derivatives in the terms
$\frac{\partial}{\partial \alpha_i} \frac{\partial}{\partial
\alpha_j} \frac{\partial}{\partial \alpha_k} \alpha_l {\cal
W}^{(s)}$. This could be expected since in many
models~\cite{r79,r44}, for instance in the anharmonic
oscillator model (for references see Ref.~\cite{r300}), there
occur third-order derivatives in the FPEs for the Wigner
function ($s=0$).

The corresponding equation of motion for the $s$-parametrized
characteristic function ${\cal
C}^{(s)}(\beta_L,\beta_S,\beta_A,t)$ can be obtained either
from the ME~(\ref{018}) by performing the transformation
(\ref{060}), or from the FPE~(\ref{061}) by means of the
Fourier transformation~(\ref{022}) with respect to the
variables $\alpha_L$, $\alpha_S$, $\alpha_A$. Finally, we
arrive at the equation of motion for ${\cal C}^{(s)}\equiv
{\cal C}^{(s)}(\beta_L,\beta_S,\beta_A,t)$:
\begin{eqnarray}
\lefteqn{ \frac{\partial }{\partial t} {\cal C}^{(s)} \:=\:
\frac{1}{2} \gamma_S \mbox{\Huge \{ \huge[} -\beta_L\beta_S
\frac{\partial}{\partial \beta_L}\ \frac{\partial}{\partial
\beta_S}\ \null} \nonumber\\ &&\null +\beta_L\left(
-\frac{1+s}{2}+ \frac{\partial}{\partial\beta_S}
\frac{\partial}{\partial\beta_S^*} \right)
\frac{\partial}{\partial\beta_L} \nonumber\\ &&\null
-\beta_S\left( \frac{1-s}{2}+ \frac{\partial}{\partial\beta_L}
\frac{\partial}{\partial\beta_L^*} \right)
\frac{\partial}{\partial\beta_S} \nonumber\\ &&\null
+\frac{1-s^2}{4}\left( |\beta_S|^2 \beta_L
\frac{\partial}{\partial\beta_L} -|\beta_L|^2 \beta_S
\frac{\partial}{\partial\beta_S} \right) + c.c. \mbox{ \huge ]}
\nonumber\\ &&\null +|\beta_L|^2 \left[ -\frac{1-s^2}{2} +(1-s)
\frac{\partial}{\partial\beta_S}
\frac{\partial}{\partial\beta_S^*} \right] \nonumber\\ &&\null
+|\beta_S|^2 \left[ \frac{1-s^2}{2} +(1+s)
\frac{\partial}{\partial\beta_L}
\frac{\partial}{\partial\beta_L^*} \right] \mbox{ \Huge \}} {\cal
C}^{(s)}
\nonumber\\ &&\null +\frac{1}{2} \gamma_A \mbox{\Huge \{ \huge[}
-\beta_L\beta_A \frac{\partial}{\partial \beta_L}\
\frac{\partial}{\partial \beta_A}\ \nonumber\\ &&\null
-\beta_L\left( \frac{1-s}{2}+ \frac{\partial}{\partial\beta_A}
\frac{\partial}{\partial\beta_A^*} \right)
\frac{\partial}{\partial\beta_L} \nonumber\\ &&\null
+\beta_A\left( -\frac{1+s}{2}+ \frac{\partial}{\partial\beta_L}
\frac{\partial}{\partial\beta_L^*} \right)
\frac{\partial}{\partial\beta_A} \nonumber\\ &&\null
+\frac{1-s^2}{4}\left( -|\beta_A|^2 \beta_L
\frac{\partial}{\partial\beta_L} +|\beta_L|^2 \beta_A
\frac{\partial}{\partial\beta_A} \right) + c.c. \mbox{ \huge ]}
\nonumber\\ &&\null +|\beta_L|^2 \left[\frac{1-s^2}{2} +(1+s)
\frac{\partial}{\partial\beta_A}
\frac{\partial}{\partial\beta_A^*} \right] \nonumber\\ &&\null
+|\beta_A|^2 \left[ -\frac{1-s^2}{2} +(1-s)
\frac{\partial}{\partial\beta_L}
\frac{\partial}{\partial\beta_L^*} \right] \mbox{ \Huge \}} {\cal
C}^{(s)} \nonumber\\ &&\null
+\mbox{ \Huge\{} \frac{1}{2} \gamma_{SA} \exp(-2 i
\Delta\Omega\Delta t) \mbox{ \huge[} \left( \beta_A
\frac{\partial}{\partial\beta_S^*} -\beta_S \beta_A -\beta_S
\frac{\partial}{\partial\beta_A^*} \right)
\frac{\partial^2}{\partial\beta_L^2} \nonumber\\ &&\null
+\beta_L^*\left( (1-s) \beta_A \frac{\partial}{\partial\beta_S^*}
+(1+s) \beta_S \frac{\partial}{\partial\beta_A^*} \right)
\frac{\partial}{\partial\beta_L} \nonumber\\ &&\null -\beta_L^{*2}
\left( \frac{1-s^2}{4} \beta_A \frac{\partial}{\partial\beta_S^*}
+\frac{\partial}{\partial\beta_S^*}
\frac{\partial}{\partial\beta_A^*} -\frac{1-s^2}{4} \beta_S
\frac{\partial}{\partial\beta_A^*} \right) \mbox{ \huge ]} + c.c.
\mbox{ \Huge \}} {\cal C}^{(s)}
\nonumber\\ &&\null +\gamma_S \langle \hat{n}_V \rangle \mbox{
\Huge \{} -\left( \frac{1}{2}\beta_L
\frac{\partial}{\partial\beta_L} +\beta_L\beta_S
\frac{\partial}{\partial\beta_L} \frac{\partial}{\partial\beta_S}
+\frac{1}{2}\beta_S \frac{\partial}{\partial\beta_S} + c.c.\right)
\nonumber\\ &&\null +|\beta_S|^2 \frac{\partial}{\partial\beta_L}
\frac{\partial}{\partial\beta_L^*} +|\beta_L|^2
\frac{\partial}{\partial\beta_S}
\frac{\partial}{\partial\beta_S^*} \mbox{ \Huge \}} {\cal C}^{(s)}
\nonumber\\ &&\null +\gamma_A \langle \hat{n}_V \rangle \mbox{
\Huge \{} -\left( \frac{1}{2}\beta_L
\frac{\partial}{\partial\beta_L} +\beta_L\beta_A
\frac{\partial}{\partial\beta_L} \frac{\partial}{\partial\beta_A}
+\frac{1}{2}\beta_A \frac{\partial}{\partial\beta_A} + c.c.\right)
\nonumber\\ &&\null +|\beta_A|^2 \frac{\partial}{\partial\beta_L}
\frac{\partial}{\partial\beta_L^*} +|\beta_L|^2
\frac{\partial}{\partial\beta_A}
\frac{\partial}{\partial\beta_A^*} \mbox{ \Huge \}} {\cal C}^{(s)}
\nonumber\\ &&\null -\mbox{ \Huge\{} \frac{1}{2} \gamma_{AS}
\langle \hat{n}_V \rangle \exp(2 i\Delta\Omega\Delta t) \mbox{
\huge[} \beta_S\beta_A \frac{\partial^2}{\partial\beta_L^{2}}
+\beta_L^{*2} \frac{\partial}{\partial\beta_S^*}
\frac{\partial^2}{\partial\beta_A^*} \nonumber\\ &&\null
-\beta_L^* \left( \beta_S \frac{\partial}{\partial\beta_A^*}
+\beta_A \frac{\partial}{\partial\beta_S^*} \right)
\frac{\partial}{\partial\beta_L} \mbox{ \huge ]} +c.c. \: \mbox{
\Huge \}} {\cal C}^{(s)}. \label{063}
\end{eqnarray}
Here, we come upon similar difficulties in the way of
obtaining an analytical solution of~(\ref{063}) as in the FPE
case~(\ref{061}), inherent in the nonlinearity of the
coefficients of the terms with first- and second-order
derivatives as well as the presence of terms with third-order
derivatives. Nevertheless, in contradistinction to ${\cal
W}^{(s)}$, the existence of a solution for ${\cal C}^{(s_1)}$
implies, in view of the property~(\ref{027}), the existence
of a solution for any other parameter $s_2$.

In view of the particularly complicated structure of Eqs.
(\ref{061}) and~(\ref{063}) or equivalent equations of motion
derived within the completely quantum model of scattering
into both the Stokes and anti-Stokes modes, it would seem
that a solution in exact closed form cannot be
obtained~\cite{r79}. It is necessary to apply further
restrictions or approximations in the model to achieve an
analytical solution of~(\ref{061}). In Sect. 6.1.2 we present
a strict analytical solution of the two-mode ME (including
pump depletion) in terms of Fock states by applying the
Laplace transform. In Sect. 5.2 we present solutions of
two-mode linearized FPEs for ${\cal W}^{(s)}(\alpha_S,
\alpha_A,t)$ and solutions of equivalent equations of motion
for ${\cal C}^{(s)}(\beta_S,\beta_A,t)$ in the Raman
scattering model under parametric approximation. In Appendix
A we give the solution of a linearized form of the three-mode
FPE (\ref{061}) for ${\cal W}^{(-1)}(\alpha_L,\alpha_S,
\alpha_A,t)$ properly describing the evolution of the
radiation fields valid only on the assumption of small
fluctuations of the fields around their mean values. There,
we restrict our considerations to the $Q$-function ($s=-1$)
to avoid problems of the existence of the quasidistribution
${\cal W}^{(s)}(\alpha_L,\alpha_S,\alpha_A,t)$ (particularly
important in the case of $s$ close or equal to 1) and to
simplify the third-order FPE~(\ref{061}) to second-order,
which takes place for $s=\pm1$.  Within a similar model of
Raman scattering from a single phonon mode, Szlachetka et
al.~\cite{r114,r114a,r113,r114b} (see also Ref.~\cite{r110})
and T\"anzler and Sch\"utte~\cite{r118} have solved the
equations of motion in the short-time approximation up to the
second power in time. Within the latter (single phonon mode)
model Pe\v{r}ina and K\v{r}epelka~\cite{r81,r81a} have
obtained approximate solutions using the approximation of
small fluctuations around a stationary solution.

\subsection{Raman scattering without pump depletion}

Here, to find a solution of the ME~(\ref{018}) we apply the
parametric approximation, so no allowance for pump depletion
is included. The trilinear Hamiltonians $\hat{H}_A$,
$\hat{H}_S$~(\ref{002}) can be reduced to bilinear functions
as a result of the replacement of the annihilation operator
$\hat{a}_L$, representing the quantum pump field, by the
classical complex amplitude of the pump field, $e_L$. This
approximation effectively linearizes our model of Raman
scattering.  Then, the Fokker-Planck equation for the
two-mode $s$-parametrized quasidistribution ${\cal
W}^{(s)}(\alpha_S,\alpha_A,t)$ takes the form
\begin{eqnarray}
\lefteqn{ \frac{\partial}{\partial t} {\cal
W}^{(s)}(\alpha_S,\alpha_A,t) \:=\: \mbox{\Huge \{} -\left[
\left(\frac{\kappa_S}{2} +i \Delta\Omega\right)
\frac{\partial}{\partial\alpha_S}\alpha_S +c.c. \right] \null}
\nonumber\\ &&\null +\left[ \left(\frac{\kappa_A}{2} -i
\Delta\Omega\right) \frac{\partial}{\partial\alpha_A}\alpha_A
+c.c. \right] \nonumber\\ &&\null - \left[
\frac{\kappa_{SA}}{2}\left(
\alpha_A^*\frac{\partial}{\partial\alpha_S}
-\alpha_S^*\frac{\partial}{\partial\alpha_A} \right) + c.c.
\right] \nonumber\\ &&\null +\kappa_{S} \left( \langle \hat{n}_V
\rangle+\frac{s+1}{2} \right)
\frac{\partial^2}{\partial\alpha_S\partial\alpha_S^*} \nonumber\\
&&\null +\kappa_{A}\left( \langle \hat{n}_V \rangle+\frac{1-s}{2}
\right) \frac{\partial^2}{\partial\alpha_A\partial\alpha_A^*}
\nonumber\\ &&\null -\left[ \kappa_{SA}\left( \langle \hat{n}_V
\rangle+\frac{1}{2} \right)
\frac{\partial^2}{\partial\alpha_S\partial\alpha_A} +c.c. \right]
\mbox{\Huge \}} {\cal W}^{(s)}(\alpha_S,\alpha_A,t) \label{064}
\end{eqnarray}
on applying the rules~(\ref{059}) to the ME~(\ref{018}) with
complex classical amplitude $e_L$ instead of the annihilation
operator $\hat{a}_L$ and transforming the variables
$\alpha_k\rightarrow \exp(-i\Delta\Omega \Delta t)\alpha_k$
and $\beta_k\rightarrow \exp(-i\Delta\Omega \Delta t)\beta_k$
with the frequency mismatch $\Delta\Omega$ defined by
(\ref{012}). For brevity, we have incorporated the complex
amplitude $e_L$ into the coupling constants $\kappa_S
=\gamma_S|e_L|^2$, $\kappa_A =\gamma_A|e_L|^2$, and
$\kappa_{SA} =\kappa_{AS}^* =\gamma_{SA} e_L^2$. The equation
(\ref{064}) is a generalization for any parameter $s$
($s\in\langle -1,1\rangle$) of the FPE given by Walls for the
$P$-function ($s=1$)~\cite{r132b} and by Pe\v{r}ina for the
$P$- and $Q$-functions ($s=\pm 1$)~\cite{r77,r78}. If we
consider production of the Stokes radiation only, neglecting
anti-Stokes scattering, then Eq.~(\ref{064}) reduces to the
$s$-parametrized FPE obtained by Pe\v{r}inov\'a et
al.~\cite{r306}.

We can interpret the FPE~(\ref{064}) in the same manner as
the ME~(\ref{018}). The first term in~(\ref{064}) describes
the amplification of the Stokes beam, whereas the second term
describes the attenuation of the anti-Stokes beam; the third
term shows the coupling between the Stokes and anti-Stokes
fields; the remaining three terms account for the noise
diffusion from the ``noisy'' (for nonzero temperature)
reservoir into the system. Contrary to the former equations
of motion~(\ref{018}),~(\ref{061}) and~(\ref{060}), we lose
all information about the depletion of the laser field. It is
seen that the FPE~(\ref{064}) for any quasidistribution
${\cal W}^{(s)}$ even if related to the field ordering $s\neq
\pm1$, does not contain third-order derivatives, contrary to
the FPE~(\ref{061}) without parametric approximation. Let us
note that~(\ref{064}) describes an Ornstein-Uhlenbeck
process~\cite{r296} since the components of the drift vector
are linear and those of the diffusion matrix are constant.
Various methods have been developed for solving the equations
of motion for Ornstein-Uhlenbeck processes~\cite{r155,r79}.
For instance, expressing the quasidistribution ${\cal
W}^{(s)}(\alpha_S,\alpha_A,t)$ by  its Fourier transform
(\ref{022}) with respect to the variables $\alpha_S$,
$\alpha_A$, we obtain the following first-order differential
equation for the Fourier transform, i.e., for the
characteristic function ${\cal C}^{(s)}(\beta_S,\beta_A,t)$:
\begin{eqnarray}
\frac{\partial }{\partial t} {\cal
C}^{(s)}&&\hspace{-7mm}(\beta_S,\beta_A,t) = \mbox{\Huge \{}
\left[ \left(\frac{\kappa_S}{2} -i \Delta\Omega\right)
\beta_S\frac{\partial}{\partial\beta_S} +c.c. \right] \nonumber\\
&&\null -\left[ \left(\frac{\kappa_A}{2} +i \Delta\Omega\right)
\beta_A\frac{\partial}{\partial\beta_A} +c.c. \right] \nonumber\\
&&\null + \left[ \frac{\kappa_{SA}}{2}\left(
\beta_A^*\frac{\partial}{\partial\beta_S}
-\beta_S^*\frac{\partial}{\partial\beta_A} \right) + c.c. \right]
\nonumber\\ &&\null -\kappa_{S} \left( \langle \hat{n}_V
\rangle+\frac{s+1}{2} \right) |\beta_S|^2 \nonumber\\ &&\null
-\kappa_{A}\left( \langle \hat{n}_V \rangle+\frac{1-s}{2} \right)
|\beta_A|^2 \nonumber\\ &&\null -\left[ \kappa_{AS}\left( \langle
\hat{n}_V \rangle+\frac{1}{2} \right) \beta_S\beta_A +c.c. \right]
\mbox{\Huge \}} {\cal C}^{(s)}(\beta_S,\beta_A,t). \label{065}
\end{eqnarray}
Again, to obtain~(\ref{022}), one might use the rules
(\ref{060}) applied to~(\ref{018}) with $\hat{a}_L\rightarrow
e_L$. Our further results presented in this Section are
mainly based on very extensive studies carried out by
Pe\v{r}ina~\cite{r77,r78}, Pe\v{r}inov\'a and
Pe\v{r}ina~\cite{r85b}, and K\'arsk\'a and
Pe\v{r}ina~\cite{r41} (see also Ref.~\cite{r79} and
references therein). However, their solutions of the
equations of motion for the Raman effect under parametric
approximation hold only for quasidistributions ${\cal
W}^{(1)}(\alpha_S,\alpha_A,t)$  or ${\cal
W}^{(-1)}(\alpha_S,\alpha_A,t)$ and characteristic functions
${\cal C}^{(\pm 1)}(\beta_S,\beta_A,t)$ related to normal
and/or antinormal ordering of the field operators. We
generalize their results to functions related to s-ordering
of the field operators, i.e., to an $s$-parametrized
quasidistribution ${\cal W}^{(s)}(\alpha_S,\alpha_A,t)$ and
$s$-parametrized characteristic function ${\cal
C}^{(s)}(\beta_S,\beta_A,t)$.  Let us use, after
Ref.~\cite{r79}, the following simplified notation for
functions characterizing the quantum noise, i.e., the Wigner
covariances and variances as well as the mean values of the
annihilation operators $\hat{a}_S$ and $\hat{a}_A$:
\begin{eqnarray}
B_{k}^{(s)}(t) &=& \frac{1}{2}\langle \left\{
\Delta\hat{a}^+_k(t), \Delta\hat{a}_k(t)
\right\}\rangle-\frac{s}{2},
\nonumber\\
D_{kl}(t) &=& D_{lk}(t) \:=\: \frac{1}{2}\langle \left\{ \Delta
\hat{a}_k(t), \Delta \hat{a}_l(t) \right\} \rangle,
\nonumber\\
\bar{D}_{kl}(t) &=& \bar{D}_{lk}^*(t) \:=\: -\frac{1}{2}\langle
\left\{ \Delta\hat{a}^+_k(t), \Delta\hat{a}_l(t) \right\}\rangle,
\nonumber\\
C_k(t) &=& \langle \left( \Delta\hat{a}_k(t) \right)^2 \rangle,
\nonumber\\
\xi_k(t) &=& \langle \hat{a}_k(t) \rangle, \label{066}
\end{eqnarray}
where $k=S,A$ and $\{..., ...\}$ is an anticommutator.
Assuming the initial condition that the Stokes and
anti-Stokes fields are stochastically independent, the
solution of~(\ref{065}) for the $s$-parametrized
characteristic function exists for any parameter $s$ and is
equal to
\begin{eqnarray}
{\cal C}^{(s)}(\beta_S,\beta_A,t) &=& \exp\mbox{\huge \{}
\sum_{k=S,A} \Big[ -B_{k}^{(s)}(t)|\beta_k|^2
\nonumber\\
&&+\left( \frac{1}{2}C^*_k(t)\beta_k^2+c.c.\right)
+\left(\beta_k\xi_{k}^{*}(t)-c.c.\right) \Big]
\nonumber\\
&&+ \left[ D_{SA}(t)\beta_S^*\beta_A^*
+\bar{D}_{SA}(t)\beta_S\beta_A^* +c.c. \right] \mbox{\huge \}},
\label{067}
\end{eqnarray}
where
\begin{eqnarray}
B_{S}^{(s)}(t)&=& \left(B_{S}^{(s)}+ \langle \hat{n}_V
\rangle+\frac{1+s}{2} \right) |U_S(t)|^2 \nonumber\\ &&+
\left(B_{A}^{(s)}-\langle \hat{n}_V \rangle-\frac{1-s}{2}
\right)|V_S(t)|^2 -\langle \hat{n}_V \rangle-\frac{1+s}{2},
\nonumber\\
B_{A}^{(s)}(t)&=& \left(B_{A}^{(s)}-\langle \hat{n}_V \rangle
-\frac{1-s}{2}\right)|U_A(t)|^2 \nonumber\\ &&+
\left(B_{S}^{(s)}+\langle \hat{n}_V \rangle+\frac{1+s}{2}
\right)|V_A(t)|^2 +\langle \hat{n}_V \rangle+\frac{1-s}{2},
\nonumber\\
D_{SA}(t) &=& \left(B_{S}^{(s)}+ \langle \hat{n}_V
\rangle+\frac{1+s}{2} \right)U_S(t)V_A(t) \nonumber\\ &&+
\left(B_{A}^{(s)}-\langle \hat{n}_V \rangle-\frac{1-s}{2}
\right)V_S(t)U_A(t),
\nonumber\\
\bar{D}_{SA}(t) &=& C_S U_S(t) V_A^*(t)+C^*_A U_A^*(t) V_S(t),
\nonumber\\
C_S(t)&= &C_S U^2_S(t)+C^*_A V^2_S(t),
\nonumber\\
C_A(t)&=& C_A U^2_A(t)+C^*_S V^2_A(t),
\nonumber\\
\xi_S(t)&=& U_S(t) \xi_S+ V_S(t) \xi_A^*,
\nonumber\\
\xi_A(t)&=& U_A(t) \xi_A+ V_A(t) \xi_S^*. \label{068}
\end{eqnarray}
The solution~(\ref{067}) for ${\cal
C}^{(s)}(\beta_S,\beta_A,t)$ with any parameter $s$ from
$\langle -1,1\rangle$ is, in view of the property
(\ref{027}), a straightforward generalization of the
solutions given by K\'arsk\'a and Pe\v{r}ina~\cite{r41} (see
also Ref.~\cite{r79}) for ${\cal C}^{(\pm1)}
(\beta_S,\beta_A,t)$ related with normal or antinormal field
operator ordering. Setting initial values $C_A=C_S=0$, which
implies that $\bar{D}_{SA}(t)=C_A(t)=C_S(t)=0$ for any time
$t$, the solution~(\ref{067}) reduces to that of
Pe\v{r}ina~\cite{r77}.  The time-dependent functions
$U_k(t)$, $V_k(t)$ ($k=S,A$) appearing in~(\ref{068}) can be
expressed as
\begin{eqnarray}
V_S(t)&=& \frac{\kappa_{SA}}{2}Q_1,
\nonumber\\
V_A(t)&=& -\frac{\kappa_{SA}}{2}Q_1^*,
\nonumber\\
U_S(t) &=& Q_2+\left(\frac{\kappa_A}{2}+i \Delta\Omega \right)Q_1,
\nonumber\\
U_A(t) &=& Q_2^*-\left(\frac{\kappa_S}{2}-i \Delta\Omega
\right)Q_1^*, \label{069}
\end{eqnarray}
in terms of the auxiliary functions
\begin{eqnarray}
Q_1 &=& \frac{\exp(P_1\Delta t)-\exp(P_2\Delta t)}{P_1-P_2},
\nonumber\\
Q_2 &=& \frac{\partial Q_1}{\partial t} \:=\:
\frac{P_1\exp(P_1\Delta t)-P_2\exp(P_2\Delta t)}{P_1-P_2},
\label{070}
\end{eqnarray}
\begin{eqnarray}
P_{1,2}= \frac{1}{2}\Big\{ \frac{\kappa_S-\kappa_A}{2}\pm[
(\frac{\kappa_S-\kappa_A}{2})^2 -4((\Delta\Omega)^2
-i\frac{\kappa_S+\kappa_A}{2} \Delta\Omega)]^{1/2}\Big\}.
\label{071}
\end{eqnarray}
It is seen that for the initial moment of time $t_0$ the
functions $V_k(t_0)$ vanish and the $U_k(t_0)$ are equal to
unity, so that the initial Wigner covariances $D_{k,l}$, and
$\bar{D}_{k,l}$ ($k,l=S,A$) also vanish as a result of the
initial condition of zero stochastical correlation between
the scattered modes. Let us note that on the assumption of
the frequency resonant condition $\Delta\Omega=0$, the
functions~(\ref{069})--(\ref{071}) simplify
considerably~\cite{r77} since
$P_1=\frac{1}{2}(\kappa_S-\kappa_A)$ and $P_2=0$. This leads,
in particular, to the relations
\begin{eqnarray}
V_S(t)+V_A(t) = 0,
\nonumber \\
U_S(t)+U_A(t) = 1 + \exp\left(\frac{\kappa_S-\kappa_A}{2}\Delta
t\right). \label{072}
\end{eqnarray}
To obtain the solution of the FPE~(\ref{064}) we perform the
Fourier transform~(\ref{021}) of ${\cal
C}^{(s)}(\beta_S,\beta_A,t)$, which leads to the
$s$-parametrized quasidistribution ${\cal
W}^{(s)}(\alpha_S,\alpha_A,t)$ in the form
\begin{eqnarray}
{\cal W}^{(s)}(\alpha_S,\alpha_A,t)&=& \frac{1}{L^{(s)}}
\exp\Big\{ (L^{(s)})^{-2}\Big[ -E_1|\alpha_S-\xi_S(t)|^2
-E_2|\alpha_A-\xi_A(t)|^2
\nonumber\\
&&+\frac{1}{2} E_3(\alpha_S^*-\xi_S^*(t))^2 +\frac{1}{2}
E_4(\alpha_A^*-\xi_A^*(t))^2
\nonumber\\
&&+E_5(\alpha_S^*-\xi_S^*(t)) (\alpha_A^*-\xi_A^*(t)) \nonumber\\
&&+ E_6(\alpha_S-\xi_S(t))(\alpha_A^*-\xi_A^*(t)) +c.c.
\Big]\Big\}, \label{073}
\end{eqnarray}
which generalizes the K\'arsk\'a and Pe\v{r}ina result for
antinormal ordering~\cite{r41} going over into s-ordering of
the field. The time-dependent functions $E_i$ ($i=1,...,6$)
and $L^{(1)}$ have been calculated by
Pe\v{r}inov\'a~\cite{r85a} in her analysis of quadratic
optical parametric processes.  Here, we have the following
generalized $s$-parametrized functions $E_i$ and $L^{(s)}$
occurring in (\ref{073}):
\begin{eqnarray}
E_1&=& B_S^{(s)}(t)K_A^{(s)}(t) - B_A^{(s)}(t) K_+(t)
\nonumber\\
&&+ (C_A^*(t)D_{SA}(t)\bar{D}_{SA}(t)+c.c.),
\nonumber\\
E_2&=& B_A^{(s)}(t)K_S^{(s)}(t) - B_S^{(s)}(t) K_+(t)
\nonumber\\
&&+ (C_S(t)D_{SA}^*(t)\bar{D}_{SA}(t)+c.c.),
\nonumber\\
E_3&=& C_S(t)K_A^{(s)}(t) +
2B_A^{(s)}(t)D_{SA}(t)\bar{D}_{SA}^*(t)
\nonumber\\
&&+ C_A^*(t)D_{SA}^2(t) +C_A(t)\bar{D}_{SA}^{*2}(t),
\nonumber\\
E_4&=& C_A(t)K_{S}^{(s)}(t) +
2B_S^{(s)}(t)D_{SA}(t)\bar{D}_{SA}(t)
\nonumber\\
&&+ C_S(t)\bar{D}_{SA}^2(t) +C_S^*D_{SA}^2(t),
\nonumber\\
E_5&=& D_{SA}(t) \left[B_S^{(s)}(t)B_A^{(s)}(t) -K_-(t)\right]
\nonumber\\
&&+B_S^{(s)}(t)C_A(t)\bar{D}_{SA}^*(t)
+B_A^{(s)}(t)C_S(t)\bar{D}_{SA}(t) \nonumber\\ &&+
C_S(t)C_A(t)D_{SA}^*(t),
\nonumber\\
E_6&=& -\bar{D}_{SA}(t) \left[B_S^{(s)}(t)B_A^{(s)}(t)
+K_-(t)\right]
\nonumber\\
&&-B_S^{(s)}(t)C_A(t)D_{SA}^*(t) -B_A^{(s)}(t)C_S^*(t)D_{SA}(t)
\nonumber\\ &&- C_S^*(t)C_A(t)\bar{D}_{SA}^*(t), \label{074}
\end{eqnarray}
\begin{eqnarray}
(L^{(s)})^2 &=& K_S^{(s)}(t) K_A^{(s)}(t) -
2B_S^{(s)}(t)B_A^{(s)}(t)K_+(t)
\nonumber\\
&&-\Big[ C_S(t)C_A(t) D_{SA}^{*2}(t) +C_S(t)C_A^*(t)
\bar{D}_{SA}^2(t)
\nonumber\\
&&+2 B_S^{(s)}(t) C_A^*(t)D_{SA}(t)\bar{D}_{SA}(t)
\nonumber\\
&&+2 B_A^{(s)}(t) C_S^*(t)D^*_{SA}(t)\bar{D}_{SA}(t)
+c.c.\Big]+K_-^2(t), \label{075}
\end{eqnarray}
with
\begin{eqnarray}
K_{S,A}^{(s)}(t) &=& \left(B_{S,A}^{(s)}(t)\right)^2
-|C_{S,A}(t)|^2,
\nonumber\\
K_±(t) &=& |D_{SA}(t)|^2 \pm |\bar{D}_{SA}(t)|^2. \label{076}
\end{eqnarray}
The two-mode functions ${\cal W}^{(s)}(\alpha_S,\alpha_A,t)$
(\ref{073}) and ${\cal
C}^{(s)}(\beta_S,\beta_A,t)$~(\ref{067}) reduce to the
single-mode functions ${\cal W}^{(s)}(\alpha_k,t)$ and ${\cal
C}^{(s)}(\beta_k,t)$ ($k=S,\,A$) simply by setting either
$\alpha_S=\beta_S=0$ or $\alpha_A=\beta_A=0$, implying that
the coefficients $V_S(t)$, $V_A(t)$, $D_{SA}(t)$, and
$\bar{D}_{SA}(t)$ vanish and, for instance, $L^{(s)}$ reduces
to $\sqrt{K_k^{(s)}(t)}$.

Contrary to the solution~(\ref{067}) for the characteristic
function ${\cal C}^{(s)}(\beta_S,\beta_A,t)$, the solution
(\ref{073}) for the quasidistribution ${\cal
W}^{(s)}(\alpha_S,\alpha_A,t)$ may be absent for some
s-ordering of the field operators depending on the choice of
initial field. The condition for the existence of the QPD
(\ref{073}), i.e., the existence of the Fourier transform
(\ref{021}) of  ${\cal C}^{(s)}(\beta_S,\beta_A,t)$
(\ref{067}), is that the function $K_A^{(s)}(t)$,
$L^{(s)}(t)$, ${\rm Re}\, C_A(t)+B_A^{(s)}(t)$, and
\begin{eqnarray}
\bar{L}^{(s)} \equiv  \left(K_A^{(s)}(t) \right)^{1/2} \left[{\rm
Re}\,C_S(t)+B_S^{(s)}\right] +\left(K_A^{(s)}(t) \right)^{-1/2}
\hspace*{17mm} \nonumber \\ \times \Big[ {\rm
Re}\,C_A^*(t)(\bar{D}_{SA}(t)-D_{SA}(t))^2 -B_A^{(s)}(t)
|\bar{D}_{SA}(t)-D_{SA}(t)|^2 \Big]>0 \label{077}
\end{eqnarray}
should be positive.  If any of the four functions
$K_A^{(s_1)}(t)$, $L^{(s_1)}(t)$, $\bar{L}^{(s_1)}(t)$, and
${\rm Re}C_A(t)+B_A^{(s_1)}(t)$ (for a particular parameter
$s_1$) is not positive definite everywhere, the equation of
motion~(\ref{064}) for the $s_1$-parametrized
quasidistribution cannot be interpreted as a FPE describing
the Brownian motion, i.e., the equation is not a ``true''
FPE. The quasidistribution ${\cal
W}^{(s_1)}(\alpha_S,\alpha_A,t)$ does not exist as a positive
well-behaved function; still it does exist as a generalized
function according to the Klauder-Sudarshan
theorem~\cite{r280}. This property is a signature of quantum
effects~\cite{r58,r121,r180}. Let us note that it is possible
to use generalized $P$-representations (positive
$P$-representations) by doubling the phase space, as has been
proposed by Drummond and Gardiner~\cite{r331}. The
generalized $P$-representations have been applied
successfully to solve master equations of various nonlinear
problems (see, e.g., Ref.~\cite{r331,r332,r333,r282}). This
method, if applied to our model, requires us to handle eight
real variables (not counting time), instead of four.

For initially coherent Stokes and anti-Stokes fields, i.e.,
satisfying $C_S=C_A=\bar{D}_{SA}=0$, the rather complicated
expressions for $\bar{L}^{(s)}$~(\ref{077}) and $L^{(s)}$
(\ref{075}) reduce to
\begin{eqnarray}
\bar{L}^{(s)} = B_S^{(s)}(t) B_A^{(s)}(t) -|D_{SA}(t)|^2,
\label{078}
\end{eqnarray}
\begin{eqnarray}
L^{(s)} = \left| B_S^{(s)}(t)B_A^{(s)}(t) -|D_{SA}(t)|^2 \right|.
\label{079}
\end{eqnarray}
It is seen that, in the case of initially coherent fields,
the sufficient condition for the existence of  ${\cal
W}^{(s)}(\alpha_S,\alpha_A,t)$~(\ref{073}) is only that the
function $\bar{L}^{(s)}$~(\ref{078}) shall be positive. One
obtains further simplification of the problem under the
assumption of negligible frequency mismatch
($\Delta\Omega=0$). The functions $B_{S,A}^{(s)}(t)$ and
$D_{SA}(t)$~(\ref{068}) now reduce to
\begin{eqnarray}
B_S^{(s)}(t) &=& \frac{\kappa_S}{\kappa_S-\kappa_A} f_- \left(
\frac{\kappa_S}{\kappa_S-\kappa_A}f_+
-2\frac{\kappa_A}{\kappa_S-\kappa_A} +\langle\hat{n}_{V}\rangle
f_+ \right) +\frac{1-s}{2} \:\geq\: 0,
\nonumber \\
B_A^{(s)}(t) &=& \frac{\kappa_A}{\kappa_S-\kappa_A} f_- \left(
\frac{\kappa_S}{\kappa_S-\kappa_A}f_- +\langle\hat{n}_{V}\rangle
f_+ \right) +\frac{1-s}{2} \:\geq\: 0,
\nonumber \\
|D_{SA}(t)| &=& \frac{\sqrt{\kappa_S\kappa_A}}{\kappa_S-\kappa_A}
f_- \left( \frac{\kappa_S}{\kappa_S-\kappa_A} f_-
+\langle\hat{n}_{V}\rangle f_+ +1 \right), \label{080}
\end{eqnarray}
with
\begin{eqnarray}
f_± = \exp\left(\frac{\kappa_S-\kappa_A}{2}\Delta t\right) \pm 1.
\label{081}
\end{eqnarray}
In particular, the Wigner function exists, since
\begin{eqnarray}
\bar{L}^{(0)} = \frac{1}{4} +\frac{1}{2}\,
\frac{\exp[(\kappa_S-\kappa_A)\Delta t]-1}{\kappa_S-\kappa_A}
[\langle\hat{n}_{V}\rangle(\kappa_S+\kappa_A)+\kappa_S] \:>\:0,
\label{082}
\end{eqnarray}
contrary to the $P$-function, which does not exist for
$t>t_0$, since~\cite{r77}
\begin{eqnarray}
\bar{L}^{(1)} = -\frac{\kappa_S \kappa_A}{(\kappa_S-\kappa_A)^2}
\left[ \exp \left( \frac{\kappa_S-\kappa_A}{2}\Delta t\right)-1
\right]^2 <0 \quad \mbox{for}\; \Delta t>0. \label{083}
\end{eqnarray}
In general the solution~(\ref{073}) at a given time $t$
exists for parameters $s$ less than
\begin{eqnarray}
s &<& B^{(1)}_S(t) + B^{(1)}_A(t) +1
-\sqrt{\left(B^{(1)}_S(t)+B^{(1)}_A(t)\right)^2-4\bar{L}^{(1)}}.
\label{084}
\end{eqnarray}
Assuming that the damping constant $\gamma_A$ is equal to the gain
constant $\gamma_S$, or equivalently $\kappa_A=\kappa_S=\kappa$, we
arrive at
\begin{eqnarray}
\bar{L}^{(s)} = \frac{1}{4} [(1-s)^2
+2(1-s)(1+2\langle\hat{n}_{V}\rangle) \kappa\Delta t -s \kappa^2
(\Delta t)^2], \label{085}
\end{eqnarray}
which is greater than zero for parameters $s$ less than
\begin{eqnarray}
s < \frac{1}{2} +\frac{(\kappa\Delta t+1)^2}{2} +\{
2\langle\hat{n}_{V}\rangle -[
(1+2\langle\hat{n}_{V}\rangle+\frac{\kappa\Delta t}{2})^2 +1
]^{1/2}\} \kappa\Delta t. \label{086}
\end{eqnarray}
In particular, $\bar{L}^{(1)}$~(\ref{083}) for the
$P$-function and $\bar{L}^{(0)}$~(\ref{082}) for the Wigner
function respectively reduce to
\begin{eqnarray}
\bar{L}^{(1)} = -\left(\frac{\kappa\Delta t}{2} \right)^2 \:<\: 0
\hspace{1cm} \mbox{for}\: \Delta t>0,
\nonumber \\
\bar{L}^{(0)} = \frac{1}{4} +
\left(\langle\hat{n}_{V}\rangle+\frac{1}{2}\right) \kappa\Delta t
\:>\: 0. \label{087}
\end{eqnarray}
The condition for $s$ fulfilling $\bar{L}^{(s)}>0$ cannot be
expressed explicitly in a simple form in cases with frequency
mismatch $\Delta\Omega\neq 0$. As another example, let us
assume that the Stokes and anti-Stokes fields are initially
chaotic, which mathematically differs from our former example
of initially coherent state by the presence of nonzero
initial coefficients
$B^{(-1)}_k=\langle\hat{n}_{ch\,k}\rangle$ ($k=S,A$). By
virtue of the relations~(\ref{068}), the functions
$B^{(s)}_k(t)$, $D_{SA}(t)$ for chaotic field are the same as
for a coherent field with extra terms. Here, the function
$\bar{L}^{(s)}$~(\ref{075}) is found to be
\begin{eqnarray}
\bar{L}^{(s)} &=& \left(
B_S^{(s)}(t)+\langle\hat{n}_{ch\,S}\rangle |U_S(t)|^2
+\langle\hat{n}_{ch\,A}\rangle |V_S(t)|^2 \right) \nonumber \\
&\times& \left( B_A^{(s)}(t)+\langle\hat{n}_{ch\,A}\rangle
|U_A(t)|^2 +\langle\hat{n}_{ch\,S}\rangle |V_A(t)|^2 \right)
\nonumber \\ &-& \bigl| |D_{SA}|+\langle\hat{n}_{ch\,S}\rangle
U_S(t)V_A(t) +\langle\hat{n}_{ch\,A}\rangle U_A(t) V_S(t)
\bigr|^2, \label{088}
\end{eqnarray}
which has a form similar to~(\ref{079}) with the same
function $B^{(s)}_k(t)$ and $D_{SA}(t)$ given by~(\ref{080}).
In the case of equal damping and gain constants we obtain
\begin{eqnarray}
\bar{L}^{(s)} &=& -\left( \frac{\gamma\Delta t}{2} \right)^2 s
(\langle \hat{n}_{ch\, A} \rangle +\langle \hat{n}_{ch\, S}
\rangle -1) + \gamma\Delta t \left[ \langle \hat{n}_{ch\, A}
\rangle \left( \langle \hat{n}_{V} \rangle+\frac{1+s}{2} \right)
\right. \nonumber \\ &+& \left. \langle \hat{n}_{ch\, S} \rangle
\left( \langle \hat{n}_{V} \rangle+\frac{1-s}{2} \right)
+\left(\langle \hat{n}_{V} \rangle+\frac{1}{2} \right) (1-s)
\right] \nonumber \\ &+& \langle \hat{n}_{ch\, A} \rangle \left(
\langle \hat{n}_{ch\, S} \rangle+\frac{1-s}{2} \right) +\langle
\hat{n}_{ch\, S} \rangle \frac{1-s}{2} +\left( \frac{1-s}{2}
\right)^2. \label{089}
\end{eqnarray}
It is seen that the Wigner function always exists, since
\begin{eqnarray}
\bar{L}^{(0)} &=& \gamma\Delta t ( \langle \hat{n}_{ch\, S}
\rangle + \langle \hat{n}_{ch\, A} \rangle+1) \left(  \langle
\hat{n}_{V} \rangle +\frac{1}{2} \right) \nonumber \\ &+&
\langle\hat{n}_{ch\,A}\rangle \left(
\langle\hat{n}_{ch\,S}\rangle+\frac{1}{2}\right)
+\frac{1}{2}\langle\hat{n}_{ch\,S}\rangle+\frac{1}{4}>0,
\label{090}
\end{eqnarray}
\begin{figure} 
\vspace*{-7mm} \hspace*{-3mm}
\centerline{\psfig{figure=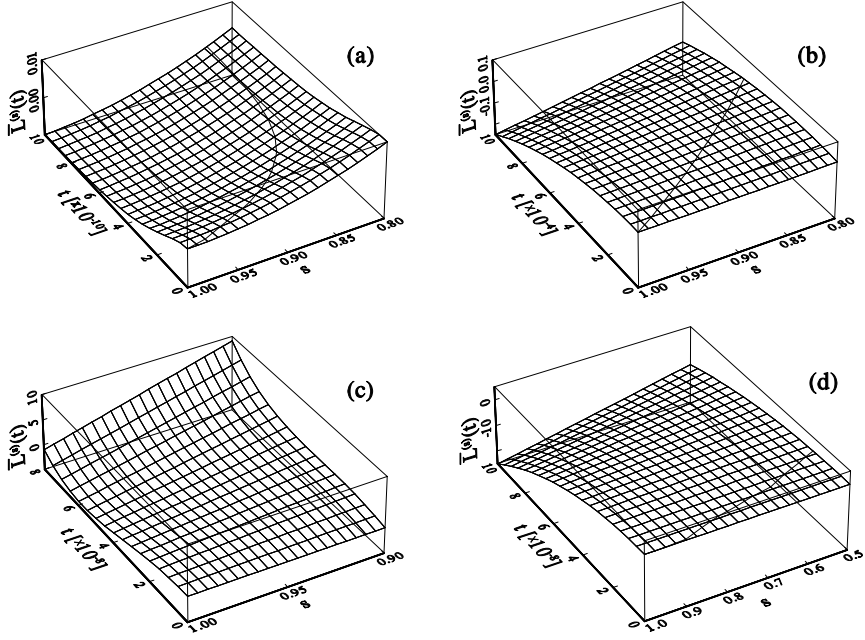,width=14cm}}
\vspace*{-10mm} \caption{The time and parameter s dependence
of the function $\bar{L}^{(s)}(t)$, related to the existence
of the QPD $W^{(s)}(\alpha_S,\alpha_A,t)$, for: (a)
$\kappa_S=10^8$, $\kappa_A=10^{10}$, $|\Delta\Omega|=1\div
10^6$ (the surfaces coincide in this range of
$|\Delta\Omega|$), $\langle\hat{n}_V\rangle=0$; (b)
$\kappa_S=\kappa_A=10^3$, $|\Delta\Omega|=1$,
$\langle\hat{n}_V\rangle=0\div 100$; (c)
$\kappa_S=\kappa_A=10^8$, $|\Delta\Omega|=10^6$,
$\langle\hat{n}_V\rangle=10$; and (d)
$\kappa_S=\kappa_A=10^8$, $|\Delta\Omega|=10^6$,
$\langle\hat{n}_V\rangle=0$. The Stokes and anti-Stokes
fields are initially coherent. The dashed lines on the
surfaces are depicted for $\bar{L}^{(s)}(t)=0$.}
\vspace*{-5mm}
\end{figure}
\noindent whereas the $P$-function exists only for times
shorter than
\begin{eqnarray}
\Delta t &<& \frac{2}{\gamma}
(\langle\hat{n}_{ch\,A}\rangle+\langle\hat{n}_{ch\,S}\rangle+1)^{-1}
\mbox{\Large \{} \langle\hat{n}_{ch\,A}\rangle^2
[\langle\hat{n}_{ch\,S}\rangle + (\langle\hat{n}_{V}\rangle+1)^2]
\nonumber \\ &&+
\langle\hat{n}_{ch\,A}\rangle\langle\hat{n}_{ch\,S}\rangle
[\langle\hat{n}_{ch\,S}\rangle+\langle\hat{n}_{V}\rangle^2
+(\langle\hat{n}_{V}\rangle+1)^2]
+\langle\hat{n}_{ch\,S}\rangle^2\langle\hat{n}_{V}\rangle^2
\mbox{\Large \}}^{1/2} \nonumber \\ &&+
\langle\hat{n}_{ch\,A}\rangle (\langle\hat{n}_{V}\rangle+1)
+\langle\hat{n}_{ch\,S}\rangle\langle\hat{n}_{V}\rangle.
\label{091}
\end{eqnarray}
The relation~(\ref{089}) is quadratic in $s$ and readily
gives an analytic expression for the largest parameter $s$
($s\leq 1$) for which the quasidistribution ${\cal
W}^{(s)}(\alpha_S,\alpha_A,t)$ exists at a given time of the
evolution $\Delta t=t-t_0$.

In Fig.1 we present the function $\bar{L}^{(s)}(t)$ for
different values of the frequency mismatch $\Delta\Omega$, of
the mean number of photons $\langle\hat{n}_V\rangle$, and of
the damping ($\kappa_A$) and gain  ($\kappa_S$) constants. We
assume that the Stokes and anti-Stokes modes are initially
coherent. Thus, for all discussed cases (Figs. 1a-1d), the
condition of a positive definite function $\bar{L}^{(s)}(t)$
is sufficient for the existence of the corresponding
$s$-parametrized QPD. For clarity, the dashed lines in Fig.1
are depicted for $\bar{L}^{(s)}(t)=0$.

We shall briefly analyze a more general situation, which
comprises the above cases and others. Let us assume after
Refs.~\cite{r41} and~\cite{r264} (for a general analysis see
Refs.~\cite{r79} and~\cite{r322}) that the Stokes (k=S) as
well anti-Stokes (A) modes are initially in squeezed states
characterized by complex amplitudes $\xi_k$, parameters
$r_k$, and phases $\phi_k$, superposed with a chaotic field,
characterized by the mean number of chaotic photons $\langle
\hat{n}_{ch\,k}\rangle$. The initial $s$-parametrized
quasidistribution ${\cal W}^{(s)}(\alpha_S,\alpha_A,t_0)$ is
then given by
\begin{eqnarray}
{\cal W}^{(s)}(\alpha_S,\alpha_A,t_0)=\prod_{k=S,A} \left(
K_{k}^{(s)}\right)^{-1/2} \exp \mbox{\huge \{}
-\frac{1}{K_{k}^{(s)}}
\nonumber\\
\times \left[ B_{k}^{(s)}|\alpha_k-\xi_k|^2 -{\rm
Re}\left(C_{k}^{*}(\alpha_k-\xi_k)^2\right) \right] \mbox{\huge
\}}, \label{092}
\end{eqnarray}
with
\begin{eqnarray}
B_{k}^{(s)}&=\,B_{k}^{(s)}(t_0)
&=\,(\cosh r)^2 + \langle n_{ch\,k}\rangle -\frac{s+1}{2},
\nonumber\\
C_k &=\,C_k(t_0) &=\, \frac{1}{2}\exp (i \phi_k) \sinh (2r_k),
\label{093}
\end{eqnarray}
which trivially reduces to the quasidistributions of a pure
squeezed state ($\langle\hat{n}_{ch k}\rangle=0$), a coherent
state ($\langle\hat{n}_{ch k}\rangle=r_k=0$), or a chaotic state
($r_k=\xi_k=0$).  In Sect. 6.2 we analyze another special case of
(\ref{093}) with $r_k=0$ describing a general superposition of
coherent and chaotic fields.

The  Raman effect model under parametric approximation is
fully specified either by the $s$-parametrized characteristic
function ${\cal C}^{(s)}(\beta_S,\beta_A,t)$~(\ref{067})  or
the $s$-parametrized quasidistribution ${\cal
W}^{(s)}(\alpha_S,\alpha_A,t)$~(\ref{073}). In particular, by
virtue of the relations presented in Section 4, one can
obtain complete information about the photon-counting
statistics and squeezing properties of the scattered fields
from~(\ref{067}) or~(\ref{073}).

One can calculate the photon-counting probability
distribution $p(n)$ from the quasidistribution ${\cal
W}^{(s)}(\alpha_S,\alpha_A,t)$ or integrated
quasidistribution ${\cal W}^{(s)}(W,t)$~(\ref{030}) by means
of~(\ref{031}), or equivalently from the generating function
$\langle\exp(-\lambda W(t))\rangle_{(s)}$~(\ref{032}) by
virtue of~(\ref{033}).  We apply the latter method, which
gives us, after insertion of ${\cal
W}^{(s)}(\alpha_S,\alpha_A,t)$~(\ref{073}) or ${\cal
C}^{(s)}(\beta_S,\beta_A,t)$~(\ref{067}) into~(\ref{032}),
the following time-dependent $s$-parametrized generating
function:
\begin{eqnarray}
\langle \exp(-\lambda W) \rangle_{(s)} = \lambda^{-2} ({\cal
L}^{(s)}_1)^{-1/2} \exp\left(\frac{{\cal L}^{(s)}_2}{{\cal
L}^{(s)}_1} \right), \label{094}
\end{eqnarray}
where the ${\cal L}^{(s)}_1$ (${\cal L}^{(s)}_2$) are polynomials  of
the fourth (third) order in $\lambda^{-1}$:
\begin{eqnarray}
{\cal L}^{(s)}_1 = \sum_{j=0}^{4} \left(\lambda^{-1}+\frac{1-s}{2}
\right)^j b_j,
\nonumber \\
{\cal L}^{(s)}_2 = \sum_{j=0}^{3} \left(\lambda^{-1}+\frac{1-s}{2}
\right)^j a_j. \label{095}
\end{eqnarray}
Adapting the results of Pe\v{r}inov\'a and
Pe\v{r}ina~\cite{r85b} for the coefficients $a_j$, $b_j$
($j=0,1,...$) occurring in~(\ref{095}), one obtains
\begin{eqnarray}
a_{0}&=& [ -B_S^{(1)} K_A^{(1)} + B_A^{(1)} K_+ +(C_A
D_{SA}\bar{D}_{AS}+c.c.)] |\xi_S|^2
\nonumber\\
&&+\mbox{\large \{} \left[
B_A^{(1)}D_{SA}\bar{D}_{SA}+\frac{1}{2}(C_S^*K_A^{(1)} +C_A
D_{SA}^2 + C_A^*\bar{D}_{SA}^2) \right] \xi_S^2
\nonumber\\
&&+ \frac{1}{2}\mbox{\large [} B_S^{(1)}B_A^{(1)}D_{SA} +2
B_S^{(1)}C_A^{*}\bar{D}_{SA} +C_S^{*}C_A^{*}D_{SA}^* -D_{SA}K_-
\mbox{\large ]} \xi_S \xi_A
\nonumber\\
&&- \frac{1}{2}\mbox{\large [} B_S^{(1)}B_A^{(1)}\bar{D}_{SA} +2
B_S^{(1)}C_A D_{SA} +C_S^{*}C_A\bar{D}_{AS}
\nonumber\\
&&+\bar{D}_{SA} K_- \mbox{\large ]} \xi_S \xi_A^* +
c.c.\mbox{\large \}}
\nonumber\\
&&+ [S \longleftrightarrow A],
\nonumber\\
a_{1}&=& [ -2B_S^{(1)} B_A^{(1)}- K_A^{(1)} +K_+)|\xi_S|^2
\nonumber\\
&&+ [(B_A^{(1)}C_S^*+ D_{SA}\bar{D}_{SA})\xi^2_S
+(B_S^{(1)}D_{SA}+ C_{S}^*\bar{D}_{AS})\xi_S\xi_A
\nonumber\\
&&-(B_S^{(1)}\bar{D}_{SA}+ C_{S}^*D^*_{SA})\xi_S\xi_A^* + c.c. ]
\nonumber\\
&&+[S \longleftrightarrow A],
\nonumber\\
a_{2}&=& -(B_S^{(1)} + 2 B_A^{(1)}) |\xi_S|^2
\nonumber\\
&&+\frac{1}{2}(C_S^* \xi_S^2 + D_{SA}\xi_S\xi_A
-\bar{D}_{SA}\xi_S\xi_A^* + c.c.)
\nonumber\\
&&+[S \longleftrightarrow A],
\nonumber\\
a_{3}&=& -|\xi_S|^2 -|\xi_A|^2, \label{096}
\end{eqnarray}
\begin{eqnarray}
b_{0}&=& \frac{1}{2}K_S^{(1)}K_A^{(1)} -B_S^{(1)}B_A^{(1)}K_+
\nonumber\\
&&-2B_S^{(1)}(C_A^*D_{SA}^*\bar{D}_{AS}+c.c.) +\frac{1}{2}K_-^2
\nonumber\\
&&-\frac{1}{2}[C_A(C_S D^2_{SA}+ C_S^*\bar{D}^2_{AS})+c.c.]
\nonumber\\
&&+[S \longleftrightarrow A],
\nonumber\\
b_{1}&=& 2 B_S^{(1)}(K_A^{(1)}-K_+) -2 (C_S D_{SA}
\bar{D}_{SA}+c.c.)
\nonumber\\
&&+[S \longleftrightarrow A],
\nonumber\\
b_{2}&=& 2 B_S^{(1)}B_A^{(1)}+K_S^{(1)}-K_+
\nonumber\\
&&+[S \longleftrightarrow A],
\nonumber\\
b_{3}&=& 2(B_S^{(1)}+B_A^{(1)}),
\nonumber\\
b_{4}&=& 1, \label{097}
\end{eqnarray}
where $[S \longleftrightarrow A]$ stands for the preceding
terms albeit with interchanged subscripts $S$ and $A$. For
brevity, we have omitted the time dependence of Eqs.
(\ref{096}) and~(\ref{097}). Our formulas~(\ref{094}) and
(\ref{095}) are generalizations of the results given in
Refs.~\cite{r79},~\cite{r85b} and~\cite{r41} for $s=-1$ to
any $s$. It is seen that the simplest form of~(\ref{095}) is
for normal ordering of the field operators; hence, here, we
use only this ordering. Pe\v{r}inov\'a and
Pe\v{r}ina~\cite{r85b} have shown that if the polynomial
${\cal L}_1^{(1)}$ has four single roots $\lambda_k=-\left(
\frac{1}{\lambda}\right)_k$, the generating function
$\langle\exp(-\lambda W(t))\rangle_{(1)}$ has the form of the
fourfold generating function for Laguerre polynomials
\begin{eqnarray}
\langle \exp(-\lambda W(t)) \rangle_{(1)} = \prod_{k=1}^{4}
(1+\lambda\lambda_k)^{-1/2} \exp\left(-\frac{\lambda
A_k}{1+\lambda \lambda_k}\right). \label{098}
\end{eqnarray}
The field is described by a superposition of signal components
\begin{eqnarray}
A_k = \prod_{l=1\atop l\neq
k}^{4}(\lambda_k^{-1}-\lambda_l^{-1})^{-1}
\sum_{l=0}^{3}a_l(-\lambda_k^{-1})^l \label{099}
\end{eqnarray}
and the noise components $\lambda_k$. With Eq.~(\ref{098})
available, one obtains~\cite{r85b,r41} the following
photocount distribution
\begin{eqnarray}
p(n,t) &=& \sum_{k_1,k_2,k_3,k_4\atop k_1+k_2+k_3+k_4=n}
\prod_{l=1}^{4} \exp\left(-\frac{A_l}{1+\lambda_l}\right)
\nonumber
\\ &&\times
\frac{\lambda_l^{k_l}}{(1+\lambda_l)^{k_l+1/2}\Gamma(k_l+1/2)}
L^{-1/2}_{k_l}
\left( \frac{-A_l}{\lambda_l(1+\lambda_l)}\right)
\label{100}
\end{eqnarray}
and its factorial moments
\begin{eqnarray}
\langle W^{k}(t) \rangle_{(1)} = k! \sum_{k_1+k_2+k_3+k_4=k}
\prod_{l=1}^{4} \frac{\lambda_l^{k_l}}{\Gamma(k_l+1/2)}
L^{-1/2}_{k_l} \left( -\frac{A_l}{\lambda_l}\right), \label{101}
\end{eqnarray}
by applying well-known properties of the generating function
of the generalized Laguerre polynomials $L^{\alpha}_k(x)$ to
the definition relations~(\ref{033}) for $p(n,t)$ with $s=1$
and to the relations~(\ref{034}) for $\langle
W^k(t)\rangle_{(1)}$. Much simpler expressions are found in
the special case when the radiation fields are initially
superpositions of coherent and chaotic fields. From relations
(\ref{093}) with $r_k=0$ and~(\ref{068}) it is seen that
$C_S(t)=C_A(t)=\bar{D}_{SA}(t)=0$. The fourfold generating
function~(\ref{098}) reduces to a twofold generating function
in the form of~(\ref{098}), where the upper limit of the
product should be replaced by 2 and the square root in
$(1+\lambda\lambda_k)^{-1/2}$ should be omitted. Then the
photocount distribution $p(n,t)$ and its factorial moments
$\langle W^k(t)\rangle_{(1)}$ become
\begin{eqnarray}
p(n,t)= (n!)^{-1} \exp \left(
-\frac{A_1}{1+\lambda_1}-\frac{A_2}{1+\lambda_2}\right)
\sum_{l=0}^{n}\left(n \atop l\right) \lambda_1^l \lambda_2^{n-l}
(1+\lambda_1)^{-(l+1)} \nonumber \\ \times
(1+\lambda_2)^{-n+(l+1)}
L_l\left(-\frac{A_1}{\lambda_1(1+\lambda_1)} \right)
L_{n-l}\left(-\frac{A_2}{\lambda_2(1+\lambda_2)}
\right),\hspace{1cm} \label{102}
\end{eqnarray}
\begin{eqnarray}
\langle W^{n}(t) \rangle_{(1)} = \sum_{l=0}^{n}\left(n \atop
l\right) \lambda_1^l \lambda_2^{n-l}
L_l\left(-\frac{A_1}{\lambda_1} \right)
L_{n-l}\left(-\frac{A_2}{\lambda_2} \right), \label{103}
\end{eqnarray}
where $L_n(x)=L_n^0(x)$ and the roots $\lambda_k$ and
coefficients $A_k$ are~\cite{r85b}
\begin{eqnarray}
\lambda_{1,2} &=& \frac{1}{2} \left\{ B_S^{(1)}(t)+B_A^{(1)}(t)
\:\mp\: [(B_S^{(1)}(t)-B_A^{(1)}(t))^2+4|D_{SA}(t)|^2]^{1/2}
\right\}
\nonumber \\
A_{1,2} &=& \pm
[(B_S^{(1)}(t)-B_A^{(1)}(t))^2+4|D_{SA}(t)|^2]^{-1/2} \nonumber \\
&&\times
\left[\frac{1}{2}(B_S^{(1)}(t)-B_A^{(1)}(t))(|\xi_A|^2-|\xi_S|^2)
-(D_{SA} \xi_S \xi_A^* +c.c.)\right] \nonumber \\ &&+
\frac{1}{2}(|\xi_S|^2+|\xi_A|^2). \label{104}
\end{eqnarray}

The photon-counting statistics of scattering either into the
Stokes or anti-Stokes mode can be calculated from formulas
(\ref{098})--(\ref{101}). In the single-mode case the moments
$D_{SA}(t)$ and $\bar{D}_{SA}(t)$ vanish, considerably
simplifying the polynomial ${\cal L}_1^{(1)}$~(\ref{095}),
with coefficients $b_j$, to the form
\begin{eqnarray}
{\cal L}_1^{(1)} = \prod_{k=S,A} \left(\lambda^{-2} +2\lambda^{-1}
B_k^{(1)}(t) +K_k^{(1)}(t) \right), \label{105}
\end{eqnarray}
with the roots $\lambda_{1,2\,S,A}=-(\lambda^{-1})_k$
\begin{eqnarray}
\lambda_{1,2\,k} = B^{(1)}_k(t) \mp |C_k(t)| \hspace{1cm} (k=S,A)
\label{106}.
\end{eqnarray}
The notation $\lambda_{1\,S,A}$ and $\lambda_{2\,S,A}$, instead of
$\lambda_{1,2,3,4}$, emphasizes the dependence on the single-mode
variables in accordance with the assumption of alternative scattering
into the Stokes or anti-Stokes mode. Analogously, it is seen that
\begin{eqnarray}
A_{1,2\,k} = \frac{1}{2} |\xi_k(t)|^2 \mp \frac{1}{4}
|C_k(t)|^{-1} (C^*_k(t) \xi^2_k(t)+c.c.). \label{107}
\end{eqnarray}
On insertion of~(\ref{106}) into twofold functions
(\ref{098})--(\ref{101}) one immediately obtains
\begin{eqnarray}
\langle \exp(-\lambda W_k) \rangle_{(1)} =
[(1+\lambda\lambda_{1k})(1+\lambda\lambda_{2k})]^{-1/2} \nonumber
\hspace{15mm} \\ \times \exp[-\lambda
A_{1k}(1+\lambda\lambda_{1k})^{-1} -\lambda
A_{2k}(1+\lambda\lambda_{2k})^{-1}], \label{108}
\end{eqnarray}
\begin{eqnarray}
p_k(n) &=& [(1+\lambda_{1k})(1+\lambda_{2k})]^{-1/2}
(1+\lambda_{2k}^{-1})^{-n} \exp\left(
-\frac{A_{1k}}{1+\lambda_{1k}} -\frac{A_{2k}}{1+\lambda_{2k}}
\right) \nonumber \\ &&\times
\sum_{l=0}^{n}\frac{1}{\Gamma(l+1/2)\Gamma(n-l+1/2)} \left(
\frac{1+\lambda_{2k}^{-1}}{1+\lambda_{1k}^{-1}} \right)^l
\nonumber \\ &&\times
L_l^{-1/2}\left(-\frac{A_{1k}}{\lambda_{1k}(1+\lambda_{1k})}
\right)
L_{n-l}^{-1/2}\left(-\frac{A_{2k}}{\lambda_{2k}(1+\lambda_{2k})}
\right), \label{109}
\end{eqnarray}
\begin{eqnarray}
\langle W^n_k(t)\rangle_{(1)} &=& n! \lambda_{2k}^n \sum_{l=0}^{n}
\frac{1}{\Gamma(l+1/2)\Gamma(n-l+1/2)} \left(
\frac{\lambda_{1k}}{\lambda_{2k}}\right)^l \nonumber \\ &&\times
L_l^{-1/2}\left(-\frac{A_{1k}}{\lambda_{1k}} \right)
L_{n-l}^{-1/2}\left(-\frac{A_{2k}}{\lambda_{2k}}\right).
\label{110}
\end{eqnarray}
To obtain the results of Refs.~\cite{r79},~\cite{r41}
and~\cite{r306} one should replace $\lambda_{1k}$ by $E_k-1$
and $\lambda_{2k}$ by $F_k-1$. In particular, assuming that a
scattered (Stokes or anti-Stokes) mode is initially in a
coherent state (thus $C_k=0$) the mean photon numbers
$\langle\hat{n}_k\rangle$ ($k=S,A$) are
\begin{eqnarray}
\langle\hat{n}_k(t)\rangle = \langle W_k(t)\rangle_{(1)} \:=\:
|\xi_k(t)|^2+B_k^{(1)}(t), \label{111}
\end{eqnarray}
or explicitly
\begin{eqnarray}
\langle\hat{n}_S(t)\rangle= |\xi_S|^2 \exp(\kappa_S\Delta t)+
(\langle\hat{n}_V\rangle+1) [\exp(\kappa_S\Delta t)-1],
\label{112}
\end{eqnarray}
\begin{eqnarray}
\langle\hat{n}_A(t)\rangle = |\xi_A|^2 \exp(-\kappa_A\Delta t)+
\langle\hat{n}_V\rangle [1-\exp(-\kappa_A\Delta t)], \label{113}
\end{eqnarray}
whereas the mean-square photon numbers
$\langle\hat{n}_k^2\rangle$ are
\begin{eqnarray}
\langle\hat{n}_k^2\rangle &=& \langle W^2_k\rangle_{(1)} + \langle
W_k\rangle_{(1)}
\nonumber\\
&=& |\xi_k(t)|^4 + |\xi_k(t)|^2(4B_k(t)+1) +2B_k^2(t)+B_k(t).
\label{114}
\end{eqnarray}
Then, the normalized second-order factorial
moments~(\ref{035}) are equal to
\begin{eqnarray}
\gamma_k^{(2)}(t) &=& B^{(1)}_k(t) \left[
|\xi_k(t)|^2+B_k^{(1)}(t) \right]^{-1} \nonumber \\ &&\times
\Big\{|\xi_k(t)|^2\left[ |\xi_k(t)|^2+B_k^{(1)}(t) \right]^{-1}+1
\Big\}. \label{115}
\end{eqnarray}

Let us proceed to analyze squeezing along the lines presented
in Section 4. We focus our attention on single- and two-mode
squeezed light according to the definition of ``usual''
squeezing and principal squeezing of Luk\v{s} et
al.~\cite{r284,r285,r303}. Using the definitions~(\ref{066})
of the functions $B_k^{(s)}(t)$, $D_{kl}(t)$,
$\bar{D}_{kl}(t)$, and $C_k(t)$  we readily obtain
expressions for the moments of the quadratures $\hat{X}_k$
and $\hat{X}_{kl}$
\begin{eqnarray}
\langle ( \Delta\hat{X}_{k1,k2})^2 \rangle = \pm 2 {\rm Re}\:
C_k(t) + 2 B_k^{(s)}(t)+s, \label{116}
\end{eqnarray}
\begin{eqnarray}
\langle ( \Delta\hat{X}_{k\pm})^2 \rangle = \pm 2 |C_k(t)| + 2
B_k^{(s)}(t)+s, \label{117}
\end{eqnarray}
\begin{eqnarray}
\langle \{\Delta\hat{X}_{k1},\Delta\hat{X}_{k2}\} \rangle &=& 4
{\rm Im}\: C_k(t),
\nonumber\\
\langle \Delta\hat{X}_{k1}\Delta\hat{X}_{l1} \rangle &=& 2 {\rm
Re} \left[D_{kl}(t) - \bar{D}_{kl}(t)\right],
\nonumber\\
\langle \Delta\hat{X}_{k2}\Delta\hat{X}_{l2}\rangle &=& - 2 {\rm
Re} \left[D_{kl}(t) + \bar{D}_{kl}(t)\right],
\nonumber\\
\langle \Delta\hat{X}_{k1}\Delta\hat{X}_{l2}\rangle &=& 2 {\rm Im}
\left[D_{kl}(t) - \bar{D}_{kl}(t)\right],
\nonumber\\
\langle \Delta\hat{X}_{k2}\Delta\hat{X}_{l1}\rangle &=& 2 {\rm Im}
\left[D_{kl}(t) + \bar{D}_{kl}(t)\right], \label{118}
\end{eqnarray}
where, as usual, $k,l=S,A$ and $k\neq l$. Thus, the two-mode quadrature
variances now have the form
\begin{eqnarray}
\left.
\begin{array}{l}
\langle ( \Delta \hat{X}_{SA1})^2\rangle \\
\langle ( \Delta \hat{X}_{SA2})^2\rangle
\end{array}
\right\} = \pm 2 {\rm Re} \left[ C_S(t) + C_A(t) + 2 D_{SA}(t)
\right] \hspace{15mm}
\nonumber\\
+2\left[ B_S^{(s)}(t) + B_A^{(s)}(t)
 -2 {\rm Re}\, \bar{D}_{SA}(t) + s\right],
\label{119}
\end{eqnarray}
and the extremal variances are
\begin{eqnarray}
\langle ( \Delta \hat{X}_{SA\pm})^2\rangle &=& \pm 2 \left| C_S(t)
+ C_A(t) + 2 D_{SA}(t) \right|
\nonumber\\
&&+2\left[ B_S^{(s)}(t) + B_A^{(s)}(t) -2 {\rm Re}\,
\bar{D}_{SA}(t) + s\right]. \label{120}
\end{eqnarray}
The single-mode squeezing, defined in standard manner, and
the single-mode principal squeezing require, respectively, that
\begin{eqnarray}
\left. |{\rm Re}C_k(t)| \atop |C_k(t)| \right\}
&>& B_k^{(s)}(t) +\frac{s}{2}
\hspace{1cm} (k=S,A),
\label{121}
\end{eqnarray}
whereas the conditions for the two-mode squeezing are,
respectively,
\begin{equation}
\left. |{\rm Re}[C_S(t)+C_A(t)+2D_{SA}(t)]|
\atop |C_S(t)+C_A(t)+2D_{SA}(t)| \right\}
> B_S^{(s)}(t)+B_A^{(s)}(t)-2{\rm Re}\bar{D}_{SA}(t) +s.
\label{122}
\end{equation}

Examples of the time evolution of
$\langle\hat{n}_S(\tau)\rangle$,
$\langle\hat{n}_S^2(\tau)\rangle$,
$\langle\hat{a}_S(\tau)\rangle$, and
$\langle\hat{a}_S^2(\tau)\rangle$ are given by curves C in
Figs. 2, 3, 7, and 8, respectively. We assume that the Stokes
fields are initially in a coherent state (stimulated Raman
scattering) or in a vacuum state (spontaneous Raman
scattering). The rescaled time $\tau$ is defined by
$t\rightarrow \tau=t\gamma_S$. Anti-Stokes scattering is
neglected. The phonon bath is at very low temperature, so we
put $\hat{n}_V=0$. In Fig.9 we present the time evolution of
the extremal variances $\langle(\Delta
X_{S\pm}(\tau))^2\rangle$ for fields initially coherent with
amplitudes equal to $\alpha_L=\sqrt{2}$,
$\alpha_S=\sqrt{0.2}$ and assuming that the heat bath is
''quiet'' (i.e., $\langle\hat{n}_V\rangle=0$). In the model
under discussion, the variance for the Stokes mode,
$\langle(\Delta X_S(\theta,\tau))^2\rangle$ (curve C in
Fig.9), is independent of $\theta$, i.e., $\langle(\Delta
X_{S+}(\tau))^2\rangle$= $\langle(\Delta
X_{S-}(\tau))^2\rangle$. Hence, squeezing is not observed if
the initial Stokes mode is in a coherent state. Even if the
Stokes field is initially squeezed and $\gamma_S>\gamma_A$
(not necessarily $\gamma_A=0$), squeezing will rapidly vanish
due to strong amplification of this mode, which leads to a
strong increase in quantum noise~\cite{r41}. The results of
this Section (curves C) are compared with the exact solutions
(without parametric approximation) derived in Sect. 6.1.2
(curves A) and the short-time solutions of Sect. 6.1.1.

\section{Master equation in Fock representation}

The parametric approximation, applied in the previous
Section, introduces linearization into our Raman scattering
model described by the Hamiltonians~(\ref{001})--(\ref{003}).
Here, we shall search for a solution to the nonlinear
problem, thus including pump depletion. The generalized
Fokker-Planck equation~(\ref{061}) and the corresponding
equation of motion~(\ref{063}) for the characteristic
function reveal the difficulties to be overcome in the
complete analysis of Raman scattering into simultaneously
both the Stokes and anti-Stokes fields from phonons treated
as a ``noisy'' ($\langle\hat{n}_V\rangle\neq 0$) reservoir.
Let us assume that the temperature of the medium is low.
Under this assumption it is quite reasonable to neglect the
anti-Stokes scattering ($\gamma_A=\gamma_{SA}=
\gamma_{AS}=0$) and, with regard to Eq.~(\ref{014}), to
assume that the reservoir is ``quiet''
($\langle\hat{n}_V\rangle=0$). Under these approximations the
master equation~(\ref{018}) reduces to the simple
form~\cite{r66}:
\begin{eqnarray}
\frac{\partial \hat{\rho}}{\partial \tau} = \frac{1}{2} \left(
[\hat{a}_L \hat{a}_S^+, \hat{\rho} \hat{a}_L^+ \hat{a}_S]
+[\hat{a}_L \hat{a}_S^+ \hat{\rho}, \hat{a}_L^+ \hat{a}_S]
\right), \label{123}
\end{eqnarray}
where we  have introduced a rescaled time
$t\rightarrow\tau=\gamma_S t$.  Let us denote the matrix elements of
the reduced density operator $\hat{\rho}$ in Fock representation by
\begin{eqnarray}
\langle n_L,n_{S}\left|\hat{\rho}(\tau)\right|n'_L,n'_{S}\rangle
\equiv \langle n,m\left|\hat{\rho}(\tau)\right|n+\nu,m+\mu \rangle
\equiv \rho_{n,m}(\nu,\mu,\tau), \label{124}
\end{eqnarray}
where for simplicity we identify $n_L=n$, and $n_S=m$; $\mu$
is the degree of off-diagonality for the elements of the
matrix $\hat{\rho}$ for the Stokes mode, whereas $\nu$ is the
degree of off-diagonality for the pump laser mode elements.
The master equation for the matrix elements~(\ref{124})
readily follows from Eq.~(\ref{123}) and can be written as
\begin{eqnarray}
\frac{\partial }{\partial \tau} \rho_{nm}(\nu\mu\tau)=
-\frac{1}{2} [n(m+1)+(n+\nu)(m+\mu+1)] \rho_{nm}(\nu\mu\tau)
\hspace{5mm}\nonumber\\ + [(n+1)(n+\nu+1)m(m+\mu)]^{1/2}
\rho_{n+1,m-1}(\nu\mu\tau). \label{125}
\end{eqnarray}
The equation~(\ref{125}) for the diagonal matrix  elements
$\rho_{nm}(00\tau)$ reduces to the rate equations of
Loudon~\cite{r57}, and McNeil and Walls~\cite{r66}.
Simaan~\cite{r100} (cf. Ref.~\cite{r57}) analyzed Raman
scattering from a gas of two-level atoms. On the assumption
that almost all the atoms are in their ground state, the
Simaan rate equation of Ref.~\cite{r100} takes the form of
Eq.~(\ref{125}) for $\nu=\mu=0$.

\subsection{Raman scattering including pump depletion}

\subsubsection{Short-time solutions}

Before proceeding to derive an exact solution of~(\ref{125})
we shall present the short-time solutions calculated with the
help of the relation $\langle\hat{A}(\tau)\rangle={\rm
Tr}\{\hat{A}
[\hat{\rho}(\tau_0)+\hat{\rho}\prime(\tau_0)\Delta\tau
+\hat{\rho}\prime\prime (\tau_0)(\Delta\tau)^2/2]\}$, where
$\hat{\rho}\prime\prime(\tau_0)$ is found by differentiating
Eq.~(\ref{125}) with respect to $\tau$. The solutions for the
mean $\langle\hat{n}\rangle$ and mean-square number of
photons $\langle\hat{n}^2\rangle$ in the laser mode up to
$\Delta\tau$ squared are
\begin{eqnarray}
\langle\hat{n}(\tau)\rangle &=& \langle\hat{n}\rangle
-\langle\hat{n}\rangle(\langle\hat{m}\rangle+1)\Delta\tau
\nonumber\\ &&- [\langle\hat{n}^2\rangle(\langle\hat{m}\rangle+1)
-\langle\hat{n}\rangle(\langle\hat{m}^2\rangle+3\langle\hat{m}\rangle+2)
]\frac{(\Delta\tau)^2}{2}, \label{126}
\end{eqnarray}
\begin{eqnarray}
\langle\hat{n}^2(\tau)\rangle &=&
\langle\hat{n}^2\rangle-(2\langle\hat{n}^2\rangle-\langle\hat{n}\rangle)
(\langle\hat{m}\rangle+1)\Delta\tau \nonumber\\ &&-
[2\langle\hat{n}^3\rangle (\langle\hat{m}\rangle+1)
-\langle\hat{n}^2\rangle(4\langle\hat{m}^2\rangle
+13\langle\hat{m}\rangle+9) \nonumber\\ &&+
3\langle\hat{n}\rangle(\langle\hat{m}^2\rangle+3\langle\hat{m}\rangle+2)]
\frac{(\Delta\tau)^2}{2}, \label{127}
\end{eqnarray}
where for brevity we set
$\langle\hat{n}^p(\tau_0)\rangle=\langle\hat{n}^p\rangle$ and
$\langle\hat{m}^p(\tau_0)\rangle=\langle\hat{m}^p\rangle$
($k=1,2,3$) as well as $\Delta\tau=\tau-\tau_0$. Then the
normalized second-order factorial moment,
$\gamma_L^{(2)}(\tau)$, defined by~(\ref{035}), is equal to
\begin{eqnarray}
\gamma_L^{(2)}(\tau) =  \gamma_L^{(2)} + \Big[
(\langle\hat{n}^2\rangle^2 -\langle\hat{n}^3\rangle
\langle\hat{n}\rangle) (1+\langle\hat{m}\rangle) \hspace{25mm}\nonumber\\
+ \langle\hat{n}\rangle
(\langle\hat{n}^2\rangle-\langle\hat{n}\rangle)
(1+\langle\hat{m}\rangle-\langle\hat{m}\rangle^2
+\langle\hat{m}^2\rangle) \Big] \langle\hat{n}\rangle^{-3}
\mbox{\large \[} (\Delta \tau)^2, \label{128}
\end{eqnarray}
which reduces to the Simaan result~\cite{r100}:
\begin{eqnarray}
\gamma_{L}^{(2)}(\tau)&=&\gamma_{L}^{(2)}
+[\langle\hat{n}^2\rangle
(\langle\hat{n}^2\rangle/\langle\hat{n}\rangle+1) \nonumber\\ &&-
\langle\hat{n}^3\rangle-\langle\hat{n}\rangle]
\langle\hat{n}\rangle^{-2}(\Delta\tau)^2 \label{129}
\end{eqnarray}
\begin{figure}  
\vspace*{-7mm} \hspace*{0mm} \hspace*{0mm}
\centerline{\psfig{figure=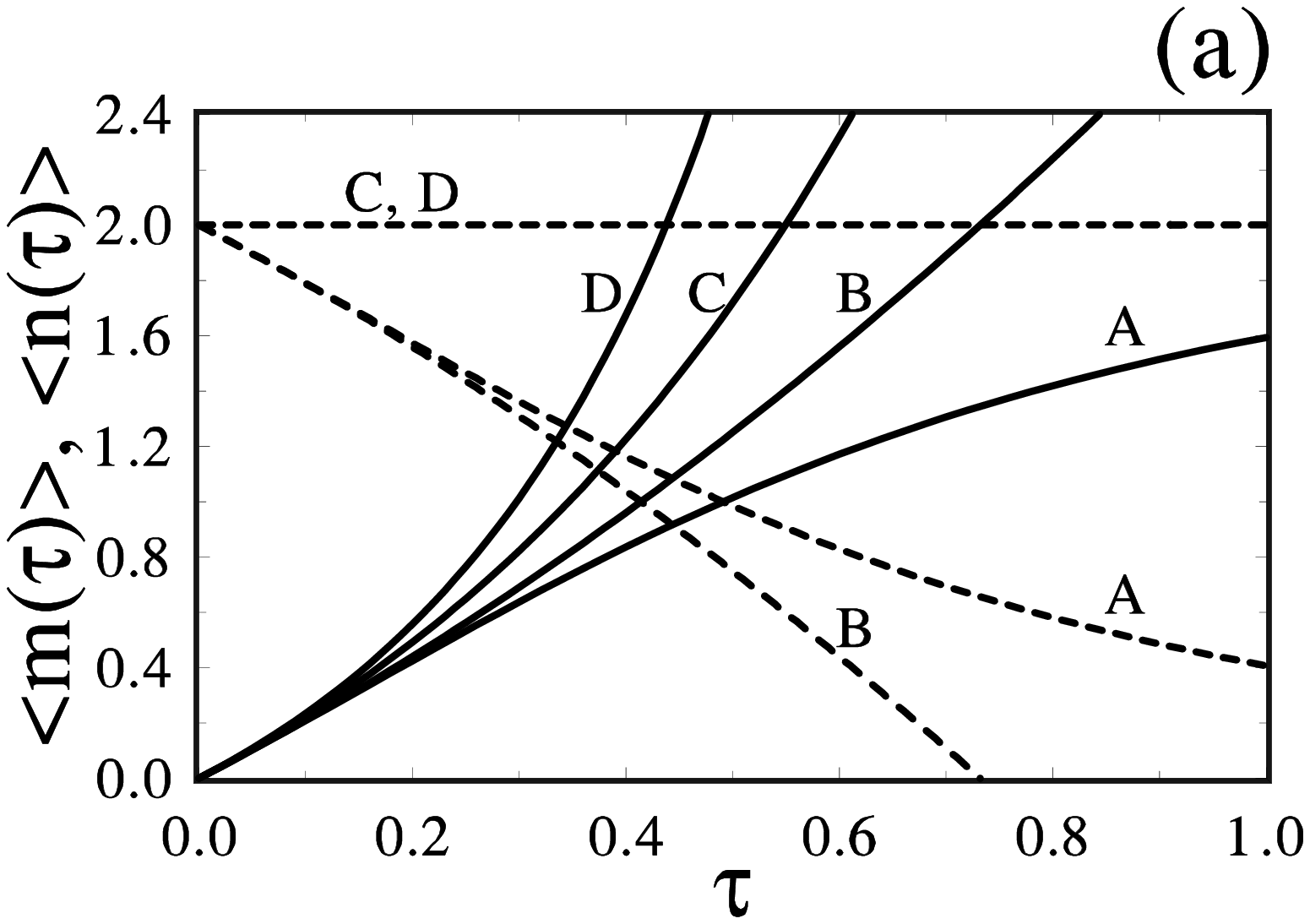,width=6.5cm}
\hspace{-5mm}\psfig{figure=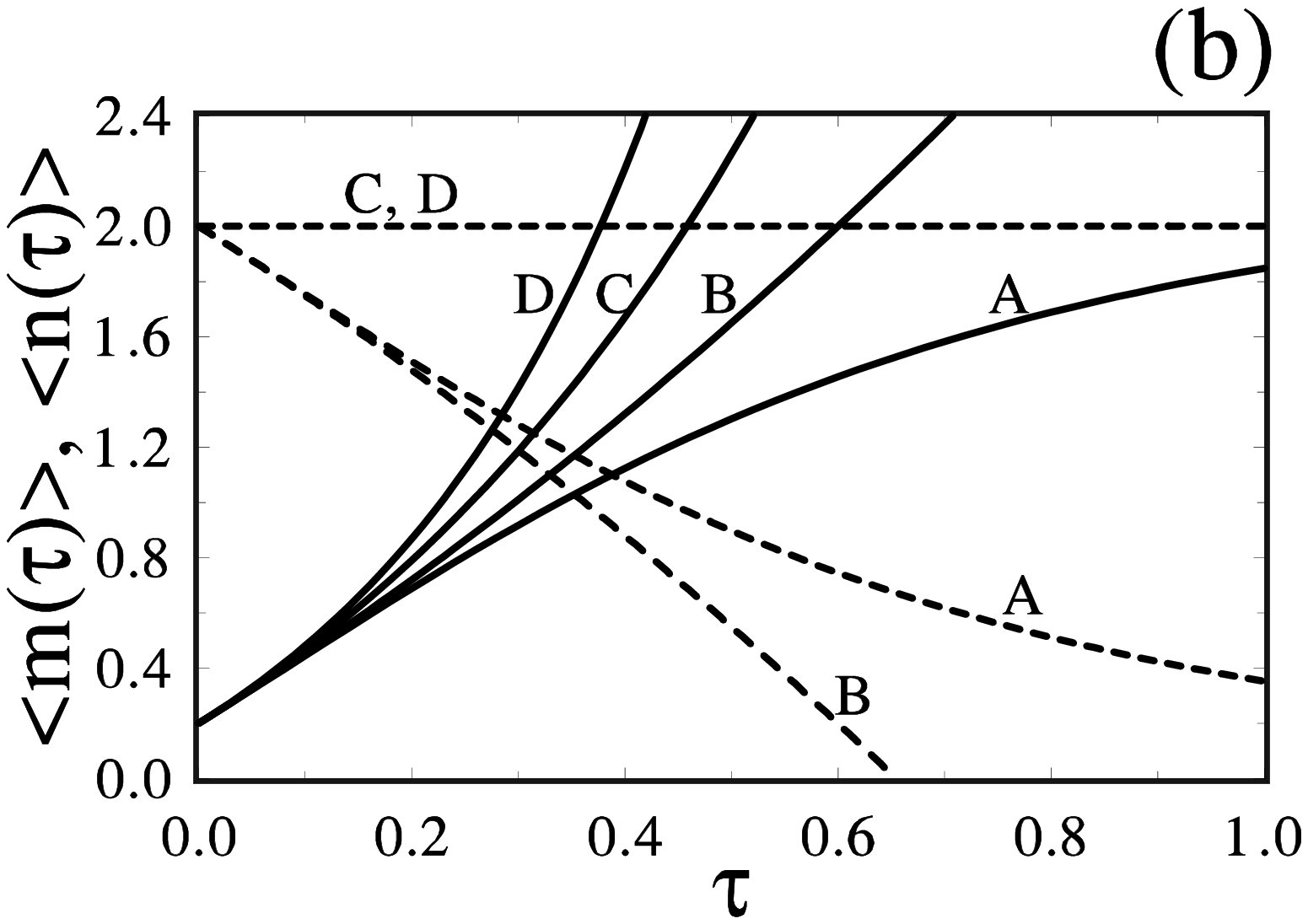,width=6.5cm}}
\vspace*{-5mm} \caption{Time behavior of the mean number of
the Stokes photons $\langle\hat{m}\rangle$ (solid lines) and
the laser photons $\langle\hat{n}\rangle$ (dashed lines) for
the initial fields: (a) $|\alpha_L=\sqrt{2}\rangle$,
$|\alpha_S=0\rangle$, and (b) $|\alpha_L=\sqrt{2}\rangle$,
$|\alpha_S=\sqrt{0.2}\rangle$. Numerical results with exact
solutions of Sect. 6.1.2 (curves A); short-time approximation
of Sect. 6.1.1 (curves B); parametric approximation of Sect.
5.2 (curves C); approximate solutions of Sect. 6.2 (curves
D).} \vspace*{-5mm} \end{figure} \noindent in the special
case in which no scattered photons are excited initially,
i.e., $\langle\hat{m}\rangle=\langle\hat{m}^2\rangle=0$. For
the initially coherent Stokes and laser modes the factorial
moment~(\ref{128}) reduces to the simple form
$\gamma_L^{(2)}\,=\,|\alpha_S|^2\,(\Delta\tau)^2$.

Our short-time solutions for the Stokes mode are
\begin{eqnarray}
\langle\hat{m}(\tau)\rangle &=& \langle\hat{m}\rangle
+\langle\hat{n}\rangle(\langle\hat{m}\rangle+1)\Delta\tau
\nonumber\\ &&- \Big[ \langle\hat{m}^2\rangle\langle\hat{n}\rangle
+\langle\hat{m}\rangle(3 \langle\hat{n}\rangle
-\langle\hat{n}^2\rangle)
+2\langle\hat{n}\rangle-\langle\hat{n}^2\rangle \Big]
\frac{(\Delta\tau)^2}{2} \nonumber\\ &&+ \Big[
\langle\hat{m}^3\rangle \langle\hat{n}\rangle
+\langle\hat{m}^2\rangle(7\langle\hat{n}\rangle
-4\langle\hat{n}^2\rangle) +\langle\hat{m}\rangle (14
\langle\hat{n}\rangle -12
\langle\hat{n}^2\rangle+\langle\hat{n}^3\rangle) \nonumber\\ &&+
8\langle\hat{n}\rangle-8\langle\hat{n}^2\rangle
+\langle\hat{n}^3\rangle \Big]\frac{(\Delta\tau)^3}{6},
\label{130}
\end{eqnarray}
\begin{eqnarray}
\langle\hat{m}^2(\tau)\rangle &=& \langle\hat{m}^2\rangle
+\langle\hat{n}\rangle
(2\langle\hat{m}^2\rangle+3\langle\hat{m}\rangle+1)\Delta\tau
\nonumber\\ &&- \Big[ 2\langle\hat{m}^3\rangle
\langle\hat{n}\rangle +\langle\hat{m}^2\rangle
(9\langle\hat{n}\rangle -4 \langle\hat{n}^2\rangle) \nonumber\\
&&+
\langle\hat{m}\rangle(13\langle\hat{n}\rangle-9\langle\hat{n}^2\rangle)
+6\langle\hat{n}\rangle -5\langle\hat{n}^2\rangle \Big]
\frac{(\Delta\tau)^2}{2} \nonumber\\ &&+ \Big[ 2
\langle\hat{m}^4\rangle \langle\hat{n}\rangle +
\langle\hat{m}^3\rangle (21 \langle\hat{n}\rangle -14
\langle\hat{n}^2\rangle) \nonumber\\ &&+
\langle\hat{m}^2\rangle(73 \langle\hat{n}\rangle -72
\langle\hat{n}^2\rangle +8\langle\hat{n}^3\rangle) \nonumber\\ &&+
\langle\hat{m}\rangle \left(102 \langle\hat{n}\rangle -118
\langle\hat{n}^2\rangle+21 \langle\hat{n}^3\rangle\right)
\nonumber\\ &&+ 48 \langle\hat{n}\rangle -60
\langle\hat{n}^2\rangle +13 \langle\hat{n}^3\rangle \Big]
\frac{(\Delta\tau)^3}{6}. \label{131}
\end{eqnarray}
On adding Eqs.~(\ref{126}) and~(\ref{130})  we note that the
sum of the mean number of photons in both the Stokes and
laser modes is constant (at least up to $(\Delta\tau)^2$):
\begin{eqnarray}
\langle \hat{n}(\tau) \rangle +\langle \hat{m}(\tau) \rangle =
\langle \hat{n} \rangle +\langle \hat{m} \rangle. \label{132}
\end{eqnarray}
\begin{figure}  
\vspace*{-1mm} \hspace*{0mm}
\centerline{\psfig{figure=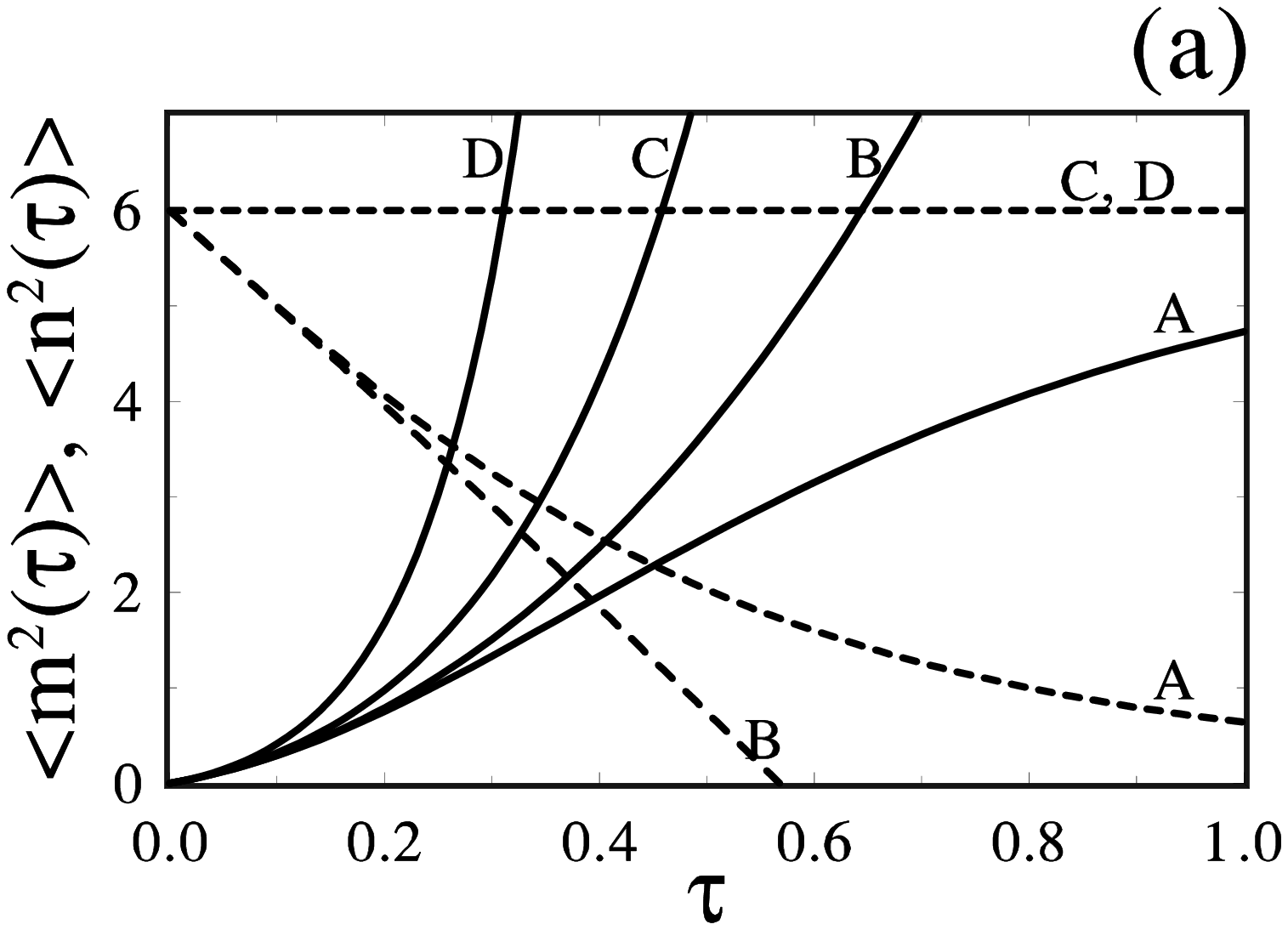,width=6.5cm}
\hspace{-5mm}\psfig{figure=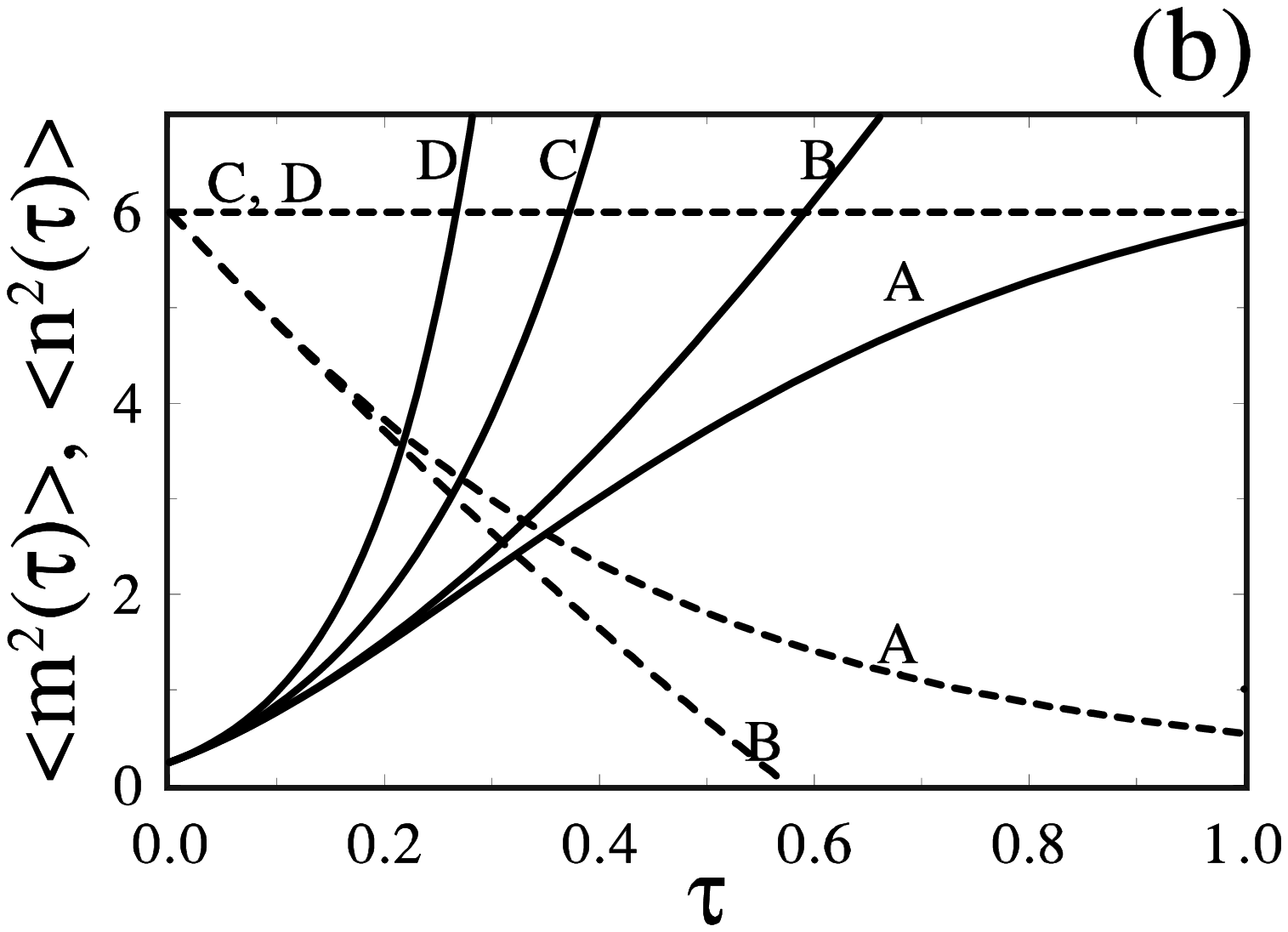,width=6.5cm}}
\vspace*{-6mm} \caption{Time behavior of the mean-square
number of the Stokes photons $\langle\hat{m}^2\rangle$ (solid
lines) and the laser photons $\langle\hat{n}^2\rangle$
(dashed lines) for the same cases as in figure 2.}
\vspace*{-5mm} \end{figure} \noindent Taking a closer look at
Eq.~(\ref{125}), which contains only terms with $\rho_{nm}$
and $\rho_{n+1,m-1}$, one can draw the more fundamental
conclusion that the property~(\ref{132})  holds for any
times, in particular for the steady solutions for
$\tau\rightarrow \infty$. Equation (\ref{132}) is a special
case of~(\ref{ex1}). Actually, we note in view of the master
equation~(\ref{123}) that the operator
$\hat{a}^+_L(\tau)\hat{a}_L(\tau)+\hat{a}^+_S(\tau)\hat{a}_S(\tau)$
is a constant of motion.

The time evolution of the mean values
$\langle\hat{n}(\tau)\rangle$, $\langle\hat{m}(\tau)\rangle$,
$\langle\hat{n}^2(\tau)\rangle$, and
$\langle\hat{m}^2(\tau)\rangle$ is shown in Figs. 2 and 3 for
initially coherent distributions. Curves B are obtained from
Eqs. (\ref{126}),~(\ref{127}),~(\ref{130}) and~(\ref{131}).

The factorial moment $\gamma_S^{(2)}(\tau)$, in the case of nonzero
$\langle\hat{m}\rangle$, is equal to
\begin{eqnarray}
\gamma_{S}^{(2)}(\tau) &=& \gamma_S^{(2)}
-2(\langle\hat{m}^2\rangle-2\langle\hat{m}\rangle^2
-\langle\hat{m}\rangle)
\langle\hat{n}\rangle\langle\hat{m}\rangle^{-3}\Delta\tau
\nonumber\\ &&- \Big[ \langle\hat{m}^3\rangle
\langle\hat{m}\rangle^2 \langle\hat{n}\rangle
-\langle\hat{m}^2\rangle^2\langle\hat{m}\rangle
\langle\hat{n}\rangle \nonumber\\ &&+ \langle\hat{m}^2\rangle
\langle\hat{m}\rangle^2 (\langle\hat{n}\rangle^2 +2
\langle\hat{n}\rangle -\langle\hat{n}^2\rangle) \nonumber\\ &&-
\langle\hat{m}^2\rangle\langle\hat{m}\rangle
(2\langle\hat{n}\rangle^2 +2\langle\hat{n}\rangle
-\langle\hat{n}^2\rangle) \nonumber\\ &&- 3
\langle\hat{m}^2\rangle \langle\hat{n}\rangle^2
+\langle\hat{m}\rangle^3(7\langle\hat{n}\rangle^2
+8\langle\hat{n}\rangle-5 \langle\hat{n}^2\rangle) \nonumber\\ &&+
\langle\hat{m}\rangle^2(10 \langle\hat{n}\rangle^2 +4
\langle\hat{n}\rangle-3 \langle\hat{n}^2\rangle) \nonumber\\ &&+ 3
\langle\hat{m}\rangle\langle\hat{n}\rangle^2 \Big]
\langle\hat{m}\rangle^{-4} (\Delta\tau)^2, \label{133}
\end{eqnarray}
whereas in the case when all moments $\langle\hat{m}^k\rangle$ (for
$k=1,2,...$) are zero, $\gamma_S^{(2)}(\tau)$ can be expressed as
\begin{eqnarray}
\gamma_{S}^{(2)}(\tau)&=&2\gamma_{L}^{(2)}+1 +\mbox{\Large (}
6\langle\hat{n}^3\rangle
-6\langle\hat{n}^2\rangle^2/\langle\hat{n}\rangle \nonumber\\ &&-
8\langle\hat{n}^2\rangle+8\langle\hat{n}\rangle \mbox{\Large )}
\langle\hat{n}\rangle^{-2} \frac{\Delta\tau}{3}. \label{134}
\end{eqnarray}
To obtain a correct time dependence of the factorial moment
(\ref{134}), it is clearly necessary to include in Eqs.
(\ref{130}) and~(\ref{131}) terms at least up to third order
in $\tau$. An equation similar to~(\ref{134}) has been
obtained by Simaan~\cite{r100}. In Fig. (4a) we compare, in
particular, our result for the factorial moments of photon
number in the Stokes mode calculated with Eq.(\ref{133})
and~(\ref{134}) (curves B) with that obtained from the exact
solution (curves A) of the master equation (\ref{125})
discussed in Sect. 6.1.2. Our Eq.~(\ref{134}) gives much
better approximation to the exact results than Simaan's
formula~(33) in Ref.~\cite{r100}. Analogously in Fig. 5, the
factorial moments for the laser mode calculated with
Eqs.~(\ref{128}) and~(\ref{129}) (curves B) are compared, in
particular, with the exact solutions (curves A).

The Eqs.~(\ref{133}) and~(\ref{134}) reduce, respectively, to
\begin{eqnarray}
\gamma_S^{(2)}(\tau) &=& 2 |\alpha_L|^2 |\alpha_S|^{-2}\Delta \tau
-\mbox{\Large (} 2|\alpha_S|^4+3|\alpha_S|^2 \nonumber\\ &&+
3|\alpha_L|^2+|\alpha_S|^2|\alpha_L|^2 \mbox{\Large )}
\:|\alpha_L|^2|\alpha_S|^{-4} (\Delta \tau)^2, \label{135}
\end{eqnarray}
\begin{eqnarray}
\gamma_S^{(2)}(\tau) = 1-\frac{2}{3} \Delta \tau. \label{136}
\end{eqnarray}
for initially coherent radiation fields.

The corresponding short-time dependence of the cross-correlation
(interbeam) function is
\begin{eqnarray}
\langle\hat{n}(\tau)\hat{m}(\tau)\rangle &=&
\langle\hat{n}\rangle\langle\hat{m}\rangle
+[\langle\hat{n}^2\rangle(\langle\hat{m}\rangle+1)
-\langle\hat{n}\rangle(\langle\hat{m}^2\rangle
+2\langle\hat{m}\rangle+1)]\Delta\tau \hspace{15mm} \nonumber\\
&&+ [\langle\hat{n}^3\rangle(\langle\hat{m}\rangle+1)
-\langle\hat{n}^2\rangle(4\langle\hat{m}^2\rangle
+11\langle\hat{m}\rangle+7) \nonumber\\ &&+ \langle\hat{n}\rangle
(\langle\hat{m}^3\rangle
+6\langle\hat{m}^2\rangle+11\langle\hat{m}\rangle+6) ]
\frac{(\Delta\tau)^2}{2} \nonumber\\ &&- \mbox {\Large [} 3
\langle\hat{n}^4\rangle (2\langle\hat{m}\rangle
+\langle\hat{m}^2\rangle+1) -\langle\hat{n}^3\rangle
(3\langle\hat{m}\rangle +4 \langle\hat{m}^2\rangle-1) \nonumber\\
&&- \langle\hat{n}^2\rangle(68 \langle\hat{m}\rangle
+43\langle\hat{m}^2\rangle+11\langle\hat{m}^3\rangle+36)
\nonumber\\ &&+ \langle\hat{n}\rangle
(66\langle\hat{m}\rangle+47\langle\hat{m}^2\rangle
+14\langle\hat{m}^3\rangle+\langle\hat{m}^4\rangle+32) \mbox
{\Large ]} \frac{(\Delta\tau)^3}{6}.\nonumber\\ \label{137}
\end{eqnarray}
On inserting~(\ref{126}),~(\ref{130}), and~(\ref{137}) into
the definition~(\ref{038}) of the interbeam degree of
second-order coherence, ${g}_{LS}^{(2)}(\tau)$, we obtain the
following relation for the case when photons are initially
present in the Stokes mode:
\begin{eqnarray}
{g}_{LS}^{(2)}(\tau) &=& \Big[
\langle\hat{n}^2\rangle(\langle\hat{m}\rangle+1)
+\langle\hat{n}\rangle(\langle\hat{m}\rangle^2-\langle\hat{m}\rangle
-\langle\hat{m}^2\rangle-1) \nonumber\\ &&-
\langle\hat{n}\rangle^2 (\langle\hat{m}\rangle+1) \Big]
(\langle\hat{n}\rangle\langle\hat{m}\rangle)^{-1}\Delta\tau
\nonumber\\ &&+ \mbox{\Large \{}
\langle\hat{n}^3\rangle\langle\hat{m}\rangle
(\langle\hat{m}\rangle+1)
+\langle\hat{n}^2\rangle\langle\hat{m}\rangle
(3\langle\hat{m}\rangle^2-6\langle\hat{m}\rangle
-4\langle\hat{m}^2\rangle -5) \nonumber\\ &&-
\langle\hat{n}^2\rangle\langle\hat{n}\rangle
(3\langle\hat{m}\rangle^2+5\langle\hat{m}\rangle+2) \nonumber\\
&&+ \langle\hat{n}\rangle\langle\hat{m}\rangle \Big[ 2
\langle\hat{m}\rangle^3-3\langle\hat{m}\rangle^2
+\langle\hat{m}\rangle(5-3\langle\hat{m}^2\rangle) \nonumber\\ &&+
4\langle\hat{m}^2\rangle +\langle\hat{m}^3\rangle +4 \Big]
\nonumber\\ &&- \langle\hat{n}\rangle^2 \Big[
2\langle\hat{m}\rangle^3 -3 \langle\hat{m}\rangle^2 -3
\langle\hat{m}\rangle
(\langle\hat{m}^2\rangle+2)-2(\langle\hat{m}^2\rangle+1) \Big]
\nonumber\\ &&+ 2\langle\hat{n}\rangle^3 (\langle\hat{m}\rangle^2
+2\langle\hat{m}\rangle+1) \mbox{\Large \}}
\langle\hat{m}\rangle^{-2} \langle\hat{n}\rangle
\frac{(\Delta\tau)^2}{2}, \label{138}
\end{eqnarray}
otherwise, for the case
\begin{figure}  
\vspace*{-9mm} \hspace*{0mm}
\centerline{\psfig{figure=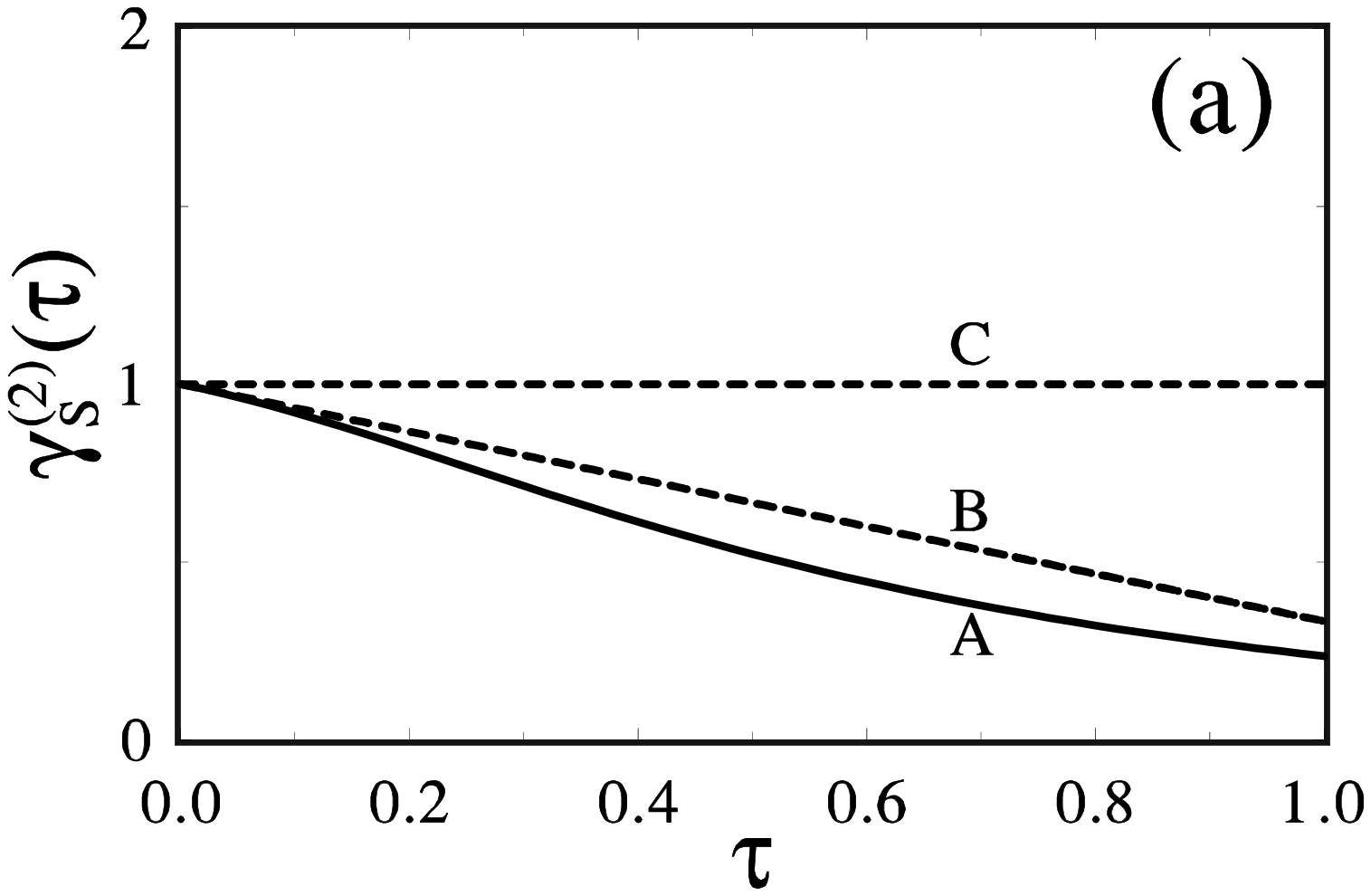,width=6.5cm}
\hspace{-5mm}\psfig{figure=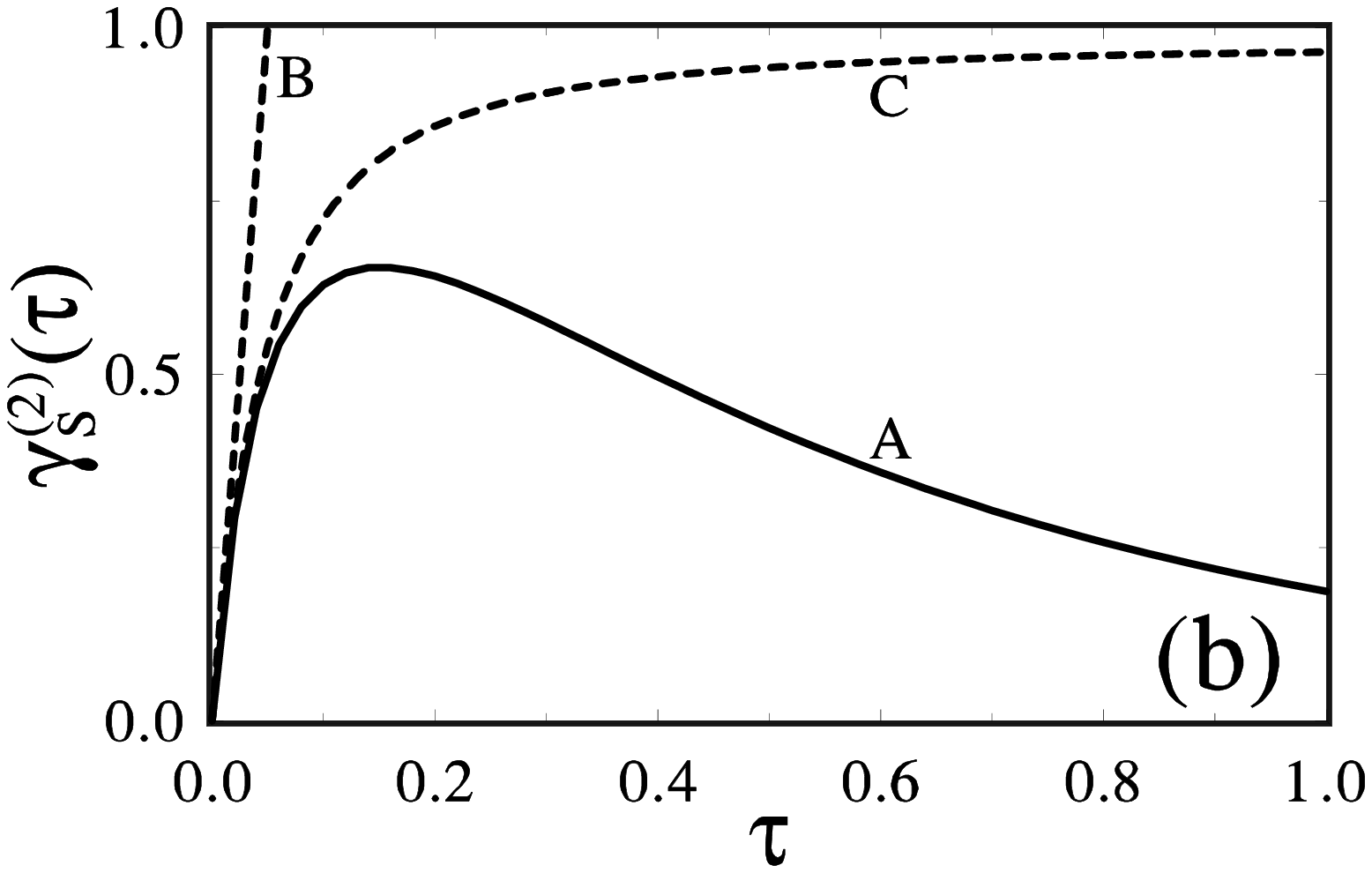,width=6.5cm}}
\caption{Time behavior of the normalized factorial moments
$\gamma_S^{(2)}(\tau)$ for the Stokes mode for the same cases
(except for curve D) as in figure 2.} \vspace*{-5mm}
\end{figure} $\langle\hat{m}\rangle=\langle\hat{m}^2\rangle
=\langle\hat{m}^3\rangle=\langle\hat{m}^4\rangle=0$, we get
\begin{eqnarray}
{g}_{LS}^{(2)}(\tau)&=& \gamma_L^{(2)} +\mbox{\Large (}
\langle\hat{n}^3\rangle
-\langle\hat{n}^2\rangle^2/\langle\hat{n}\rangle \nonumber\\ &&- 2
\langle\hat{n}^2\rangle+2 \langle\hat{n}\rangle \mbox{\Large )}
\langle\hat{n}\rangle^{-2} \frac{\Delta\tau}{2} \nonumber\\ &&-
\mbox{\Large (} 6\langle\hat{n}^4\rangle \langle\hat{n}\rangle^2
+5\langle\hat{n}^3\rangle \langle\hat{n}^2\rangle
\langle\hat{n}\rangle \nonumber\\ &&- 12
\langle\hat{n}^3\rangle\langle\hat{n}\rangle^2 -3
\langle\hat{n}^2\rangle^3 -22
\langle\hat{n}^2\rangle^2\langle\hat{n}\rangle \nonumber\\ &&+ 26
\langle\hat{n}^2\rangle\langle\hat{n}\rangle^2 \mbox{\Large )}
\langle\hat{n}\rangle^{-4} \frac{(\Delta\tau)^2}{12}. \label{139}
\end{eqnarray}
Assuming that the Stokes and laser modes are initially
coherent, Eqs.~(\ref{138}) and~(\ref{139}) reduce
respectively to
\begin{eqnarray}
\gamma_{LS}^{(2)}(\tau) = -\Delta\tau
+\left(|\alpha_S|^2-2|\alpha_S|^2|\alpha_L|^2+|\alpha_L|^2\right)
|\alpha_S|^{-2} \frac{(\Delta\tau)^2}{2}, \label{140}
\end{eqnarray}
\begin{eqnarray}
\gamma_{LS}^{(2)}(\tau) = -\frac{\Delta\tau}{2}
+(1-13|\alpha_L|^2-8|\alpha_L|^4)\, \frac{(\Delta\tau)^2}{12}.
\label{141}
\end{eqnarray}

Equation~(\ref{137}), calculated up to the third order in
$\Delta\tau$, enables us to determine the
relation~(\ref{139}) correct up to $\Delta\tau$ squared,
only. Simaan~\cite{r100} has calculated an expression similar
to Eq. (\ref{139}).  Examples of the time evolution of
$g_{LS}^{(2)}(\tau)$ for initially coherent fields are
presented in Fig. 6. Curves B in Figs. 6a and b are
calculated with Eqs.~(\ref{139}) and~(\ref{138}) (including
terms up to $\Delta\tau$ only). Curve S in Fig. 6a is
calculated from the Simaan short-time approximate solution
(32) of Ref.~\cite{r100}. One can compare these results
(curves B and S) with $g_{LS}^{(2)}(\tau)$ obtained from our
numerical calculations utilizing the exact solution of the
master equation~(\ref{123}) (curves A). We note the supremacy
of our short-time approximation~(\ref{141}).

By analogy to the photon-number moments we calculate, in the
short-time approximation, the mean and mean square of the
annihilation operators, $\langle\hat{a}^+_k(\tau)\rangle$ and
$\langle\hat{a}^{+2}_k(\tau)\rangle$ for both fields
($k=L,S$), as well as the cross-correlation functions
$\langle\hat{a}^{+}_L(\tau)\hat{a}^{+}_S(\tau)\rangle$ and
$\langle\hat{a}_L(\tau)\hat{a}^{+}_S(\tau)\rangle$. After
some algebra, we arrive at
\begin{figure}  
\vspace*{-11mm} \hspace*{0mm}
\centerline{\psfig{figure=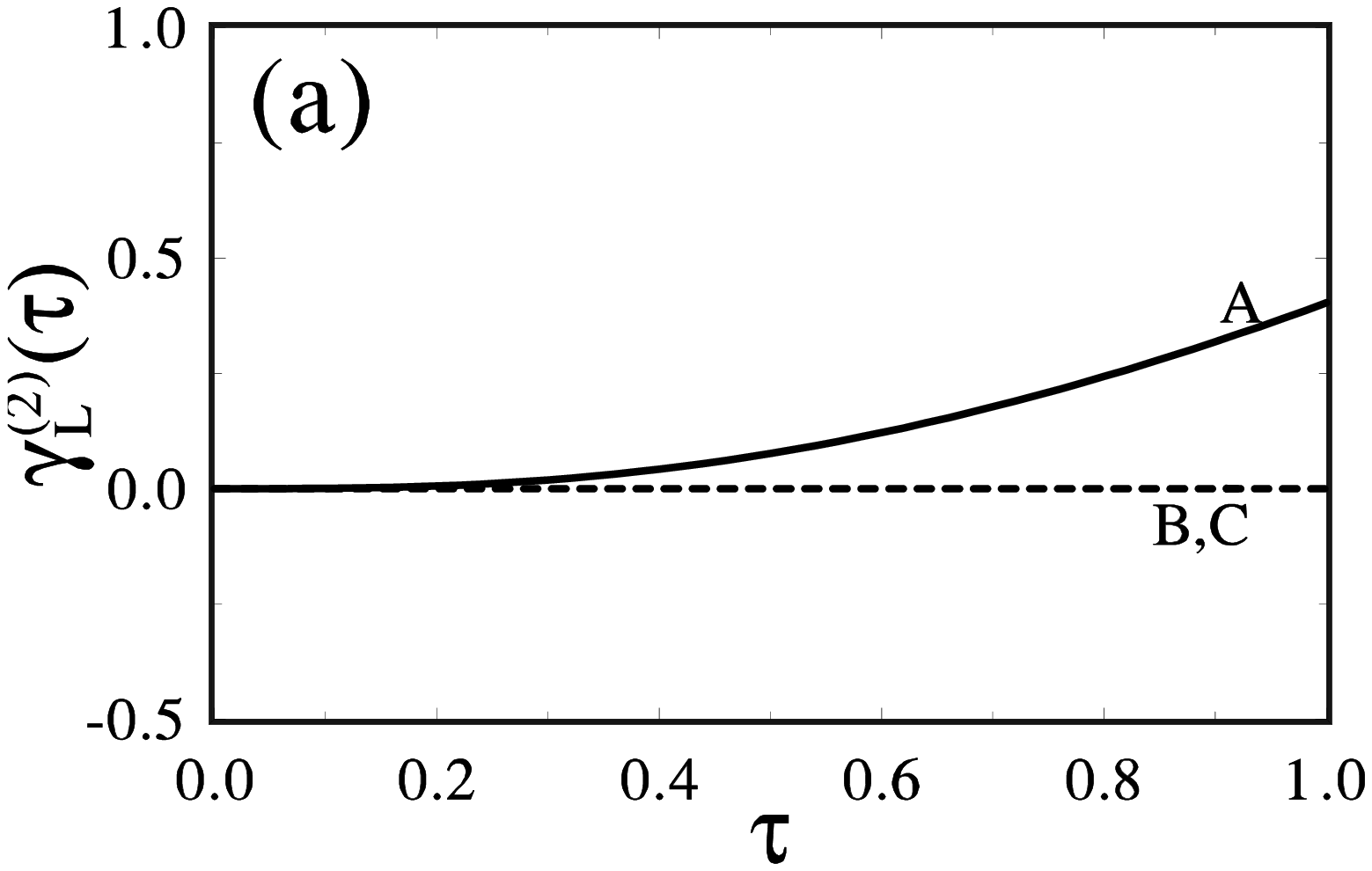,width=6.5cm}
\hspace{-5mm}\psfig{figure=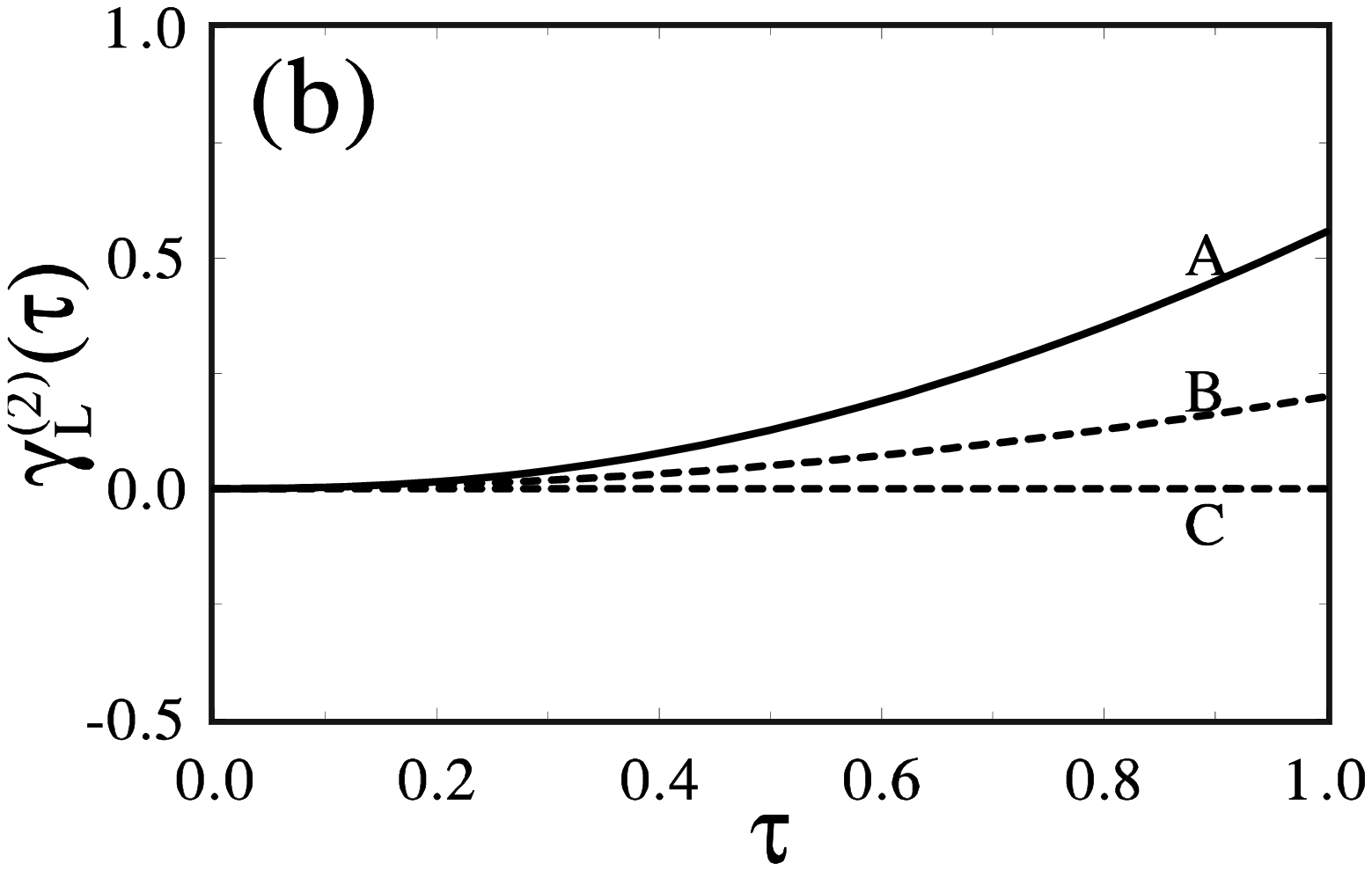,width=6.5cm}} \caption{ Same
as figure 4, but for the normalized factorial moments
$\gamma_L^{(2)}(\tau)$ of the laser mode.} \vspace*{-7mm}
\end{figure}
\begin{eqnarray}
\langle \hat{a}^+_S(\tau)\rangle = \langle \hat{a}^+_S\rangle
+\langle \hat{a}^+_L \hat{a}_L\rangle\langle\hat{a}^+_S\rangle
\frac{\Delta\tau}{2} \hspace{5cm}\nonumber \\ + (\langle
\hat{a}^{+2}_L \hat{a}^2_L\rangle \langle \hat{a}^+_S\rangle
-2\langle \hat{a}^{+}_L \hat{a}_L\rangle
\langle\hat{a}^{+2}_S\hat{a}_S\rangle -3\langle \hat{a}^{+}_L
\hat{a}_L\rangle \langle \hat{a}^{+}_S\rangle)
\frac{(\Delta\tau)^2}{8}, \label{142}
\end{eqnarray}
\begin{eqnarray}
\langle \hat{a}^+_L(\tau)\rangle = \langle \hat{a}^+_L\rangle
-\langle \hat{a}^{+}_L\rangle(\langle\hat{a}^+_S
\hat{a}_S\rangle+1) \frac{\Delta\tau}{2}+ [\langle
\hat{a}^{+}_L\rangle\langle\hat{a}^{+2}_S\hat{a}^2_S\rangle
\hspace{1cm}\nonumber\\ - 2\langle \hat{a}^{+2}_L \hat{a}_L\rangle
(\langle\hat{a}^{+}_S\hat{a}_S\rangle+1) +\langle
\hat{a}^{+}_L\rangle (3\langle\hat{a}^{+}_S\hat{a}_S\rangle+1)]
\frac{(\Delta\tau)^2}{8},\label{143}
\end{eqnarray}
\begin{eqnarray}
\langle \hat{a}^{+2}_S(\tau)\rangle = \langle
\hat{a}^{+2}_S\rangle +\langle \hat{a}^{+}_L \hat{a}_L \rangle
\langle\hat{a}^{+2}_S\rangle \Delta\tau  \hspace{4.5cm}\nonumber\\
+ (\langle \hat{a}^{+2}_L \hat{a}^{2}_L \rangle
\langle\hat{a}^{+2}_S\rangle -\langle \hat{a}^{+}_L
\hat{a}_L\rangle \langle\hat{a}^{+3}_S\hat{a}_S\rangle -2\langle
\hat{a}^{+}_L \hat{a}_L\rangle \langle \hat{a}^{+2}_S\rangle)
\frac{(\Delta\tau)^2}{2}, \label{144}
\end{eqnarray}
\begin{eqnarray}
\langle \hat{a}^{+2}_L(\tau)\rangle = \langle
\hat{a}^{+2}_L\rangle -\langle \hat{a}^{+2}_L \rangle
(\langle\hat{a}^{+}_S\hat{a}_S\rangle +1)\Delta\tau +[\langle
\hat{a}^{+2}_L \rangle \langle\hat{a}^{+2}_S\hat{a}^2_S\rangle
\hspace{1cm}\nonumber\\ - \langle \hat{a}^{+3}_L \hat{a}_L \rangle
(\langle\hat{a}^{+}_S\hat{a}_S\rangle+1) +\langle \hat{a}^{+2}_L
\rangle (3\langle\hat{a}^{+}_S\hat{a}_S\rangle+1)]
\frac{(\Delta\tau)^2}{2}, \label{145}
\end{eqnarray}
\begin{eqnarray}
\langle \hat{a}^{+}_L(\tau)\hat{a}^{+}_S(\tau)\rangle = \langle
\hat{a}^{+}_L\rangle\langle\hat{a}^{+}_S\rangle +(\langle
\hat{a}^{+2}_L\hat{a}_L\rangle \langle \hat{a}^{+}_S\rangle
-\langle \hat{a}^{+}_L\rangle \langle
\hat{a}^{+2}_S\hat{a}_S\rangle -2\langle
\hat{a}^{+}_L\rangle\langle\hat{a}^{+}_S\rangle )
\frac{\Delta\tau}{2} \nonumber \\ + (\langle
\hat{a}^{+3}_L\hat{a}^2_L\rangle \langle \hat{a}^{+}_S\rangle
-11\langle \hat{a}^{+2}_L\hat{a}_L\rangle \langle
\hat{a}^{+}_S\rangle -6\langle \hat{a}^{+2}_L\hat{a}_L\rangle
\langle \hat{a}^{+2}_S\hat{a}_S\rangle \nonumber \\ + 5\langle
\hat{a}^{+}_L\rangle \langle \hat{a}^{+2}_S\hat{a}_S\rangle
+\langle \hat{a}^{+}_L\rangle \langle
\hat{a}^{+3}_S\hat{a}^2_S\rangle +4\langle \hat{a}^{+}_L\rangle
\langle \hat{a}^{+}_S\rangle) \frac{(\Delta\tau)^2}{8},\nonumber\\
\label{146}
\end{eqnarray}
\begin{eqnarray}
\langle \hat{a}_L(\tau)\hat{a}^{+}_S(\tau)\rangle = \langle
\hat{a}_L\rangle\langle\hat{a}^{+}_S\rangle +(\langle
\hat{a}^{+}_L\hat{a}^2_L\rangle \langle \hat{a}^{+}_S\rangle
-\langle \hat{a}_L\rangle \langle \hat{a}^{+2}_S\hat{a}_S\rangle
-2\langle \hat{a}_L\rangle\langle\hat{a}^{+}_S\rangle)
\frac{\Delta\tau}{2} \nonumber \\ + (\langle
\hat{a}^{+2}_L\hat{a}^3_L\rangle \langle \hat{a}^{+}_S\rangle
-11\langle \hat{a}^+_L\hat{a}^2_L\rangle \langle
\hat{a}^{+}_S\rangle -6\langle \hat{a}^+_L\hat{a}^2_L\rangle
\langle \hat{a}^{+2}_S\hat{a}_S\rangle \nonumber \\ + 9\langle
\hat{a}_L\rangle \langle \hat{a}^{+2}_S\hat{a}_S\rangle +\langle
\hat{a}_L\rangle \langle \hat{a}^{+3}_S\hat{a}^2_S\rangle
+12\langle\hat{a}_L\rangle \langle\hat{a}^{+}_S\rangle)
\frac{(\Delta\tau)^2}{8}. \nonumber\\ \label{147}
\end{eqnarray}
\begin{figure}  
\vspace*{-6mm} \hspace*{0mm}
\centerline{\psfig{figure=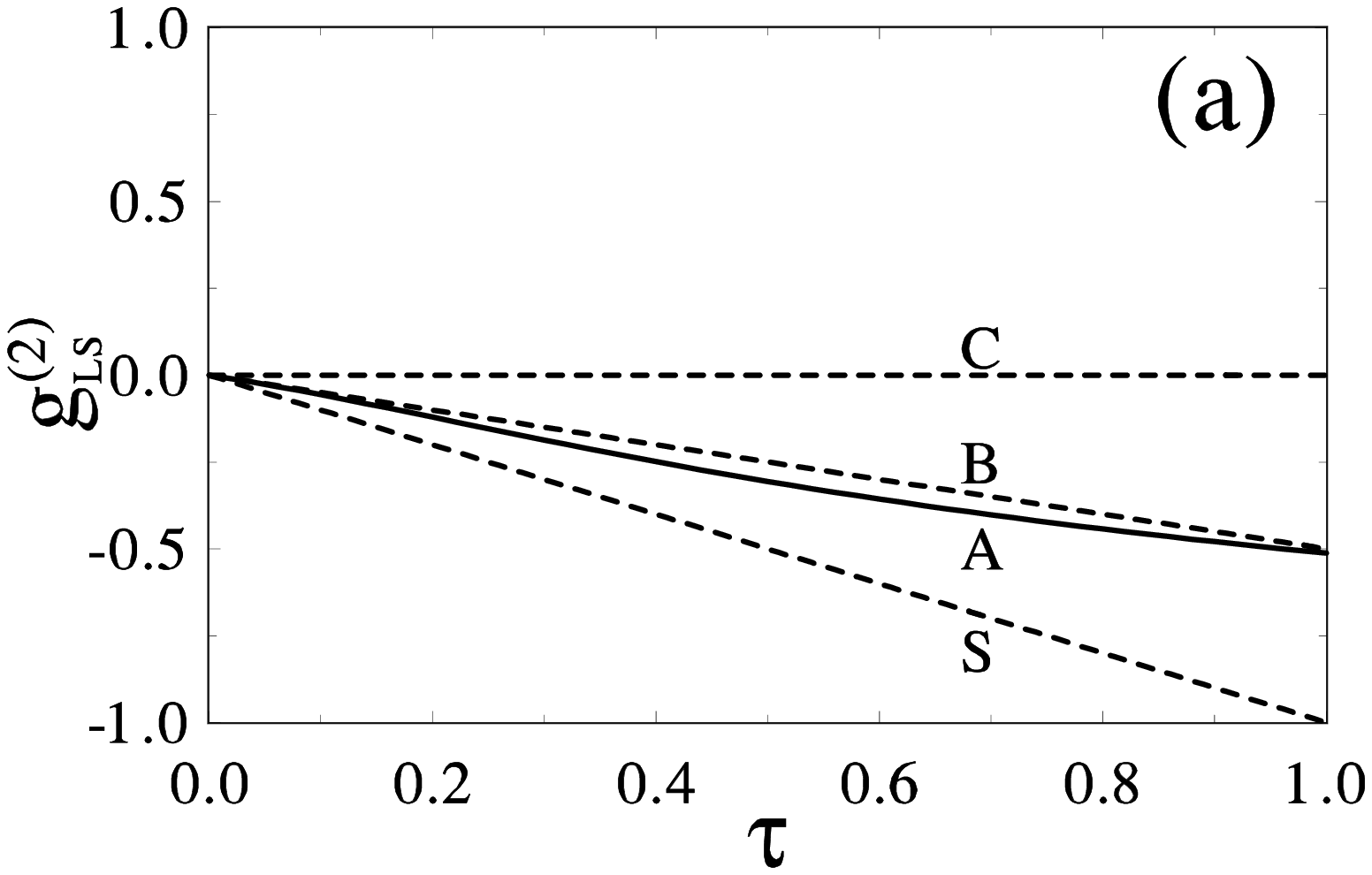,width=6.5cm}
\hspace{-5mm}\psfig{figure=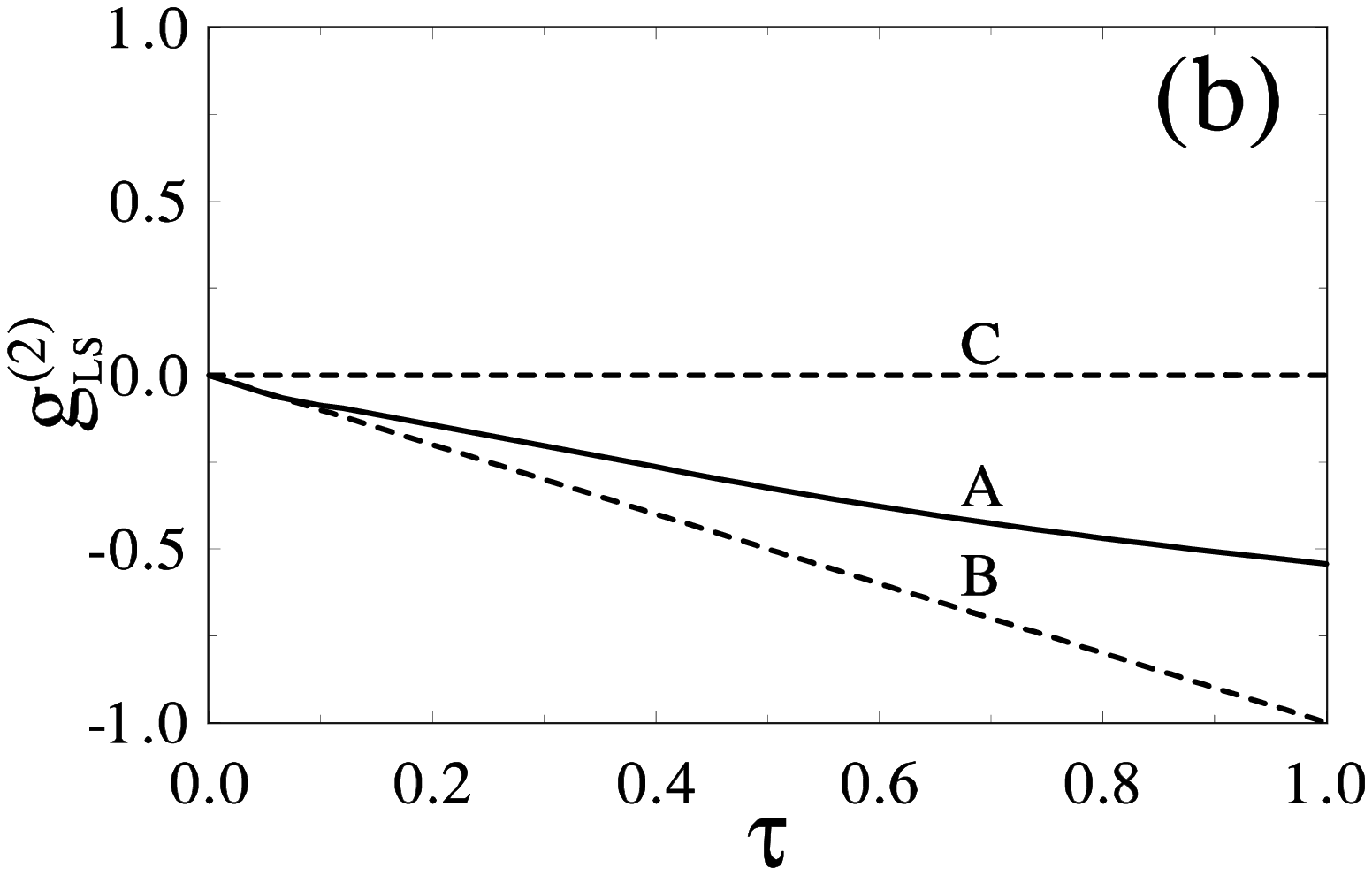,width=6.5cm}}
 \caption{Same as figure 4, but for the interbeam degree of
coherence $g_{LS}^{(2)}(\tau)$. Additional curve S is
calculated with the Simaan short-time approximation (Eq. (32)
of Ref.~\cite{r100}).} \vspace*{-5mm} \end{figure}

For brevity, here, we shall restrict our considerations to initially
coherent states for the Stokes mode denoted as $\alpha_S=|\alpha_S|
\exp(i\phi_S)$ and for the laser mode $\alpha_L=|\alpha_L|
\exp(i\phi_L)$. In Figs. 7 and 8 we demonstrate the evolution of our
short-time approximations for $\langle\hat{a}_S(\tau)\rangle$ (solid
line B in Fig.  7), $\langle\hat{a}_L(\tau)\rangle$ (dashed line B in
Fig. 7), $\langle\hat{a}^2_S(\tau)\rangle$ (solid line B in Fig. 8),
and $\langle\hat{a}^2_L(\tau)\rangle$ (dashed line B in Fig. 8) for
initially coherent radiation modes.

From the general relations~(\ref{130}) and~(\ref{126}), under
the condition of initial coherent Stokes and laser fields, we
get
\begin{eqnarray}
\langle \hat{m}(\tau)\rangle = |\alpha_S|^2
+|\alpha_L|^2(|\alpha_S|^2+1) \Delta\tau \hspace{4cm}\nonumber\\
+ |\alpha_L|^2[(|\alpha_L|^2+1)(|\alpha_S|^2+1)
-(|\alpha_S|^4+4|\alpha_S|^2+2)] \frac{(\Delta\tau)^2}{2}
\label{148},
\end{eqnarray}
\begin{eqnarray}
\langle \hat{n}(\tau)\rangle = |\alpha_L|^2 + |\alpha_S|^2 -
\langle \hat{m}(\tau)\rangle. \label{149}
\end{eqnarray}
Inserting~(\ref{148}) as well as~(\ref{142}) and~(\ref{144})
with
$\langle\hat{a}_k^{+p}\hat{a}_k^{q}\rangle=|\alpha_k|^{p+q}\exp
[i(p-q)\phi_k]$ ($k=S,A$), into~(\ref{043}) we obtain the
$\theta$-dependent variance for the Stokes mode
\begin{eqnarray}
\langle (\Delta\hat{X}_S(\theta))^2\rangle &=&
1+2|\alpha_L|^2\Delta\tau +|\alpha_L|^2\{|\alpha_L|^2 \nonumber\\
&&- [1+ \cos^2(\theta-\phi_S)]|\alpha_S|^2-1\}(\Delta\tau)^2.
\label{150}
\end{eqnarray}
The minimal variance $\langle(\Delta\hat{X}_{S-})^2\rangle$,
which follows from~(\ref{046}), or directly from~(\ref{150}),
is equal to
\begin{eqnarray}
\langle (\Delta\hat{X}_{S-})^2\rangle = 1+2|\alpha_L|^2\Delta\tau
+|\alpha_L|^2 (|\alpha_L|^2-2|\alpha_S|^2-1)(\Delta\tau)^2.
\label{151}
\end{eqnarray}
Analogously, for the laser field we obtain the following
$\theta$-dependent variance:
\begin{eqnarray}
\langle (\Delta\hat{X}_{L}(\theta))^2\rangle =
1+\frac{1}{2}[\cos(2\theta-2\phi_L)+1]
|\alpha_L|^2|\alpha_S|^2(\Delta\tau)^2, \label{152}
\end{eqnarray}
\begin{figure}  
\vspace*{-15mm} \hspace*{0mm}
\centerline{\psfig{figure=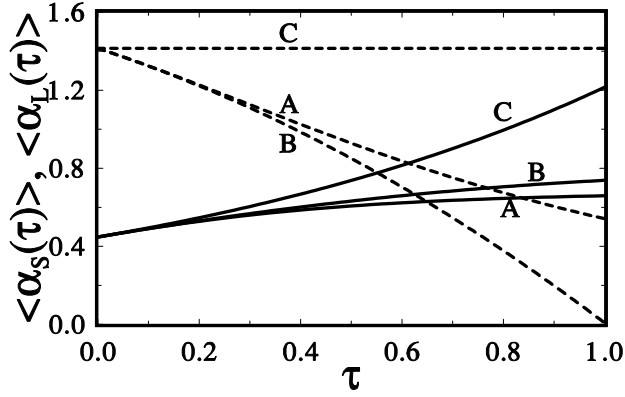,width=9cm}}
\vspace*{-5mm} \caption{Time dependence of the expectation
values of the field amplitudes
$\langle\hat{a}_S(\tau)\rangle=\langle\alpha_S(\tau)\rangle$
(solid lines) and
$\langle\hat{a}_L(\tau)\rangle=\langle\alpha_L(\tau)\rangle$
(dashed lines) for fields initially coherent
$|\alpha_L=\sqrt{2}\rangle$ and
$|\alpha_S=\sqrt{0.2}\rangle$. Curves A, B, C are calculated
within the formalisms of Sects. 6.1.2, 6.1.1 and 5.2,
respectively.} \vspace*{-7mm} \end{figure} \noindent on
insertion of Eqs.~(\ref{149}),~(\ref{143}), and (\ref{145})
into~(\ref{043}). With regard to the relation (\ref{046}),
the minimal variance for the field,
$\langle(\Delta\hat{X}_{L-})^2\rangle$, is constant up to the
second order in time:
\begin{eqnarray}
\langle (\Delta\hat{X}_{L-})^2\rangle = 1. \label{153}
\end{eqnarray}
One can readily deduce the maximal variances
$\langle(\Delta\hat{X}_{L,S\pm})^2\rangle$ from~(\ref{150})
and~(\ref{152}) or from~(\ref{046}). The time evolution of
the single-mode extremal variances obtained from
(\ref{150})-(\ref{153}) is presented in Figs. 9 and 10 (for
$\phi_L=0$):
$\langle(\Delta\hat{X}_{S-}(\tau))^2\rangle=\langle(\Delta\hat{X}_{S2}(\tau))^2\rangle$
(solid line B in Fig. 9),
$\langle(\Delta\hat{X}_{S+}(\tau))^2\rangle=\langle(\Delta\hat{X}_{S1}(\tau))^2\rangle$
(dashed line B in Fig. 9),
$\langle(\Delta\hat{X}_{L-}(\tau))^2\rangle=\langle(\Delta\hat{X}_{L1}(\tau))^2\rangle$
(solid line B in Fig. 10),
$\langle(\Delta\hat{X}_{L+}(\tau))^2\rangle=\langle(\Delta\hat{X}_{L2}(\tau))^2\rangle$
(dashed line B in Fig. 10).

The covariances for quadratures in the Stokes and laser mode,
according to~(\ref{052}), are, respectively,
\begin{eqnarray}
\langle \{ \Delta\hat{X}_{S1},\Delta\hat{X}_{S2}\}\rangle =
|\alpha_L|^2|\alpha_S|^2\sin(2\phi_S) (\Delta\tau)^2, \label{154}
\end{eqnarray}
\begin{eqnarray}
\langle \{ \Delta\hat{X}_{L1},\Delta\hat{X}_{L2}\}\rangle =
-|\alpha_L|^2|\alpha_S|^2\sin(2\phi_L) (\Delta\tau)^2. \label{155}
\end{eqnarray}
The generalized Heisenberg uncertainty relation~(\ref{051})
with the covariances~(\ref{154}) and~(\ref{155}) inserted,
takes the following form for the Stokes mode in our
short-time approximation:
\begin{eqnarray}
4|\alpha_L|^2\Delta\tau+|\alpha_L|^2(6|\alpha_L|^2-3|\alpha_S|^2-2)
(\Delta\tau)^2 \ge 0,
\label{156}
\end{eqnarray}
and for the laser mode
\begin{eqnarray}
2 |\alpha_L|^2 |\alpha_S|^2 (\Delta\tau)^2 \ge 0.
\label{157}
\end{eqnarray}
\begin{figure}  
\vspace*{-10mm} \hspace*{0mm}
\centerline{\psfig{figure=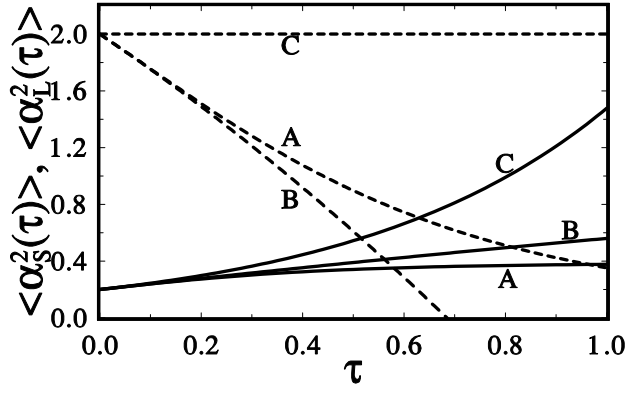,width=9cm}} \vspace*{-5mm}
\caption{Time dependence of
$\langle\hat{a}^2_S(\tau)\rangle=\langle\alpha_S^2(\tau)\rangle$
(solid lines) and
$\langle\hat{a}^2_L(\tau)\rangle=\langle\alpha_L^2(\tau)\rangle$
(dashed lines) for the same cases as in figure 7. } \vspace*{-5mm}
\end{figure}
To obtain the two-mode variances and covariances of the
quadratures one has to calculate, apart from the single-mode
functions~(\ref{150}),~(\ref{152}),~(\ref{154}), and
(\ref{155}), the cross-correlations~(\ref{057}), which are
obtained in the following form:
\begin{eqnarray}
\langle \Delta\hat{X}_{L1}\Delta\hat{X}_{S1} \rangle =
-|\alpha_L|\,|\alpha_S| \mbox{\Large \{} 2\cos\phi_L \cos\phi_S
\Delta\tau + [ \cos(\phi_L-\phi_S)\hspace{2cm} \nonumber\\ \times
(4 n_L-6 n_S-11) + \cos(\phi_L+\phi_S) (4
|\alpha_L|^2-2|\alpha_S|^2-3)\frac{(\Delta\tau)^2}{4} \mbox{\Large
\}},\hspace{1cm} \label{158}
\end{eqnarray}
\begin{eqnarray}
\langle \Delta\hat{X}_{L2}\Delta\hat{X}_{S2} \rangle =
-|\alpha_L|\,|\alpha_S| \mbox{\Large \{} 2\sin\phi_L \sin\phi_S
\Delta\tau + [ \cos(\phi_L-\phi_S) \hspace{1cm}\nonumber\\\times
(4 n_L-6 n_S-11)  - \cos(\phi_L+\phi_S) (4
|\alpha_L|^2-2|\alpha_S|^2-3)\frac{(\Delta\tau)^2}{4} \mbox{\Large
\}},\label{159}
\end{eqnarray}
\begin{eqnarray}
\langle \Delta\hat{X}_{L1}\Delta\hat{X}_{S2} \rangle =
|\alpha_L|\,|\alpha_S| \mbox{\Large \{} 2\cos\phi_L \sin\phi_S
\Delta\tau + [ \sin(\phi_S-\phi_L)  \hspace{1cm}\nonumber\\ \times
(4 n_L-6 n_S-11) + \sin(\phi_S+\phi_L) (4
|\alpha_L|^2-2|\alpha_S|^2-3)\frac{(\Delta\tau)^2}{4} \mbox{\Large
\}}, \label{160}
\end{eqnarray}
\begin{eqnarray}
\langle \Delta\hat{X}_{L2}\Delta\hat{X}_{S1} \rangle =
|\alpha_L|\,|\alpha_S| \mbox{\Large \{} 2\sin\phi_L \cos\phi_S
\Delta\tau + [ \sin(\phi_L-\phi_S) \hspace{1cm} \nonumber\\
\times(4 n_L-6 n_S-11) + \sin(\phi_L+\phi_S) (4
|\alpha_L|^2-2|\alpha_S|^2-3)\frac{(\Delta\tau)^2}{4} \mbox{\Large
\}}.\label{161}
\end{eqnarray}
Thus, the two-mode Wigner covariance~(\ref{056}) of the
quadratures $\hat{X}_{LS1}$ and $\hat{X}_{LS2}$ is
\begin{eqnarray}
\langle \{ \Delta\hat{X}_{LS1},\Delta\hat{X}_{LS2}\}\rangle &=&
4|\alpha_L||\alpha_S|\sin(\phi_L+\phi_S)\Delta\tau \nonumber\\ &&+
\Big[ 4 |\alpha_L|^2|\alpha_S|^2 \cos\frac{1}{2}(\phi_S+\phi_L)
\sin\frac{1}{2}(\phi_S-\phi_L) \nonumber\\&&+
|\alpha_L||\alpha_S|\sin(\phi_L+\phi_S)
(4|\alpha_L|^2-2|\alpha_S|^2-3) \Big] (\Delta\tau)^2,\nonumber\\
\label{162}
\end{eqnarray}
\begin{figure} 
\vspace*{-15mm} \hspace*{0mm}
\centerline{\psfig{figure=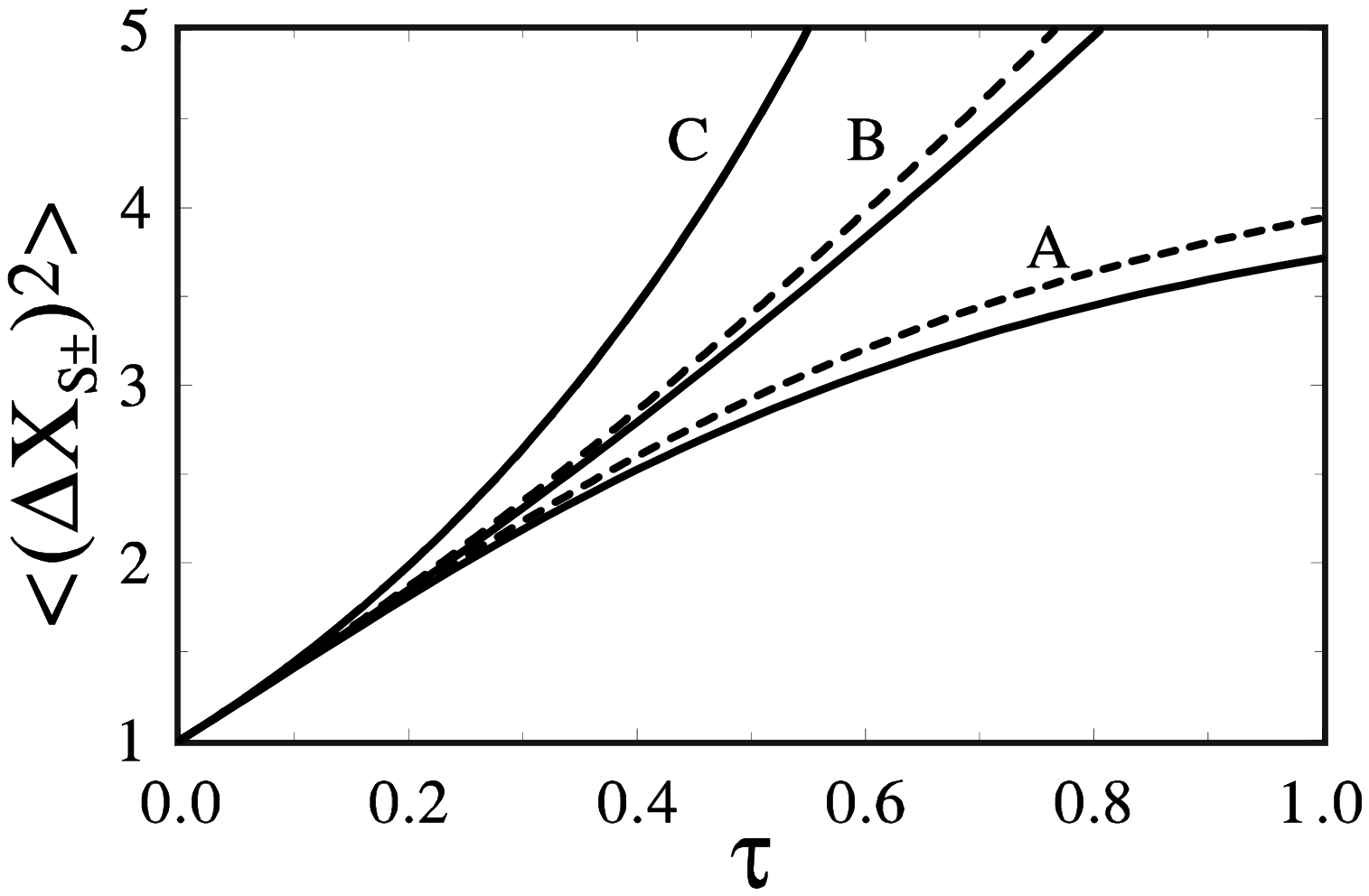,width=9cm}} \vspace*{-5mm}
\caption{Time dependence of the extremal variances
$\langle(\Delta\hat{X}_{S-})^2\rangle$ (solid lines) and
$\langle(\Delta\hat{X}_{S+})^2\rangle$ (dashed lines) for the same
cases as in figure 7. } \vspace*{-5mm} \end{figure}
\begin{figure} 
\vspace*{-10mm} \hspace*{0mm}
\centerline{\psfig{figure=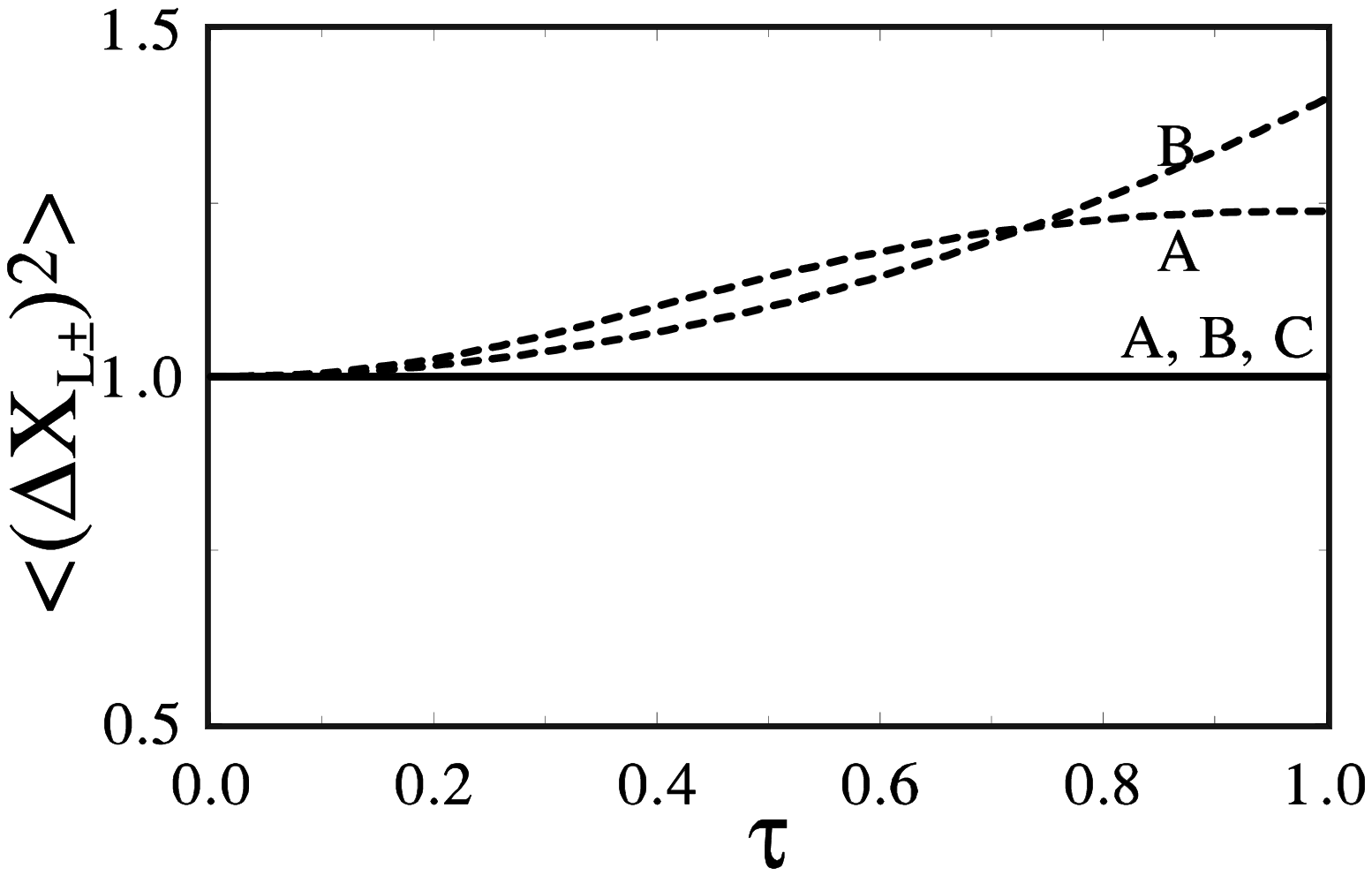,width=9cm}} \vspace*{-5mm}
\caption{Time dependence of the extremal variances
$\langle(\Delta\hat{X}_{L-})^2\rangle$  (solid lines) and
$\langle(\Delta\hat{X}_{L+})^2\rangle$  (dashed lines) for the
same cases as in figure 7. } \vspace*{-5mm} \end{figure}

The two-mode variances~(\ref{055}) of $\hat{X}_{LS1}$ and
$\hat{X}_{LS2}$ are
\begin{eqnarray}
\langle (\Delta\hat{X}_{LS\,1,2})^2\rangle &=& 2+2\{
|\alpha_L|^2-|\alpha_L||\alpha_S| [ \cos(\phi_L-\phi_S) \pm
\cos(\phi_L+\phi_S) ] \} \Delta\tau \nonumber\\ &&+ \mbox{\Large
\{} 2 |\alpha_L|^2 (|\alpha_L|^2-|\alpha_S|^2-1) \pm |\alpha_L|^2
|\alpha_S|^2 (\cos 2\phi_L-\cos 2\phi_S) \nonumber\\ &&-
|\alpha_S||\alpha_L| [\cos(\phi_L-\phi_S) (4
|\alpha_L|^2-6|\alpha_S|^2-11) \nonumber\\ &\pm&
\cos(\phi_L+\phi_S) (4|\alpha_L|^2-2|\alpha_S|^2-3) ] \mbox{\Large
\}} \frac{(\Delta\tau)^2}{2}, \label{163}
\end{eqnarray}
whereas the extremal variances are
\begin{eqnarray}
\langle (\Delta\hat{X}_{LS\,\pm})^2\rangle &=& 2 + 2\left[
|\alpha_L|^2-|\alpha_L||\alpha_S|\cos(\phi_L-\phi_S)
\right]\Delta\tau \nonumber\\ &&+ \Big[
|\alpha_L|^2(|\alpha_L|^2-|\alpha_S|^2-1) \nonumber\\ &&-
|\alpha_L||\alpha_S|\cos(\phi_L-\phi_S)
\left(2|\alpha_L|^2-3|\alpha_S|^2-\frac{11}{2}\right) \Big]
(\Delta\tau)^2 \nonumber\\ &\pm& \left| 2 \alpha_L
\alpha_S\Delta\tau + \Big[ |\alpha_L|^2
\alpha_S^2-|\alpha_S|^2\alpha_L^2 \right. \nonumber\\ &&+ \left.
\alpha_L\alpha_S (4|\alpha_L|^2-2 |\alpha_S|^2-3) \Big]
\frac{(\Delta\tau)^2}{2} \right|, \label{164}
\end{eqnarray}
according to the general expression~(\ref{058}).

Equations~(\ref{158})--(\ref{164}) can be readily generalized
to any initial distribution of the radiation fields.

\subsubsection{Exact solutions}

Let us now proceed to the exact solution of the master
equation~(\ref{125}). We apply the Laplace transform method.
The method is readily applicable to nonlinear master
equations for a variety of nonlinear optical phenomena
(Refs.~\cite{r177,r66,r57} and references therein); in
particular, it has been applied successfully to different
multiphoton Raman processes in
Refs.~\cite{r66,r100,r101,r318,r320,r144,r161,r162}.  The
solution of~(\ref{125}) for diagonal terms of the density
matrix $\rho_{nm}(00\tau)$ (i.e., for $\nu=\mu=0$) was
derived by McNeil and Walls~\cite{r66} and then, in a more
general form, by Simaan~\cite{r100}. As the chief result of
the present work we derive the time-dependence  of the
complete density matrix $\rho_{nm}(\nu\mu\tau)$, where the
degrees of off-diagonality $\nu$, $\mu$ are arbitrary. To the
best of our knowledge, ours is the first derivation of a
complete analytical solution to the Raman scattering model
including depletion of the pump field.

As usual, we assume that the Stokes and laser beams are mutually
independent at the initial time $\tau=\tau_0$. Thus, the
initial joint distribution $\rho_{nm}(\nu\mu\tau_0)$ is a product of
the distributions for the separate beams,
\begin{eqnarray}
\rho_{nm}(\nu\mu\tau_0) = \rho^L_{n}(\nu\tau_0) \:
\rho^S_{m}(\mu\tau_0). \label{165}
\end{eqnarray}
Let us define the coefficient $\lambda$ in terms of the
integer-value function $[[x]]$ (the maximum integer $\leq
x$):
\begin{eqnarray}
\lambda = \left[\left[ \frac{m-n+1}{2}+\frac{\mu-\nu}{4}
\right]\right]. \label{166}
\end{eqnarray}
The exact solution of~(\ref{125}) under the condition
(\ref{165}), derived in Appendix B, reads as follows for
$\lambda<0$:
\begin{eqnarray}
\rho_{nm}(\nu\mu\tau) &=& \left[
\frac{m!(m+\mu)!}{n!(n+\nu)!}\right]^{1/2} \sum_{l=0}^{m}
\rho_{n+l}^{L}(\nu\tau_0) \rho_{m-l}^{S}(\mu\tau_0) \nonumber\\
&&\times \left[
\frac{(n+l)!(n+l+\nu)!}{(m-l)!(m-l+\mu)!}\right]^{1/2} \nonumber\\
&&\times \sum_{q=0}^{l} \exp [-f(q)\Delta\tau] \prod_{p=0\atop
p\neq q}^{l} [f(p)-f(q)]^{-1}, \label{167}
\end{eqnarray}
whereas for $\lambda\geq 0$ it is
\begin{eqnarray}
\rho_{nm}(\nu\mu\tau) &=& \left[
\frac{m!(m+\mu)!}{n!(n+\nu)!}\right]^{1/2} \left\{
\sum_{l=0}^{\lambda} \rho_{n+l}^{L}(\nu\tau_0)
\rho_{m-l}^{S}(\mu\tau_0) \right. \nonumber\\ &&\times \left[
\frac{(n+l)!(n+l+\nu)!}{(m-l)!(m-l+\mu)!}\right]^{1/2} \nonumber\\
&&\times \sum_{q=0}^{l} \exp [-f(q)\Delta\tau] \prod_{p=0\atop
p\neq q}^{l} [f(p)-f(q)]^{-1} \nonumber\\ &&+ (1-\delta_{m0})
\sum_{l=\lambda+1}^{m} \rho_{n+l}^{L}(\nu\tau_0)
\rho_{m-l}^{S}(\mu\tau_0) \nonumber\\ &&\times \left[
\frac{(n+l)!(n+l+\nu)!}{(m-l)!(m-l+\mu)!}\right]^{1/2} \nonumber\\
&&\times \sum_{q=0}^{\lambda}\sum_{q'=\lambda+1}^{l}
\prod_{p=0\atop p\neq q}^{\lambda} [f(p)-f(q)]^{-1}
\prod_{p'=\lambda+1\atop p'\neq q'}^{l} [f(p')-f(q')]^{-1}
\nonumber\\ &&\times \mbox{\huge (}
\delta_{f(q)f(q')}\Delta\tau\exp[-f(q)\Delta\tau] \nonumber\\ &&+
\left. (\delta_{f(q)f(q')}-1)
\frac{\exp[-f(q)\Delta\tau]-\exp[-f(q')\Delta\tau]}{f(q)-f(q')}
\mbox{\huge )} \right\},\nonumber\\ \label{168}
\end{eqnarray}
where the function $f(x)$ is given by
\begin{eqnarray}
f(x) = \frac{1}{2} [(n+x)(m-x+1)+(n+x+\nu)(m-x+\mu+1]. \label{169}
\end{eqnarray}
The coefficient $\lambda$ can be alternatively defined as
\begin{eqnarray}
\lambda = \left[\left[ \frac{m-n}{2}+\frac{\mu-\nu}{4}
\right]\right]. \label{170}
\end{eqnarray}
The solution~(\ref{167}) and~(\ref{168}) with consequent
application of the coefficient $\lambda$ of Eq.~(\ref{170})
reduces, as it should, to the Simaan solution (45) and (49)
of Ref.~\cite{r100} for the diagonal matrix elements
($\nu=\mu=0$); McNeil and Walls have also obtained a solution
of~(\ref{125}) for the diagonal matrix elements (Eqs. (6.5),
(6.6) and (4.14) in Ref.~\cite{r66}); however, their solution
is not in full agreement with Simaan's solution and is not a
special case of ours for reasons given by Simaan~\cite{r100}.

The solution~(\ref{167}) and~(\ref{168}) is very well adapted
to numerical analysis; nonetheless, it is of a rather
complicated form. Our solution of~(\ref{125}) can be
rewritten more compactly. Following the method of
Malakyan~\cite{r162}, we find (for details, see Appendix B)
\begin{eqnarray}
\rho_{nm}(\nu\mu\tau) &=& \left[
\frac{m!(m+\mu)!}{n!(n+\nu)!}\right]^{1/2} \sum_{l=0}^{m}
\rho_{n+l}^{L}(\nu\tau_0) \rho_{m-l}^{S}(\mu\tau_0) \nonumber\\
&&\times \left[
\frac{(n+l)!(n+l+\nu)!}{(m-l)!(m-l+\mu)!}\right]^{1/2} \\
&&\times \hat{\cal D} \sum_{q=0\atop q\neq q'_1,q'_2,...,q'_d}^{l}
\exp [-f(q)\Delta\tau] \prod_{p=0\atop p\neq
q,q'_1,q'_2,...,q'_d}^{l} [f(p)-f(q)]^{-1}.\nonumber \label{171}
\end{eqnarray}
The differential operator  of the $d$-th order, $\hat{\cal
D}$, is defined as follows:
\begin{eqnarray}
\hat{\cal D} = (-1)^d \prod_{r=1}^{d} \frac{\partial}{\partial
f(q_r)} \label{172}.
\end{eqnarray}
The order $d$ of the differential operator~(\ref{172}) is
equal to the number of pairs of mutually equal factors
occurring in the product of Eq.~(\ref{171}),
$f(q_1)=f(q_1')$, $f(q_2)=f(q_2')$, ..., $f(q_d)=f(q_d')$. If
there are no pairs of equal factors then the operator
$\hat{\cal D}$ is defined to be unity (see Appendix~B). The
solutions~(\ref{167}),~(\ref{168}), and~(\ref{171}) represent
the chief result of our paper. In Sect. 5.2, in the Raman
effect model under the parametric approximation we have
analyzed, in particular, the single-mode solutions for either
the Stokes mode or for the anti-Stokes mode. For
completeness, we give in Appendix~C the solution for
anti-Stokes scattering without the parametric approximation.
The degree of off-diagonality $\mu$ is assumed to be
nonnegative (contrary to $\nu$); nonetheless, the time
dependence of the complete density matrix
$\rho_{nm}(\nu\mu\tau)$ is determined by the simple relation
for the inverse matrix elements:
\begin{eqnarray}
\rho_{nm}^*(\nu\mu\tau) = \rho_{n+\nu,m+\mu}(-\nu,-\mu,\tau).
\label{173}
\end{eqnarray}
Thus solutions~(\ref{167}),~(\ref{168}), and/or~(\ref{171})
provide an entire specification for all measurable properties
of the light field under consideration.

The two-mode (joint) density matrix with elements
$\rho_{nm}(\nu\mu\tau)$~(\ref{167}) and~(\ref{168}), enables
the calculation of the single-mode (separate) density matrix
with elements $\rho_m^{S}(\mu\tau)$ and
$\rho_m^{L}(\mu\tau)$. The Stokes mode matrix elements
$\rho_m^{S}(\mu\tau)$ can be calculated from
\begin{eqnarray}
\rho_{m}^{S}(\mu\tau)  = \sum_{n=0}^{\infty}\sum_{\nu=-n}^{\infty}
\rho_{nm}(\nu\mu\tau) \label{174}
\end{eqnarray}
and the laser mode matrix elements $\rho_m^{L}(\mu\tau)$ can
be found analogously with the exception that for terms
$\rho_{nm}(\nu\mu\tau)$ with $\mu<0$ the property~(\ref{173})
must be used. The already mentioned solution of McNeil and
Walls~\cite{r66} corresponds to the separate diagonal density
matrix~(\ref{174}).

There is yet another manner of expressing the two-mode
solutions of the master equation~(\ref{125}),
$\rho_{nm}(00\tau)$, for any initial distributions, via the
density matrix elements for the initial number states in the
Stokes and laser fields:
\begin{eqnarray}
\rho_{nm}(00\tau) = \sum_{n_0=0}^{\infty} \sum_{m_0=0}^{\infty}
\rho_{nm}^{(n_0,m_0)}(\tau) \rho_{n_0}^{L}(0\tau_0)
\rho_{m_0}^{S}(0\tau_0), \label{175}
\end{eqnarray}
where $\rho^{(n_0,m_0)}_{nm}(00\tau)$ is the solution
(\ref{167}) and~(\ref{168}) for $\rho_{nm}(00\tau)$, under
the initial conditions that the laser field is in the number
state $|n_0\rangle$ and the Stokes mode is in the number
state $|m_0\rangle$. The weighting functions in~(\ref{175})
are arbitrary initial distributions of the laser,
$\rho^L_n(0\tau_0)$, and the Stokes field,
$\rho^S_n(0\tau_0)$.  Here, for brevity, we restrict our
considerations to diagonal terms (with $\nu=\mu=0$).
Otherwise, instead of $\rho^{(n_0,m_0)}_{nm}(00\tau)$ we
would have to use the solution
$\rho^{(n_0,m_0,\nu_0,\mu_0)}_{nm}(\nu\mu\tau)$ and perform
two extra summations in Eq.~(\ref{175}) over $\nu$, $\mu$.
McNeil and Walls~\cite{r66} have presented their solution of
Eq.~(\ref{125}) in this manner. Analogously, we can express
the single-mode distributions $\rho^L_n(0\tau)$
($\rho^S_m(0\tau)$) for arbitrary initial states using the
solutions $\rho^{(n_0)}_n(\tau)$ ($\rho^{(m_0)}_m(\tau)$) for
the initial photon-number states $|n_0\rangle$
($|m_0\rangle$). For instance, for the Stokes mode solution
we apply the formula
\begin{eqnarray}
\rho_{n}^L(0\tau) = \sum_{n_0=0}^{\infty} \rho_{n}^{(n_0)}(\tau)
\rho_{n_0}^{L}(0\tau_0). \label{176}
\end{eqnarray}
We shall make use of this procedure for the diagonal
approximate solutions(\ref{194}).

Having the solutions~(\ref{167}),~(\ref{168}), or~(\ref{171})
available we can, at least numerically, analyze, e.g., the
single- and two-mode photocount statistics and quadrature
squeezing. The expectation values in the relations describing
squeezing and photocount statistics (see Section 4) are
readily expressed in terms of the density matrix elements
$\rho_{nm}(\nu\mu\tau)$ by way of
\begin{eqnarray}
\langle \hat{n}^k(t)\rangle \:=\:
\sum_{n,m=0}^{\infty}n^k \rho_{n,m}(0,0,t),
\label{177}
\end{eqnarray}
\begin{eqnarray}
\langle \hat{m}^k(t)\rangle \:=\:
\sum_{n,m}m^k \rho_{n,m}(0,0,t),
\label{178}
\end{eqnarray}
\begin{eqnarray}
\langle \hat{a}_L^{+k}(t)\rangle \:=\:
\sum_{n,m}
\left[ \frac{(n+k)!}{n!}\right]^{1/2}
\rho_{n,m}(k,0,t),
\label{179}
\end{eqnarray}
\begin{eqnarray}
\langle \hat{a}_S^{+k}(t)\rangle \:=\:
\sum_{n,m}
\left[ \frac{(m+k)!}{m!}\right]^{1/2}
\rho_{n,m}(0,k,t),
\label{180}
\end{eqnarray}
\begin{eqnarray}
\langle \hat{a}_L^+(t)\hat{a}^+_S(t)\rangle \:=\:
\sum_{n,m}
[(n+1)(m+1)]^{1/2}
\rho_{n,m}(1,1,t),
\label{181e}
\end{eqnarray}
\begin{eqnarray}
\langle \hat{a}_L^+(t)\hat{a}_S(t)\rangle \:=\:
\sum_{n,m}
[(n+1)m]^{1/2}
\rho_{n,m}(1,-1,t).
\label{182}
\end{eqnarray}

In Section 3 we defined the $s$-parametrized quasiprobability
distribution ${\cal W}^{(s)}(\{\alpha_k\})$ and the
$s$-parametrized characteristic function ${\cal
C}^{(s)}(\{\beta_k\})$ and adduced relations between them for
any parameter $s$.  Here, we deal with matrix elements in
Fock basis of the density operator, $\rho_{nm}(\nu\mu\tau)$.
To achieve consistency between our analysis of the Raman
effect presented in this Section with the analysis of Section
5, we shall present some relations between the functions
${\cal W}^{(s)}(\{\alpha_k\})$ or ${\cal
C}^{(s)}(\{\beta_k\})$ and the density operator
$\hat{\rho}(\{\hat{a}_k\})$. We restrict the general formulas
for the $M$-mode fields to our two-mode situations, so that
$(\{\alpha_k \})=(\alpha_L,\alpha_{S,A})$. These formulas are
in complete analogy with the results of Cahill and
Glauber~\cite{r278} for the single-mode case. By virtue of
the operator $\hat{T}(\alpha_L,\alpha_{S,A})$, which is the
Fourier transform of the $s$-parametrized displacement
operator $\hat{D}^{(s)}(\beta_L,\beta_{S,A})$ (see
Eq.~(\ref{019})),
\begin{eqnarray}
\hat{T}^{(s)}(\alpha_L,\alpha_{S,A}) = \frac{1}{\pi^2}\int
\hat{D}^{(s)}(\beta_L,\beta_{S,A})\hspace{4cm}
\nonumber \\
\times\exp\left(
\alpha_L\beta_L^*+\alpha_{S,A}\beta_{S,A}^*-c.c.\right) {\rm d}^2
\beta_L {\rm d}^2 \beta_{S,A}, \label{183}
\end{eqnarray}
the density matrix $\hat{\rho}(\hat{a}_L,\hat{a}_{S,A})$ can
be obtained from the $s$-parametrized quasidistribution
${\cal W}^{(s)}(\alpha_L,\alpha_{S,A})$,~(\ref{021}),
\begin{eqnarray}
\hat{\rho}\left( \hat{a}_L, \hat{a}_{S,A}\right) =
\frac{1}{\pi^2}\int {\cal W}^{(s)}(\alpha_L,\alpha_{S,A})
\hspace{2cm}
\nonumber \\
\times\; \hat{T}^{(-s)}(\alpha_L,\alpha_{S,A}) {\rm d}^2\alpha_L
{\rm d}^2\alpha_{S,A}. \label{184}
\end{eqnarray}
The inverse relation,
\begin{eqnarray}
{\cal W}^{(s)}(\alpha_L,\alpha_{S,A}) = {\rm Tr}\left[
\hat{\rho}\left( \hat{a}_L, \hat{a}_{S,A}\right)
\hat{T}^{(s)}(\alpha_L,\alpha_{S,A}) \right], \label{185}
\end{eqnarray}
resembles expression~(\ref{020}) for the characteristic
function ${\cal C}^{(s)}(\beta_L,\beta_{S,A})$, which is the
average value of the displacement operator
$\hat{D}^{(s)}(\alpha_L,\alpha_{S,A})$. We are interested in
the relations for the matrix elements of
$\hat{\rho}(\hat{a}_L,\hat{a}_{S,A})$. They immediately
follow from~(\ref{184}) and~(\ref{185}):
\begin{eqnarray}
\rho_{n,m}(\nu,\mu)= \frac{1}{\pi^2}\int {\cal
W}^{(s)}(\alpha_L,\alpha_{S,A}) \hspace{4cm}
\nonumber\\
\times\; \langle n,m | \hat{T}^{(s)}(\alpha_L,\alpha_{S,A})
|n+\nu,m+\mu\rangle {\rm d}^2\alpha_L {\rm d}^2\alpha_{S,A},
\label{186}
\end{eqnarray}
\begin{eqnarray}
{\cal W}^{(s)}(\alpha_L,\alpha_{S,A}) =
\sum_{n=0}^{\infty}\sum_{m=0}^{\infty}
\sum_{\nu=-n}^{\infty}\sum_{\mu=-m}^{\infty} \rho^*_{n,m}(\nu,\mu)
\hspace{2cm}
\nonumber\\
\times\; \langle n,m| \hat{T}^{(s)}(\alpha_L,\alpha_{S,A})
|n+\nu,m+\mu \rangle, \label{187}
\end{eqnarray}
The Fock matrix elements for the two-mode field,
\begin{eqnarray}
\langle n,m | \hat{T}^{(s)}(\alpha_L,\alpha_{S,A}) |n+\nu,m+\mu
\rangle = \langle n| \hat{T}^{(s)}(\alpha_L) |n+\nu\rangle
\hspace{1cm}
\nonumber\\
\times \langle m| \hat{T}^{(s)}(\alpha_{S,A}) |m+\mu \rangle,
\label{188}
\end{eqnarray}
are simply products of the two single-mode Fock matrix
elements given by  Cahill and Glauber~\cite{r278}:
\begin{eqnarray}
\langle n| \hat{T}^{(s)}(\alpha_L) |n+\nu \rangle =
\sqrt{\frac{n!}{(n+\nu)!}}\left( \frac{2}{1-s}\right)^{\nu+1}
\left( \frac{s+1}{s-1}\right)^{n} \hspace{1cm}\nonumber\\ \times
\exp\left(-\frac{2}{1-s}|\alpha_L|^2 \right) L_{n}^{(\nu)}
\left(\frac{4|\alpha_L|^2}{1-s^2} \right) \left(
\alpha_{L}^{*}\right)^{\nu}, \label{189}
\end{eqnarray}
where $L_n^{(\nu)}(x)$ is the generalized Laguerre polynomial. The
above equations show equivalency of the two apparently different
formalisms we have been dealing with: on the one side, the
$s$-parametrized quasiprobability distribution functions ${\cal
W}^{(s)}$ obtained within the Fokker-Planck equation formalism
presented in Section 5, and, on the other side, the density matrix
operator formalism under discussion in this Section.

Solutions~(\ref{167}),~(\ref{168}), or~(\ref{171}) reduce to
rather simple expressions in special cases, for instance, on
the one side, for long periods of time when the laser beam is
totally depleted, and on the other, for an intense laser beam
almost unaffected (undepleted) during the process of
scattering. We now discuss these two cases.

\subsubsection{Long-time solutions}

After a sufficiently long time, the system settles down to a
steady state as a result of the total depletion of the laser
pump. The steady-state solutions can be readily deduced from
(\ref{167}) and~(\ref{168}). Indeed, in the time limit
($\tau\rightarrow\infty$), the nonzero matrix elements
$\rho_{nm}(\nu\mu\infty)$ must satisfy the condition for the
function $f(x)$~(\ref{169}) that $f(q)=0$, which implies that
$q=0$. Hence, we have
\begin{eqnarray}
\rho_{nm}(\nu\mu,\tau=\infty)=0 \hspace{2cm}\mbox{for}\hspace{5mm}
n,\nu\neq 0 \label{190}
\end{eqnarray}
for arbitrary $m$, $m+\mu$ ranging from zero to infinity.
All photon-number and annihilation operator moments for the laser
beam vanish in the time limit
\begin{eqnarray}
\langle\hat{n}^p(\infty)\rangle &=& 0,
\nonumber\\
\langle\hat{a}_L^p(\infty)\rangle &=& 0
\hspace{2cm}\mbox{for}\hspace{5mm} p>0, \label{191}
\end{eqnarray}
which reflects the fact that the laser beam is totally
depleted. The normalization condition takes the form
\begin{eqnarray}
\sum_{m=0}^{\infty} \rho_{0m}(00\infty)=1, \label{192}
\end{eqnarray}
In the model of hyper-Raman scattering, as was shown by
Malakyan~\cite{r162}, there intervene in the limit
$\tau\rightarrow\infty$ the nonzero density matrix elements
$\rho_{0m}(0\mu\infty)$, $\rho_{1m}(0\mu\infty)$,
$\rho_{0m}(1\mu\infty)$, and $\rho_{1m}(-1,\mu\infty)$.
Hence, the photon-number moments and annihilation operator
moments of the laser mode do not vanish, contrary to the
model of Raman scattering under consideration in view of
(\ref{191}).  If we assume that initially there are no
photons in the Stokes mode, then $\rho_{n}^L(0\tau_0) \:=\:
\rho_{n}^S(0\infty)$ (Ref.~\cite{r100}), which implies that
an arbitrary photon-number moment
$\langle\hat{m}^{p}(\infty)\rangle$ (with any $p$) for the
Stokes mode in the time limit is identical with the
corresponding moment for the laser mode,
$\langle\hat{n}^{p}(\tau_0)\rangle$, at the time
$\tau=\tau_0$.

\subsection{Raman scattering without pump depletion}

Compact approximate solutions can be obtained from Eqs.
(\ref{167}) under the condition that the initial laser beam
is much more intense than the Stokes beam, i.e.,
$\langle\hat{n}\rangle>>\langle\hat{m}\rangle $. The
depletion of the laser beam and amplification of the Stokes
beam restrict the validity of this approximation to short
evolution times $\tau$ ($\tau<<1$). This approximation
implies that the density matrix elements
$\rho_{nm}(\nu\mu\tau)$ for $\lambda\ge 0$ given by Eq.
(\ref{168}) are negligible. Moreover, we can simplify the
remaining solution~(\ref{167}) by setting $n\approx n\pm m$.
In the analysis of the phenomena described by the density
$\rho_{nm}(\nu\mu\tau)$ with small degree of off-diagonality
$\nu$ (such as quadrature squeezing), we can set $n\approx
n+\nu$. Alternatively, in order to, for instance, investigate
phase properties~\cite{r311,r304,r305} (which require
summation over $\nu$ ranging from zero to infinity) one might
assume that the fluctuations in the laser beam are small in
comparison to their mean value, i.e.,
$\langle\hat{n}\rangle>>\sqrt{\langle(\Delta\hat{n})^2\rangle}$.
Under these approximations the solution of~(\ref{125}) takes
the form
\begin{eqnarray}
\rho_{nm}(\nu\mu\tau)&\approx& [m!(m+\mu)!]^{1/2} \nonumber\\
&\times& \sum_{l=0}^{m}
\rho_{n}^L(\nu\tau_0)\rho_{m-l}^S(\mu\tau_0)
[(m-l)!(m-l+\mu)!]^{-1/2} \nonumber\\ &\times&
\sum_{q=0}^{l}\exp[-n (m-q+1+\mu/2) \Delta\tau] \prod_{p=0 \atop
p\neq q}^{l} (q-p)^{-1}. \label{193}
\end{eqnarray}
Applying the binomial theorem we rewrite~(\ref{193}) as
\begin{eqnarray}
\rho_{nm}(\nu\mu\tau)&\approx& \sum_{l=0}^{m} \left[ {m\choose l}
{m+\mu\choose l} \right]^{1/2} \left( e^{n\Delta\tau}-1\right)^l
\nonumber\\ &&\times \exp [-n(m+1+\mu/2)\Delta\tau] \;
\rho_{n}^L(\nu\tau_0)\rho_{m-l}^S(\mu\tau_0), \label{194}
\end{eqnarray}
which, for $\mu=0$ and $\nu=0$, goes over into Simaan's
equation of Ref.~\cite{r100}.  The density matrix
(\ref{193}), applied to relations~(\ref{177})--(\ref{182}),
enables the calculation of the expectation values and
variances for the Stokes mode and the laser mode; however, in
the latter case, as a result of the approximations assumed,
we find no time dependence of the laser field photon-number
moments for a Stokes beam initially in a number state
containing
\begin{eqnarray}
\left< f[\hat{n}(\tau)] \right> &=& \sum_{n=0}^{\infty} f(n)
\rho_n^L(0\tau_0) \exp[-n (m_0+1)] \sum_{m=0}^{\infty} \left(
m+m_0 \atop m_0 \right) (1-e^{-n \tau})^m \nonumber \\ &=&
\sum_{n=0}^{\infty} f(n) \rho_n^L(0\tau_0) \:=\: \left< f(\hat{n})
\right>. \label{195}
\end{eqnarray}
The result~(\ref{195}a) is valid for any initial number-state
Stokes beam, so we conclude that the pump beam is
time-independent for arbitrary initial Stokes beam. The
photon-number moments for the Stokes mode calculated from
(\ref{194}) are of particularly simple form. For instance, we
have
\begin{eqnarray}
\langle\hat{m}(\tau)\rangle &=& \langle\hat{m}\rangle
\sum_{n=0}^{\infty} \exp(n\Delta\tau) \rho^L_n(0\tau_0)
\nonumber\\  &&+ \mbox{\huge (} \sum_{n=0}^{\infty}
\exp(n\Delta\tau) \rho^L_n(0\tau_0)-1 \mbox{\huge )}, \label{196}
\end{eqnarray}
\begin{eqnarray}
\langle\hat{m}^2(\tau)\rangle &=&
(\langle\hat{m}^2\rangle+3\langle\hat{m}\rangle+2)
\sum_{n=0}^{\infty} \exp(2n\Delta\tau) \rho^L_n(0\tau_0)
\nonumber\\ &&- 3(\langle\hat{m}\rangle+1) \sum_{n=0}^{\infty}
\exp(n\Delta\tau) \rho^L_n(0\tau_0)+1, \label{197}
\end{eqnarray}
\begin{eqnarray}
\langle \hat{m}(\tau) \hat{n}(\tau)\rangle = (\langle \hat{m}
\rangle +1) \sum_{n=0}^{\infty} n \exp(n\Delta\tau)
\rho_n^L(0\tau_0) -\langle \hat{n} \rangle. \label{198}
\end{eqnarray}
Equations~(\ref{196}) and~(\ref{197}) were obtained by
Simaan~\cite{r100}. Equation~(\ref{196}) is in agreement with
the Shen relation in Ref.~\cite{r97}. Equations
(\ref{196})--(\ref{198}) reduce to Loudon's results of
Ref.~\cite{r57} for the simpler special case in which no
scattered photons are excited initially. The sum of the mean
photon numbers for the laser and Stokes mode~(\ref{196}) is
not a constant of motion, contrary to our former
considerations~(\ref{132}). Nonetheless, in view of the
intense laser field approximation, the conservation of the
total number of photons is at least approximately fulfilled.
It is easy to find a physical interpretation of Eq.
(\ref{196}).  The first term of~(\ref{196}) describes the
amplification of the initial Stokes beam with
$\langle\hat{m}\rangle$ photons at the time $\tau_0$ and can
be identified as $sensu$ $stricto$ stimulated Raman
scattering. The second term of~(\ref{196}) corresponds to an
amplification of the vacuum fluctuations and can be
interpreted as spontaneous Raman scattering, which occurs
even in the case when the Stokes field contains initially no
photons ($\langle\hat{m}\rangle=0$). Note that even in the
model of scattering from phonons at zero temperature
(``quiet'' reservoir), spontaneous scattering does take
place. The coefficients $\gamma_S^{(2)}(\tau)$ and
$g_{LS}^{(2)}(\tau)$, readily obtained from~(\ref{035}) and
(\ref{038}) by insertion of Eqs.~(\ref{196})--(\ref{198}),
can be explicitly compared to the coefficients calculated
from other, corresponding relations.  Assuming that initially
there are no photons in the Stokes beam,
$\langle\hat{m}\rangle=\langle\hat{m}^2\rangle=0$, we obtain
from small-time expansions of the exponential functions in
Eqs.(\ref{196}) and~(\ref{197}) the following simple
expressions for the normalized factorial moment
$\gamma_S^{(2)}(\tau)$:
\begin{eqnarray}
\gamma_S^{(2)}(\tau) = 2\gamma_L^{(2)}+1 +2(\langle
\hat{n}^3\rangle-\langle \hat{n}^2 \rangle^2/ \langle
\hat{n}\rangle) \langle \hat{n}\rangle^{-2} \Delta\tau,
\label{199}
\end{eqnarray}
as well as the normalized cross-correlation function
$g_{LS}^{(2)}(\tau)$:
\begin{eqnarray}
{g}_{LS}^{(2)}(\tau) = \gamma_L^{(2)} +(\langle \hat{n}^3\rangle
-\langle \hat{n}^2 \rangle^2/ \langle \hat{n}\rangle) \langle
\hat{n}\rangle^{-2} \Delta\tau/2. \label{200}
\end{eqnarray}
Equations~(\ref{199}) and~(\ref{200}) can be equivalently
obtained from the short-time expansions~(\ref{134}) and
(\ref{139}), respectively, on omitting  the expressions
$1/\langle\hat{n}\rangle$ and
$\langle\hat{n}^2\rangle/\langle\hat{n}\rangle^2$ in the
terms proportional to $\Delta\tau$, which are negligible in
comparison with the terms
$\langle\hat{n}^3\rangle/\langle\hat{n}\rangle^2$ and
$\langle\hat{n}^2\rangle^2/\langle\hat{n}\rangle^3$. The
Simaan approximate relations for $g_{LS}^{(2)}(\tau)$ (32)
and $\gamma_{S}^{(2)}(\tau)$ (33) in Ref.~\cite{r100},
rewritten in our notation (with extra $-1$ in view of
(\ref{038}) and~(\ref{035})), do not reduce exactly to our
Eqs.~(\ref{199}) and~(\ref{200}), respectively.

By substituting Eq.~(\ref{194}) with $\nu=\mu=0$ into
(\ref{174}) one can obtain solution~(\ref{176}), for any
initial distribution of the laser mode, with the following
distribution $\rho_m^{(n_0)}(\tau)$:
\begin{eqnarray}
\rho_{m}^{(n_0)}(\tau)  &=& \exp[-(m+1)n_0 \Delta\tau] \nonumber\\
&&\times \sum_{l=0}^{m}\left( m \atop l\right) \left( e^{n_0
\Delta\tau}-1\right)^{l} \rho_{m-l}^{S}(0\tau_0), \label{201}
\end{eqnarray}
calculated for the laser field initially in a number state
containing $n_0$ photons. In this case the mean
($\langle\hat{m}(\tau)\rangle$) and mean-square number of
Stokes photons ($\langle\hat{m}^2(\tau)\rangle$),
\begin{eqnarray}
\langle\hat{m}(\tau)\rangle = (\langle\hat{m}\rangle
+1)\exp(n_0\Delta\tau)-1, \label{202}
\end{eqnarray}
\begin{eqnarray}
\langle\hat{m}^2(\tau)\rangle &=&
(\langle\hat{m}^2\rangle+3\langle\hat{m}\rangle +2 ) \exp(2
n_0\Delta\tau) \nonumber\\  &&- 3(\langle\hat{m}\rangle
+1)\exp(n_0\Delta\tau)+1 \label{203}
\end{eqnarray}
can be immediately obtained either from~(\ref{196}) and
(\ref{197}) or from~(\ref{201}). Assuming that the Stokes
beam is initially in a coherent state $|\alpha\rangle$, we
can perform summation in~(\ref{201})  which leads to

\begin{eqnarray}
\rho_{m}^{(n_0)}(\tau)  &=& \exp[-|\alpha|^2-n_0\Delta\tau]
\left(1 - e^{-n_0\Delta\tau}\right)^m \nonumber\\ &&\times
_1F_1\left[-m;1;\,
-|\alpha|^2\left(e^{n_0\Delta\tau}-1\right)^{-1} \right],
\label{204}
\end{eqnarray}
where $_1F_1$ is a confluent hypergeometric function. The
density matrix elements $\rho_m^{(n_0)}(\tau_0)$~(\ref{204})
describe a superposition of coherent and chaotic
fields~\cite{r79,r116}. This will be more transparent if we
rewrite Eq.~(\ref{204}) in terms of the average number of
Stokes photons in the chaotic part,
\begin{eqnarray}
\langle\hat{m}_{ch}(\tau)\rangle = \exp(n_0\Delta\tau)-1,
\label{205}
\end{eqnarray}
and the mean number of photons in the coherent part alone,
\begin{eqnarray}
\langle\hat{m}_c(\tau)\rangle = |\alpha|^2  \exp(n_0\Delta\tau).
\label{206}
\end{eqnarray}
Then, one obtains, using the Laguerre polynomial $L_m$, the
standard form of the
distribution~(\ref{204})~\cite{r100,r313,r71,r79}:
\begin{eqnarray}
\rho_{m}^{(n_0)}(\tau) &=&
\frac{\langle\hat{m}_{ch}(\tau)\rangle^m}
{(1+\langle\hat{m}_{ch}(\tau)\rangle)^{1+m}}
\exp\left(-\frac{\langle\hat{m}_c(\tau)\rangle}
{1+\langle\hat{m}_{ch}(\tau)\rangle}\right) \nonumber\\  &&\times
L_m\left(- \frac{\langle\hat{m}_c(\tau)\rangle}
{\langle\hat{m}_{ch}(\tau)\rangle(1+\langle\hat{m}_{ch}(\tau)\rangle)}
\right). \label{207}
\end{eqnarray}

Similarly, by expressing the relation~(\ref{202}) in terms of
the mean values~(\ref{205}) and~(\ref{206}) it is seen that
\begin{eqnarray}
\langle\hat{m}(\tau)\rangle & =&
\langle\hat{m}_{c}(\tau)\rangle
+\langle\hat{m}_{ch}(\tau)\rangle.
\label{204e1}
\end{eqnarray}
The general moment of the p-th order
$\langle\hat{m}^p\rangle$ can be found  by repeated use of
the recursion relation~\cite{r313}:
\begin{eqnarray}
\langle m^{r+1}(\tau)\rangle = \langle m_{ch}(\tau)\rangle(\langle
m_{ch}(\tau)\rangle+1) \frac{\partial \langle m^r(\tau)\rangle}
{\partial \langle m_{ch}(\tau)\rangle}\hspace{2.5cm} \nonumber \\
+ \langle m_{c}(\tau)\rangle(2\langle m_{ch}(\tau)\rangle+1)
\frac{\partial \langle m^r(\tau)\rangle} {\partial \langle
m_{c}(\tau)\rangle} +\langle m^{r}(\tau)\rangle\langle
m(\tau)\rangle \label{204e2}
\end{eqnarray}
with the help of~(\ref{204e1}) or using the following
explicit expression~\cite{r71,r79}:
\begin{eqnarray}
\langle m^r(\tau)\rangle = r! \langle m^r_{ch}(\tau)\rangle\,
L_r\left(-\frac{\langle m_{c}(\tau)\rangle} {\langle
m_{ch}(\tau)\rangle}\right). \label{210}
\end{eqnarray}
The factorial moments of p-th  order can readily be
calculated from~(\ref{204e2}) or~(\ref{210}). In particular,
the second-order factorial moment reads as follows:
\begin{eqnarray}
\gamma_S^{(2)}(\tau) = 1-\left( \frac{\langle
m_c(\tau)\rangle}{\langle m(\tau)\rangle}\right)^2, \label{204e3}
\end{eqnarray}
which takes the minimal value, equal to zero, for the initial
time $\tau_0$, since only then $\langle\hat{m}_{ch}\rangle=0$.

For the Stokes beam initially in a vacuum state $|0\rangle$
the distributions~(\ref{204}) and~(\ref{207})  reduce to the
Bose-Einstein distribution
\begin{eqnarray}
\rho_{m}^{(n_0)}(\tau) = \frac{\langle\hat{m}_{ch}(\tau)\rangle^m}
{(1+\langle\hat{m}_{ch}(\tau)\rangle)^{1+m}}, \label{212}
\end{eqnarray}
describing a chaotic field (cf.~(\ref{014})). In this case,
in the absence of stimulated scattering
($\langle\hat{m}\rangle=0$), the chaotic field is generated
in spontaneous Raman scattering as an amplification of the
vacuum fluctuations.

To compare the results for the expectation values of the
Stokes mode obtained in Sect. 5.2 with the present results,
we assume that the laser and Stokes beams are initially in a
coherent state $|\alpha_L\rangle$  and $|\alpha_S\rangle$,
respectively. Performing summation in Eqs.
(\ref{196})--(\ref{198}) with the coherent weight function
$\rho_n^L(0\tau_0)$ one readily arrives at
\begin{eqnarray}
\langle\hat{m}(\tau)\rangle = (|\alpha_S|^2+1)
\exp[|\alpha_L|^2(e^{\Delta\tau}-1)]-1, \label{213}
\end{eqnarray}
\begin{eqnarray}
\langle\hat{m}^2(\tau)\rangle &=& (|\alpha_S|^4+4|\alpha_S|^2+2)
\exp[|\alpha_L|^2(e^{2\Delta\tau}-1)] \nonumber\\ &&-
3(|\alpha_S|^2+1) \exp[|\alpha_L|^2(e^{\Delta\tau}-1)]+1,
\label{214}
\end{eqnarray}
\begin{eqnarray}
\langle\hat{m}(\tau)\hat{n}(\tau)\rangle =
|\alpha_L|^2(|\alpha_S|^2+1)
\exp[|\alpha_L|^2(e^{\Delta\tau}-1)+\Delta\tau]-|\alpha_L|^2.
\label{215}
\end{eqnarray}
Within the Fokker-Planck equation approach under parametric
approximation (Sect. 5.2) we have obtained Eqs.~(\ref{111})
and~(\ref{114}), which can be rewritten, using the notation
of this Section, i.e. $\kappa_S t=|e_L|^2 \gamma_S
t=|\alpha_L|^2\tau$, and assuming that the mean number of
phonons is zero, in the following form:
\begin{eqnarray}
\langle\hat{m}(\tau)\rangle = (|\alpha_S|^2+1)
\exp(|\alpha_L|^2\Delta\tau)-1, \label{216}
\end{eqnarray}
\begin{eqnarray}
\langle\hat{m}^2(\tau)\rangle &=& (|\alpha_S|^4+4|\alpha_S|^2+2)
\exp(2|\alpha_L|^2\Delta\tau) \nonumber\\ &&- 3(|\alpha_S|^2+1)
\exp(|\alpha_L|^2\Delta\tau)+1. \label{217}
\end{eqnarray}
We also note that
\begin{eqnarray}
\langle\hat{m}(\tau)\hat{n}(\tau)\rangle =
\langle\hat{m}(\tau)\rangle\langle\hat{n}\rangle. \label{218}
\end{eqnarray}
For short times of evolution, $\Delta\tau<<1$, and intense
pump beams, $|\alpha_L|^2>>1$, Eqs.~(\ref{213}),~(\ref{214}),
and~(\ref{215}) go over into Eqs.~(\ref{216}),~(\ref{217}),
and~(\ref{218}), respectively. Indeed, the short-time
expansions of~(\ref{213}) and~(\ref{214}) are:
\begin{eqnarray}
\langle\hat{m}(\tau)\rangle &=&
|\alpha_S|^2+|\alpha_L|^2(1+|\alpha_S|^2)\Delta\tau \nonumber \\
&&+ |\alpha_L|^2(|\alpha_L|^2+1)(1+|\alpha_S|^2)
\frac{(\Delta\tau)^2}{2}, \label{219}
\end{eqnarray}
\begin{eqnarray}
\langle\hat{m}^2(\tau)\rangle &=& |\alpha_S|^2(1+|\alpha_S|^2)+
|\alpha_L|^2(1+5|\alpha_S|^2+2|\alpha_S|^4)\Delta\tau \nonumber \\
&&+ |\alpha_L|^2(|\alpha_L|^2+1)(5+13|\alpha_S|^2+4|\alpha_S|^4)
\frac{(\Delta\tau)^2}{2}, \label{220}
\end{eqnarray}
whereas Eqs.~(\ref{216}) and~(\ref{217}) obtained within the
formalism of Sect. 5.2 reduce to
\begin{eqnarray}
\langle\hat{m}(\tau)\rangle &=&
|\alpha_S|^2+|\alpha_L|^2(1+|\alpha_S|^2)\Delta\tau \nonumber \\
&&+ |\alpha_L|^4(1+|\alpha_S|^2) \frac{(\Delta\tau)^2}{2},
\label{221}
\end{eqnarray}
\begin{eqnarray}
\langle\hat{m}^2(\tau)\rangle &=& |\alpha_S|^2(1+|\alpha_S|^2)+
|\alpha_L|^2(1+5|\alpha_S|^2+2|\alpha_S|^4)\Delta\tau \nonumber \\
&&+ |\alpha_L|^4(5+13|\alpha_S|^2+4|\alpha_S|^4)
\frac{(\Delta\tau)^2}{2}, \label{222}
\end{eqnarray}
respectively. It is seen that for high intensity of the pump
field,~(\ref{219}) goes over into~(\ref{221}), and
(\ref{220}) into~(\ref{222}) by setting
$|\alpha_L|^2(|\alpha_L|^2+1)\approx |\alpha_L|^4$. Hence,
the factorial moment $\gamma_S^{(2)}(\tau)$,
\begin{eqnarray}
\gamma_S^{(2)}(\tau) &=& 2 |\alpha_L|^2 |\alpha_S|^{-2} \Delta\tau
- \Big[ |\alpha_L|^2 (3+|\alpha_S|^2)
\nonumber \\
&&- |\alpha_S|^4-5 |\alpha_S|^2-2 \Big]
|\alpha_L|^2|\alpha_S|^{-4} (\Delta\tau)^2, \label{223}
\end{eqnarray}
calculated from~(\ref{219}) and~(\ref{220}) in the case of
nonzero $\alpha_S$ and an intense pump beam, goes over into
\begin{eqnarray}
\gamma_S^{(2)}(\tau) = 2 |\alpha_L|^2|\alpha_S|^{-2} \Delta\tau
-(3+|\alpha_S|^2) |\alpha_L|^4|\alpha_S|^{-4} (\Delta\tau)^2
\label{224}
\end{eqnarray}
obtained from~(\ref{221}) and~(\ref{222}). If the initial
field contains no Stokes photons, we obtain the following
factorial moments $\gamma_S^{(2)}(\tau)$:
\begin{eqnarray}
\gamma_S^{(2)}(\tau) = 1+2 |\alpha_L|^{-2} +2\Delta\tau+ \left(
2+\frac{5}{6}|\alpha_L|^2\right) (\Delta\tau)^2, \label{225}
\end{eqnarray}
\begin{eqnarray}
\gamma_S^{(2)}(\tau) =1, \label{226}
\end{eqnarray}
within the formalisms of this Section and Sect. 5.2,
respectively. The differences between the factorial moments
$\gamma_S^{(2)}(\tau)$ are more pronounced in the case
$\alpha_S=0$ since  the expansion of
$\langle\hat{m}(\tau)\rangle$ and
$\langle\hat{m}^2(\tau)\rangle$ correct to the third order in
$\tau$ is required in the derivation of~(\ref{225}). The
interbeam degree of coherence $g_{LS}^{(2)}(\tau)$
(\ref{038}), as expected, vanishes for the model of Sect.
5.2. The short-time expansion of $g_{LS}^{(2)}(\tau)$
obtained from Eqs.~(\ref{213})--(\ref{215}), for
$\alpha_S\neq0$, is
\begin{eqnarray}
g_{LS}^{(2)}(\tau) &=& (1+|\alpha_S|^{-2})\Delta\tau \nonumber\\
&&- (1+|\alpha_S|^2) (2 |\alpha_L|^2-|\alpha_S|^2) |\alpha_S|^{-4}
\frac{(\Delta\tau)^2}{2}, \label{227}
\end{eqnarray}
otherwise, $\alpha_S=0$, we get
\begin{eqnarray}
g_{LS}^{(2)}(\tau) =
|\alpha_L|^{-2}+\frac{1}{2}\Delta\tau+\frac{1}{12}
(\langle\hat{n}\rangle+3)\frac{(\Delta\tau)^2}{2}. \label{228}
\end{eqnarray}
It is seen that the approaches of Sects. 5.2 and 6.2 give similar
predictions for the Stokes beam.

The evolution of the photon-number moments is demonstrated in
Figs. 2 and~3: $\langle\hat{m}(\tau)\rangle$ calculated with
Eq.~(\ref{221}) is depicted by solid line C or with
(\ref{219}) by solid line D in Fig. 2;
$\langle\hat{m}^2(\tau)\rangle$ obtained from Eq.~(\ref{222})
is given by solid line C  or from~(\ref{220}) by solid line D
in Fig. 3. No time dependence of
$\langle\hat{n}(\tau)\rangle$ and
$\langle\hat{n}^2(\tau)\rangle$ is observed for the results
of this Section and Sect. 5.2, i.e., we obtain straight lines
C and D in Figs. 2 and 3. Similar notation is used in Figs.
4-6 for the normalized factorial moments
$\gamma_S^{(2)}(\tau)$ (Fig. 4),  $\gamma_L^{(2)}(\tau)$
(Fig. 5), and the degree of interbeam coherence
$g_{LS}^{(2)}(\tau)$ (Fig. 6). We have chosen rather small
initial numbers of laser photons ($|\alpha_L|^2=2$) for
numerical reasons. In this case, the factorial moments,
calculated from~(\ref{223}),~(\ref{225}),~(\ref{227}), and
(\ref{228}), differ significantly from the exact numerical
results.  So we omit them (curves D) in Figs 4 and 6.

\section{Conclusions}

In this paper we have presented in a systematic way
quantum-statistical theory of the standing-wave Raman
scattering within  the $s$-parametrized quasidistribution
formalism in Section 5 and the density-matrix formalism in
Section 6. Particular attention has been paid to quantum
properties of light such as squeezing and sub-Poissonian
photon-counting statistics.

In Figs. 2-9 we compared various statistical moments obtained
(i) from numerical calculations utilizing the exact solutions
without the parametric approximation - (Eqs.~(\ref{167}),
(\ref{168}), and/or~(\ref{171}) derived in Appendix B), (ii)
from the short-time solutions of Sect. 6.1.1, (iii) from the
solutions obtained in Sect.  5.2 within the framework of the
FPE approach under the parametric approximation, and (iv)
from the  approximate solutions derived in Sect. 6.2 within
the density-matrix formalism.

In Figs. 2,3,7 and 8 we demonstrated that the initial and
short-time behavior of the approximate functions (curves B,C,
and D) is consistent with the exact evolution (curves A). We
have shown analytically that our expressions for the Stokes
scattering formalisms presented in Sects. 5.2, 6.1.1, 6.1.2,
6.2 are equivalent for short times and high initial
intensities of the pump field. Nevertheless, it is seen that
the equations derived in Sect. 6.1.1 give the best, whereas
those derived in Sect. 6.2 give the worst approximation to
the exact solution of Sect. 6.1.2 for small initial
intensities of the laser field.

We showed in Fig.4 (curve A) that the Stokes mode
photon-number fluctuations vary from initially chaotic to
Poissonian in asymptotics (for $\langle\hat{m}\rangle=0$ and
$\langle\hat{n}\rangle=|\alpha_L|^2$) or from Poissonian,
through super-Poissonian, to Poissonian for large times (if
$\langle\hat{m}\rangle=|\alpha_S|^2\neq 0$ and
$\langle\hat{n}\rangle=|\alpha_L|^2$). The asymptotic
behavior of $\gamma_S^{(2)}(\tau)$ is consistent with our
predictions in Sect. 6.1.3. The short time behavior of
$\gamma_S^{(2)}(\tau)$ for hyper-Raman scattering~\cite{r162}
is similar to that presented in Fig.4 for Raman scattering.
However, for long times the Stokes hyper-Raman photon-number
fluctuations become sub-Poissonian for reasons given in Sect.
6.1.3.

In Fig. 5 we demonstrated that the normalized factorial
moment for the laser mode, $\gamma_L^{(2)}(\tau)$, varies
from initially Poissonian to super-Poissonian. The
differences in $\gamma_L^{(2)}(\tau)$ between Fig. 5a
(spontaneous Stokes scattering) and Fig. 5b (stimulated
Stokes scattering) are only quantitative.  We note that for
hyper-Raman scattering  the photon-number fluctuations in the
initially coherent laser mode become
sub-Poissonian~\cite{r162}. The time behavior of the
interbeam degree of coherence was presented in Figs. 6a and
b. Curve A in Fig. 6a coincides with the Simaan exact
solution~\cite{r100}. Sub-Poissonian statistics in the
compound laser-Stokes mode is observed.

It is thought (see, for instance Ref.~\cite{r91}, p.~192)
that the Simaan approach~\cite{r100} is the most rigorous
application of master equations to the Raman problem.
However, Simaan's solution is restricted to the diagonal
matrix elements in number representation.  Only these terms
are needed to obtain the mean photon numbers and their higher
moments, whereby the photon correlation effects can be
investigated. To investigate squeezing properties and phase
correlations it is necessary to obtain the off-diagonal
elements of the density matrix. We generalized the solution
of the master equation obtained by McNeil and
Walls~\cite{r66} and Simaan~\cite{r100} to comprise all the
off-diagonal matrix elements as well. Our derivation of the
complete density matrix represents the main result of this
paper.

Our intention was to cite an extensive literature related to
our Raman scattering approaches. Nevertheless, we realize
that the cited literature is not complete. We include only
those references that are most relevant for the purposes of
our article.

{\bf Acknowledgments}. In the preparation of this paper we
have benefitted greatly from contacts with Ryszard Tana\'s,
Przemys{\l}aw Szlachetka, and Krzysztof Grygiel. Our warm
appreciation must also go to Jan Pe\v{r}ina.

\appendix

\section{Solution of generalized Fokker-Planck equation}

Here, we give a simple formal solution of the generalized FPE
(\ref{061}) for the quasidistribution ${\cal
W}^{(-1)}(\alpha_L, \alpha_S,\alpha_A,t)$  ($Q$-functions) as
well the corresponding characteristic function ${\cal
C}^{(-1)}(\beta_L ,\beta_S,\beta_A,t)$  - the solution of the
simplified equation of motion~(\ref{063}). We choose
antinormal order ($s=-1$) to avoid the problems of existence
of the quasiprobability distributions and to reduce the FPE
(\ref{061}) containing the third-order derivatives (for
$s\neq \pm 1$) to a second-order FPE. It is by no means easy
to find an exact solution of~(\ref{061}) or~(\ref{063}) even
for particular orders, because the drift coefficients are not
linear and the diffusion coefficients are not constant. An
often employed method to solve problems of this kind is to
assume that the fluctuations of the radiation fields are
small compared to their mean values; i.e., the
quasidistribution describing the fields is sharply
peaked~\cite{r44,r155,r79,r56,r291}. This will be the case
for suitably chosen input state and the initial output
states. Under these restrictions we can rewrite our
FPE~(\ref{061}) related to antinormal order in the linearized
form\begin{eqnarray} \lefteqn{ \frac{\partial}{\partial
t}{\cal W}^{(-1)}(\alpha_L,\alpha_S,\alpha_A,t)\:=\:
\frac{1}{2}\gamma_S \mbox{\huge \{ [} -(D_{LS}+\xi_L \xi_S)
\frac{\partial}{\partial
\alpha_L} \frac{\partial}{\partial \alpha_S} \null} \nonumber\\
&&\null +\langle\hat{n}_S\rangle\frac{\partial}{\partial
\alpha_L}\alpha_L
-(\langle\hat{n}_L\rangle-1)\frac{\partial}{\partial
\alpha_S}\alpha_S + c.c. \mbox{\huge ]} \nonumber\\ &&\null +2
\langle\hat{n}_S\rangle\frac{\partial}{\partial \alpha_L}
\frac{\partial}{\partial \alpha_L^*} \mbox{\huge \}} {\cal
W}^{(-1)}
\nonumber\\ &&\null +\frac{1}{2}\gamma_A \mbox{\huge \{ [}
-(D_{LA}+\xi_L\xi_A) \frac{\partial}{\partial \alpha_L}
\frac{\partial}{\partial \alpha_A} \nonumber\\ &&\null
-(\langle\hat{n}_A\rangle-1)\frac{\partial}{\partial
\alpha_L}\alpha_L +\langle\hat{n}_L\rangle\frac{\partial}{\partial
\alpha_A}\alpha_A + c.c. \mbox{\huge ]} \nonumber\\ &&\null +2
\langle\hat{n}_L\rangle\frac{\partial}{\partial \alpha_A}
\frac{\partial}{\partial \alpha_A^*} \mbox{\huge \}} {\cal
W}^{(-1)} \nonumber\\ &&\null
-\mbox{\huge \{ } \frac{1}{2}\gamma_{SA} e^{-2i\Delta\Omega\Delta
t} \mbox{\huge [} (C_{L}+\xi_L^2) \left( \alpha_A^*
\frac{\partial}{\partial \alpha_S} +\frac{\partial}{\partial
\alpha_S} \frac{\partial}{\partial \alpha_A} -\alpha_S^*
\frac{\partial}{\partial \alpha_A} \right) \nonumber\\ &&\null
-2(\bar{D}_{SL}-\xi_L\xi_S^*) \frac{\partial}{\partial \alpha_L^*}
\frac{\partial}{\partial \alpha_A} -(D^*_{SA}+\xi_S^*\xi_A^*)
\frac{\partial^2}{\partial \alpha_L^{*2}} \mbox{\huge ]} + c.c.
\mbox{\huge \}} {\cal W}^{(-1)}
\nonumber\\ &&\null + \gamma_S \langle\hat{n}_V\rangle \mbox{\huge
\{} \left( \frac{1}{2}\frac{\partial}{\partial \alpha_L}\alpha_L
-(D_{LS}+\xi_L\xi_S) \frac{\partial}{\partial \alpha_L}
\frac{\partial}{\partial \alpha_S} +\frac{1}{2}
\frac{\partial}{\partial \alpha_S} \alpha_S + c.c. \right)
\nonumber\\ &&\null +\langle\hat{n}_S\rangle
\frac{\partial}{\partial \alpha_L} \frac{\partial}{\partial
\alpha_L^*} +\langle\hat{n}_L\rangle \frac{\partial}{\partial
\alpha_S} \frac{\partial}{\partial \alpha_S^*} \mbox{\huge \}}
{\cal W}^{(-1)}
\nonumber\\ &&\null + \gamma_A \langle\hat{n}_V\rangle \mbox{\huge
\{} \left( \frac{1}{2}\frac{\partial}{\partial \alpha_L}\alpha_L
-(D_{LA}+\xi_L\xi_A) \frac{\partial}{\partial \alpha_L}
\frac{\partial}{\partial \alpha_A} +\frac{1}{2}
\frac{\partial}{\partial \alpha_A} \alpha_A + c.c. \right)
\nonumber\\ &&\null +\langle\hat{n}_A\rangle
\frac{\partial}{\partial \alpha_L} \frac{\partial}{\partial
\alpha_L^*} +\langle\hat{n}_L\rangle \frac{\partial}{\partial
\alpha_A} \frac{\partial}{\partial \alpha_A^*} \mbox{\huge \}}
{\cal W}^{(-1)}
\nonumber\\ &&\null -\mbox{\huge \{} \gamma_{AS}
\langle\hat{n}_V\rangle e^{2i\Delta\Omega\Delta t} \mbox{\Large (}
(D_{SA}^*+\xi_S^*\xi_A^*) \frac{\partial^2}{\partial
\alpha_L^{*2}} +(C_L+\xi_L^2) \frac{\partial}{\partial \alpha_S}
\frac{\partial}{\partial \alpha_A} \nonumber\\ &&\null
+(\bar{D}_{AL}-\xi_L\xi_A^*) \frac{\partial}{\partial \alpha_L^*}
\frac{\partial}{\partial \alpha_S} +(\bar{D}_{SL}-\xi_L\xi_S^*)
\frac{\partial}{\partial \alpha_L^*} \frac{\partial}{\partial
\alpha_A} \mbox{\Large )} + c.c. \mbox{\huge \}} {\cal W}^{(-1)},
\nonumber\\ \label{a01}
\end{eqnarray}
where the coefficients $D_{kl}$, $\hat{D}_{kl}$, $C_{k}$,
$\xi_{k}$ for $k,l=L,S,A$ are defined by~(\ref{066}) at the
initial moment $t_0$. Equation~(\ref{a01}) is the
generalization of the FPE given in Ref.~\cite{r68} for the
case of nonzero $\gamma_S$ and $\gamma_{AS}$. It is seen that
the Raman effect under the approximations applied can be
treated as an Ornstein-Uhlenbeck process~\cite{r296}.  The
FPE~(\ref{a01}) can be solved exactly by various techniques;
see, e.g., Ref.~\cite{r155}. For instance, using the inverse
Fourier transform~(\ref{022}), one can transform the FPE
(\ref{a01}) into the corresponding equation of motion for the
characteristic function ${\cal
C}^{(-1)}(\beta_L,\beta_S,\beta_A,t)$, which is a first-order
differential equation. The method of characteristics applied
to the latter equation leads to the solution
\begin{eqnarray}
\lefteqn{ {\cal C}^{(-1)}(\beta_L,\beta_S,\beta_A,t)= \left<
\exp\mbox{\Large \{} \right. -\sum_{k=L,S,A} \mbox{\Large [}
B_{k}^{(-1)}(t)|\beta_k|^2 \null} \nonumber\\ && +\left(
\frac{1}{2}C^*_k(t)\beta_k^2+c.c.\right)
+\left(\beta_k\xi_{k}^{*}(t)-c.c.\right) \mbox{\Large ]}
\nonumber\\ && +\mbox{\Large [}
D_{LS}(t)\beta_L^*\beta_S^*+\bar{D}_{LS}\beta_L\beta_S^*
\nonumber\\ &&\null
+D_{LA}(t)\beta_L^*\beta_A^*+\bar{D}_{LA}\beta_L\beta_A^*
\nonumber\\ &&\null \left.
+D_{SA}(t)\beta_S^*\beta_A^*+\bar{D}_{SA}\beta_S\beta_A^* +c.c.
\mbox{\Large ]} \mbox{\Large \}} \right>. \label{a02}
\end{eqnarray}
The angle brackets mean averaging over the complex amplitudes
$\xi_k$ ($k=L,S,A$) with the initial distribution ${\cal
W}^{(-1)}(\alpha_L, \alpha_S, \alpha_A, t_0)$. They represent
the influence of the initial photon statistics of the  pump
and scattered fields on the evolution of the system.  The
solution of Eq.~(\ref{a01}) can be readily obtained by
applying the Fourier transform~(\ref{021}) to
solution~(\ref{a02}), and has the form of a shifted
seven-dimensional (including time) Gaussian distribution
involving correlation between the radiation fields,
\begin{eqnarray}
\lefteqn{ {\cal W}^{(-1)}(\alpha_1,\alpha_2,\alpha_3,t)\:=\:
\left< \frac{1}{L^{(-1)}} \exp \mbox{\huge \{}
-(L^{(-1)})^{-2}\sum_{j=1}^{3} \right. \null} \nonumber\\ &&\null
\left[ |\alpha_j-\xi_j(t)|^2 E_j^{(-1)} +\frac{1}{2} \left(
\left(\alpha_j^*-\xi_j^*(t)\right)^2 E_{j+3}^{(-1)} + c.c. \right)
\right] \nonumber\\ &&\null
+(L^{(-1)})^{-2}\sum_{j=1}^{2}\sum_{k=j+1}^{3} \mbox{ \large [}
(\alpha_j^*-\xi_j^*(t)) (\alpha_k^*-\xi_k^*(t)) E_{j+k+4}^{(-1)}
\nonumber\\ &&\null \left. + (\alpha_j-\xi_j(t))
(\alpha_k^*-\xi_k^*(t)) E_{j+k+7}^{(-1)} \mbox{\large ]} + c.c.
\mbox{\huge \}} \right>, \label{a03}
\end{eqnarray}
where, for simplicity, we have identified the subscripts in
$\alpha_1=\alpha_L$, $\alpha_2 = \alpha_S$, $\alpha_3 =
\alpha_A$, $\xi_1=\xi_L$, $\xi_2=\xi_S$, $\xi_3=\xi_A$. The
functions $E_1^{(-1)}$, ..., $E_{12}^{(-1)}$, and $L^{(-1)}$,
which are time-dependent, are connected with the functions
$B^{(-1)}_k$, $C_k$, $D_{kl}$, $\bar{D}_{kl}$ appearing in
(\ref{a02}) in a manner similar to~(\ref{074}) and
(\ref{075}), respectively.  We do not adduce explicit
formulas for the coefficients listed, since solutions
(\ref{a02}) and~(\ref{a03}) serve only as an example of how
one can deal with Eqs.~(\ref{061}) and~(\ref{063}). The
validity of solutions~(\ref{a02}) and~(\ref{a03}) is
restricted by strong approximations, which are actually
equivalent to the parametric approximation and the short-time
approximation.

\section{Solution of master equation in Fock basis}

Here, we solve the equation of motion~(\ref{125}). To
eliminate the square root appearing in Eq.~(\ref{125}) for
off-diagonal terms, it is convenient to introduce the
transformation
\begin{eqnarray}
\psi_{nm}(\nu\mu\tau) = \left[ \frac{n!(n+\nu)!}{m!(m+\mu)!}
\right]^{1/2} \rho_{nm}(\nu\mu\tau), \label{b01}
\end{eqnarray}
where the degree of off-diagonality $\mu$ is restricted to
nonnegative integers, whereas the degree $\nu$ is $\ge -m$.
On insertion of~(\ref{b01}) into~(\ref{125}), the equation of
motion for the transformed matrix elements
$\psi_{nm}(\nu\mu\tau)$ takes the form
\begin{eqnarray}
\psi_{nm}(\nu\mu\tau) &=& -\frac{1}{2} [n(m+1)+(n+\nu)(m+\mu+1)]
\psi_{nm}(\nu\mu\tau) \nonumber \\ &&+ \psi_{n+l,m-l}(\nu\mu\tau).
\label{b02}
\end{eqnarray}
We apply the Laplace transform method to~(\ref{b02}), which
readily leads to the solution
\begin{eqnarray}
{\overline \psi}_{nm}(\nu\mu s) = \sum_{l=0}^{m}
{\psi}_{n+l,m-l}(\nu\mu\tau_0) \prod_{p=0}^{l}[s+f(p)]^{-1}
\label{b03}
\end{eqnarray}
for ${\overline \psi}_{nm}(\nu\mu\tau)$, the Laplace
transform of $\psi_{nm} (\nu\mu\tau)$. The function $f(p)$
occurring in~(\ref{b03}) is given by~(\ref{169}). If there
are no equal terms among the elements of the set
$f(0),f(1),...,f(l)$ the inverse transform, after retaining
the $\rho_{nm}(\nu\mu\tau)$ notation, yields~(\ref{167}).  If
there are repeated elements in the denominator of
(\ref{b03}), the inverse transforms will involve
convolutions. We apply two general procedures essentially
equivalent to that of Simaan~\cite{r100} and
Malakyan~\cite{r162}. It is convenient to split~(\ref{b03})
into two terms as follows:
\begin{eqnarray}
{\overline \psi}_{nm}(\nu\mu s) &=& \sum_{l=0}^{\lambda}
{\psi}_{n+l,m-l}(\nu\mu\tau_0) \prod_{p=0}^{l}[s+f(p)]^{-1}
\nonumber\\ &&+ (1-\delta_{m0}) \sum_{l=\lambda+1}^{m}
\psi_{n+l,m-l}(\nu\mu\tau_0) \nonumber\\ &&\times
\prod_{p=0}^{\lambda}[s+f(p)]^{-1}
\prod_{p'=\lambda+1}^{l}[s+f(p')]^{-1} \label{b04}
\end{eqnarray}
with $\lambda$ defined by~(\ref{166}) (or equivalently by
(\ref{170})).  Let us note that a parabola $f(q)$=const takes
its maximum value for $q_0=(2m-2n+\mu-\nu+2)/4$.  This value,
$q_0$ or better $\lambda$, the maximum integer $\leq q_0$,
can serve as a criterion to split~(\ref{b03}) in such a
manner that a convolution theorem can be easily applied.  The
first term in~(\ref{b04}) has no mutually equal factors in
the denominator, so the inverse Laplace transform has the
form of~(\ref{167}) with the proper upper limit of summation.
The denominator of the second term of~(\ref{b04}) contains
repeated factors, which are separated, so that we can readily
apply the convolution theorem finally obtaining the solution
(\ref{168}). Equations~(\ref{167}) and~(\ref{168}) have a
rather complicated structure. We can rewrite~(\ref{167}) and
(\ref{168}) in a more compact form.  If we assume that there
is only one pair of equal factors among the elements of the
set $f(0),f(1),...,f(l)$, i.e., if
\begin{eqnarray*}
\bigvee_{q_1\neq q_1' \atop q_1,q_1' \in \{0,...,l\}}
f(q_1)\:=\:f(q_1')
\bigwedge_{q=0,...,l \atop q \neq q_1,q_1'}
f(q_1) \neq f(q),
\end{eqnarray*}
then we can express the solution~(\ref{b03}) for ${\overline
\psi}_{nm}(\nu\mu\tau)$ as
\begin{eqnarray}
{\overline \psi}_{nm}(\nu\mu s) = \sum_{l=0}^{m}
{\psi}_{n+l,m-l}(\nu\mu\tau_0) [s+f(q_1)]^{-2} \prod_{p=0 \atop
p\neq q_1,q_1'}^{l}[s+f(p)]^{-1}. \label{b05}
\end{eqnarray}
The inverse transform of~(\ref{b05}) is
\begin{eqnarray}
{\psi}_{nm}(\nu\mu \tau) &=& \sum_{l=0}^{m}
{\psi}_{n+l,m-l}(\nu\mu\tau_0) \nonumber\\ &&\times \mbox{\huge
\{} \sum_{q=0 \atop q\neq q_1,q_1'}^{l} \exp[-f(q) \Delta\tau]
\prod_{p=0 \atop p\neq q}^{l} [f(p)-f(q)]^{-1} \nonumber\\ &&+
\mbox{\huge (} \Delta\tau-\sum_{k=0 \atop k\neq q_1,q_1'}^{l}
[f(k)-f(q_1)]^{-1} \mbox{\huge )} \exp[-f(q_1) \Delta\tau]
\nonumber\\ &&\times \prod_{p=0 \atop p\neq q_1,q_1'}^{l}
[f(p)-f(q_1)]^{-1} \mbox{\huge \}}, \label{b06}
\end{eqnarray}
which is a derivative of the solution~(\ref{167}) over
$f(q_1)$ with extra minus,
\begin{eqnarray}
{\psi}_{nm}(\nu\mu \tau) = \sum_{l=0}^{m}
{\psi}_{n+l,m-l}(\nu\mu\tau_0) \hspace{4cm} \nonumber\\ \times
\left(- \frac{\partial}{\partial f(q_1)}\right) \sum_{q=0 \atop
q\neq q_1'}^{l} \exp[-f(q) \Delta\tau] \prod_{p=0 \atop p\neq
q,q_1'}^{l} [f(p)-f(q)]^{-1}. \label{b07}
\end{eqnarray}
In the case of $d$ equal pairs, i.e.,
$f(q_1)=f(q_1'),...,f(q_d)=f(q_d')$, the Laplace transform
solution~(\ref{b03}) can be rewritten as
\begin{eqnarray}
{\overline \psi}_{nm}(\nu\mu s) &=& \sum_{l=0}^{m}
{\psi}_{n+l,m-l}(\nu\mu\tau_0) \nonumber\\ &&\times
\prod_{r=1}^{d}[s+f(q_r)]^{-2} \prod_{p=0 \atop p\neq
q_1,q_1',...,q_d,q_d'}^{l}[s+f(p)]^{-1}, \label{b08}
\end{eqnarray}
finally leading to
\begin{eqnarray}
{\psi}_{nm}(\nu\mu \tau) = \sum_{l=0}^{m}
{\psi}_{n+l,m-l}(\nu\mu\tau_0) \left((-1)^d \prod_{r=1}^{d}
\frac{\partial}{\partial f(q_r)}\right) \hspace{2cm} \nonumber\\
\times \sum_{q=0 \atop q\neq q_1',...,q_d'}^{l} \exp[-f(q)
\Delta\tau] \prod_{p=0 \atop p\neq q,q_1',...,q_d'}^{l}
[f(p)-f(q)]^{-1}, \label{b09}
\end{eqnarray}
or equivalently to the solution~(\ref{171}) with the $d$-th
order differential operator $\hat{\cal D}$~(\ref{172}).

\section{Solution for anti-Stokes scattering}

In Section 6 we have given an analysis of Stokes scattering.
For completeness, in this Appendix, we present the solution
describing the anti-Stokes effect including laser depletion,
but neglecting the Stokes generation and assuming that the
reservoir is `quiet', i.e.,
\begin{eqnarray}
\gamma_S \,=\, \gamma_{AS} \,=\, \gamma_{SA} &=& 0,
\nonumber\\
\langle \hat{n}_V\rangle = 0. \label{c01}
\end{eqnarray}
Under these conditions the master equation~(\ref{018}) in
Fock representation is
\begin{eqnarray}
\frac{\partial }{\partial \tau} \rho_{nm}(\nu\mu\tau)&=&
-\frac{1}{2} [(n+1)m+(n+\nu+1)(m+\mu)] \rho_{nm}(\nu\mu\tau)
\nonumber\\ &&+ [n(n+\nu)(m+1)(m+\mu+1)]^{1/2}
\rho_{n-1,m+1}(\nu\mu\tau), \label{c02}
\end{eqnarray}
where, for brevity, we have set $n_L=n$, $n_L'=n+\nu$ (as in Section
6) and $n_A=m, n_A'=m+\mu$. If we define $\lambda$ as follows:
\begin{eqnarray}
\lambda = \left[\left[ \frac{n-m+1}{2}+\frac{\nu-\mu}{4}
\right]\right], \label{c03}
\end{eqnarray}
then the solution of~(\ref{c02}) for $\lambda<0$ becomes
\begin{eqnarray}
\rho_{nm}(\nu\mu\tau) &=& \left[
\frac{n!(n+\nu)!}{m!(m+\mu)!}\right]^{1/2} \sum_{l=0}^{n}
\rho_{n-l}^{L}(\nu\tau_0) \rho_{m+l}^{A}(\mu\tau_0) \nonumber\\
&&\times \left[
\frac{(m+l)!(m+l+\mu)!}{(n-l)!(n-l+\nu)!}\right]^{1/2} \nonumber\\
&&\times \sum_{q=0}^{l} \exp [-g(q)\Delta\tau] \prod_{p=0\atop
p\neq q}^{l} [g(p)-g(q)]^{-1}, \label{c04}
\end{eqnarray}
whereas for $\lambda\ge0$ it becomes
\begin{eqnarray}
\rho_{nm}(\nu\mu\tau) &=& \left[
\frac{n!(n+\nu)!}{m!(m+\mu)!}\right]^{1/2} \left\{
\sum_{l=0}^{\lambda} \rho_{n-l}^{L}(\nu\tau_0)
\rho_{m+l}^{A}(\mu\tau_0) \right. \nonumber\\ &&\times \left[
\frac{(m+l)!(m+l+\mu)!}{(n-l)!(n-l+\nu)!}\right]^{1/2} \nonumber\\
&&\times \sum_{q=0}^{l} \exp [-g(q)\Delta\tau] \prod_{p=0\atop
p\neq q}^{l} [g(p)-g(q)]^{-1} \nonumber\\ &&+ (1-\delta_{n0})
\sum_{l=\lambda+1}^{n} \rho_{n-l}^{L}(\nu\tau_0)
\rho_{m+l}^{A}(\mu\tau_0) \nonumber\\ &&\times \left[
\frac{(m+l)!(m+l+\mu)!}{(n-l)!(n-l+\nu)!}\right]^{1/2} \nonumber\\
&&\times \sum_{q=0}^{\lambda}\sum_{q'=\lambda+1}^{l}
\prod_{p=0\atop p\neq q}^{\lambda} [g(p)-g(q)]^{-1}
\prod_{p'=\lambda+1\atop p'\neq q'}^{l} [g(p')-g(q')]^{-1}
\nonumber\\ &&\times \mbox{\huge (}
\delta_{g(q)g(q')}\Delta\tau\exp[-g(q)\Delta\tau] \nonumber\\ &&+
\left. (\delta_{g(q)g(q')}-1)
\frac{\exp[-g(q)\Delta\tau]-\exp[-g(q')\Delta\tau]}{g(q)-g(q')}
\mbox{\huge )} \right\} \label{c05}
\end{eqnarray}
with
\begin{eqnarray}
g(x) = \frac{1}{2} [(m+x)(n-x+1)+(m+x+\mu)(n-x+\nu+1]. \label{c06}
\end{eqnarray}
Alternatively, we can express solution~(\ref{c04}) and
(\ref{c05}) as
\begin{eqnarray}
\rho_{nm}(\nu\mu\tau) &=& \left[
\frac{n!(n+\nu)!}{m!(m+\mu)!}\right]^{1/2} \sum_{l=0}^{n}
\rho_{n-l}^{L}(\nu\tau_0) \rho_{m+l}^{A}(\mu\tau_0) \nonumber\\
&&\times \left[
\frac{(m+l)!(m+l+\mu)!}{(n-l)!(n-l+\nu)!}\right]^{1/2} \nonumber\\
&&\times \hat{\cal D} \sum_{q=0\atop q\neq q'_1,q'_2,...,q'_d}^{l}
\exp [-g(q)\Delta\tau] \prod_{p=0\atop p\neq
q,q'_1,q'_2,...,q'_d}^{l} [g(p)-g(q)]^{-1},\nonumber\\ \label{c07}
\end{eqnarray}
using the differentiation operator $\hat{\cal D}$ given by
(\ref{172}).


\end{document}